\documentclass[prx,twocolumn,amsmath,amssymb,floatfix,superscriptaddress, longbibliography]{revtex4-1}

\usepackage[pdftex]{graphicx} 
\usepackage{booktabs}
\usepackage{color}
\usepackage[version=4]{mhchem}
\usepackage{xspace, bm, braket}
\usepackage{siunitx}
\usepackage[colorlinks, citecolor=blue, linkcolor=blue, urlcolor=blue, breaklinks]{hyperref}
\usepackage[svgnames]{xcolor}

\newif\ifshowcomments\showcommentstrue
\usepackage[final]{changes}
\definechangesauthor[color=Red]{2nd}

\usepackage{verbatim} 

\newcommand{\SRO}{\ce{Sr2RuO4}}

\newcommand{\etal}{\textit{et al.}}

\newcommand{\bk}{{\bm k}}
\newcommand{\Hzero}{$\hat{H}^0$}
\newcommand{\Hdft}{$\hat{H}^{\mathrm{DFT}}$}

\newcommand{\Hsoclam}{$\hat{H}^{\mathrm{SOC}}_{\lambda}$}
\newcommand{\Hsocdft}{$\hat{H}^{\mathrm{SOC}}_{\lambda_\mathrm{DFT}}$}
\newcommand{\Hsocdlam}{$\hat{H}^{\mathrm{SOC}}_{\lambda_\mathrm{DFT}+\Delta\lambda}$}

\newcommand{\Dcf}{$\Delta \epsilon_{\mathrm{cf}}$}
\newcommand{\epscf}{$\epsilon_{\mathrm{cf}}$}

\newcommand{\leff}{$\lambda_{\mathrm{eff}}$}

\newcommand{\psiknu}{\psi_\nu\left({\bk}\right)}

\newcommand{\chikm}{\chi_m\left(\bk\right)}

\newcommand{\epsknu}{\varepsilon_\nu\left(\bk\right)}

\newcommand{\nup}{\nu^\prime}
\newcommand{\omqp}{\omega^{\mathrm{qp}}}
\newcommand{\kmaxnuom}{\bk_{\mathrm{max}}^{\nu}(\omega)}

\def\beq{\begin{equation}}
\def\eeq{\end{equation}}

\begin{document}

\title{High-resolution photoemission on \SRO{} reveals correlation-enhanced effective spin-orbit coupling 
and dominantly local self-energies}

\author{A. Tamai}
\affiliation{Department of Quantum Matter Physics, University of Geneva, 24 Quai Ernest-Ansermet, 1211 Geneva 4, Switzerland}
\author{M. Zingl}
\affiliation{Center for Computational Quantum Physics, Flatiron Institute, 162 5th Avenue, New York, NY 10010, USA}
\author{E. Rozbicki}
\affiliation{SUPA, School of Physics and Astronomy, University of St Andrews, St Andrews, Fife KY16 9SS, United Kingdom}
\author{E. Cappelli}
\affiliation{Department of Quantum Matter Physics, University of Geneva, 24 Quai Ernest-Ansermet, 1211 Geneva 4, Switzerland}
\author{S. Ricc\`o}
\affiliation{Department of Quantum Matter Physics, University of Geneva, 24 Quai Ernest-Ansermet, 1211 Geneva 4, Switzerland}
\author{A. de la Torre}
\affiliation{Department of Quantum Matter Physics, University of Geneva, 24 Quai Ernest-Ansermet, 1211 Geneva 4, Switzerland}
\author{S. McKeown-Walker}
\affiliation{Department of Quantum Matter Physics, University of Geneva, 24 Quai Ernest-Ansermet, 1211 Geneva 4, Switzerland}
\author{F. Y. Bruno}
\affiliation{Department of Quantum Matter Physics, University of Geneva, 24 Quai Ernest-Ansermet, 1211 Geneva 4, Switzerland}
\author{P.D.C. King}
\affiliation{SUPA, School of Physics and Astronomy, University of St Andrews, St Andrews, Fife KY16 9SS, United Kingdom}
\author{W. Meevasana}
\affiliation{School of Physics, Suranaree University of Technology and Synchrotron Light Research Institute, Nakhon Ratchasima, 30000, Thailand and Thailand Center of Excellence in Physics, CHE, Bangkok, 10400, Thailand}
\author{M. Shi}
\affiliation{Swiss Light Source, Paul Scherrer Institut, CH-5232 Villigen PSI, Switzerland}
\author{M. Radovi\'{c}}
\affiliation{Swiss Light Source, Paul Scherrer Institut, CH-5232 Villigen PSI, Switzerland}
\author{N.C. Plumb}
\affiliation{Swiss Light Source, Paul Scherrer Institut, CH-5232 Villigen PSI, Switzerland}
\author{A.S. Gibbs}
\altaffiliation[present address: ]{ISIS Facility, Rutherford Appleton Laboratory, Chilton, Didcot OX11 OQX, United Kingdom}
\affiliation{SUPA, School of Physics and Astronomy, University of St Andrews, St Andrews, Fife KY16 9SS, United Kingdom}
\author{A.P. Mackenzie}
\affiliation{Max Planck Institute for Chemical Physics of Solids, D-01187 Dresden, Germany}
\affiliation{SUPA, School of Physics and Astronomy, University of St Andrews, St Andrews, Fife KY16 9SS, United Kingdom}
\author{C. Berthod}
\affiliation{Department of Quantum Matter Physics, University of Geneva, 24 Quai Ernest-Ansermet, 1211 Geneva 4, Switzerland}
\author{H. Strand}
\affiliation{Center for Computational Quantum Physics, Flatiron Institute, 162 5th Avenue, New York, NY 10010, USA}
\author{M. Kim}
\affiliation{Department of Physics and Astronomy, Rutgers, The State University of New Jersey, Piscataway, NJ 08854, USA}
\affiliation{Centre de Physique Th\'{e}orique Ecole Polytechnique, CNRS, Universite Paris-Saclay, 91128 Palaiseau, France}
\author{A. Georges}
\affiliation{Coll\`ege de France, 11 place Marcelin Berthelot, 75005 Paris, France}
\affiliation{Center for Computational Quantum Physics, Flatiron Institute, 162 5th Avenue, New York, NY 10010, USA}
\affiliation{Centre de Physique Th\'{e}orique Ecole Polytechnique, CNRS, Universite Paris-Saclay, 91128 Palaiseau, France}
\affiliation{Department of Quantum Matter Physics, University of Geneva, 24 Quai Ernest-Ansermet, 1211 Geneva 4, Switzerland}
\author{F. Baumberger}
\affiliation{Department of Quantum Matter Physics, University of Geneva, 24 Quai Ernest-Ansermet, 1211 Geneva 4, Switzerland}
\affiliation{Swiss Light Source, Paul Scherrer Institut, CH-5232 Villigen PSI, Switzerland}

\date{\today}

\begin{abstract}
We explore the interplay of electron-electron correlations and spin-orbit coupling in the model Fermi liquid \SRO{}
using laser-based angle-resolved photoemission spectroscopy.
Our precise measurement of the Fermi surface confirms the importance of spin-orbit
coupling in this material and reveals that its effective value is enhanced
by a factor of about two, due to electronic correlations.
The self-energies for the $\beta$ and $\gamma$ sheets are found to display significant angular dependence. 
By taking into account the multi-orbital composition of quasiparticle states, we 
determine self-energies associated with each orbital component directly from the experimental data. 
This analysis demonstrates that the perceived angular dependence does not imply momentum-dependent 
many-body effects, but arises from a substantial orbital mixing induced by spin-orbit coupling. 
A comparison to single-site dynamical mean-field theory further supports the notion of dominantly local orbital self-energies,  
and provides strong evidence for an electronic origin of the observed non-linear frequency dependence of the 
self-energies, leading to `kinks' in the quasiparticle dispersion of \SRO{}.
\end{abstract}

\maketitle

\section{Introduction}

The layered perovskite \SRO{} is an important model system for correlated electron physics. Its intriguing superconducting ground state, sharing similarities with superfluid $^{3}$He~\cite{Maeno1994,Rice1995,Mackenzie2003a}, has attracted much interest and continues to stimulate advances in unconventional superconductivity~\cite{Mackenzie2017}. Experimental evidence suggest odd-parity spin-triplet pairing, yet questions regarding the proximity of other order parameters, the nature of the pairing mechanism and the apparent absence of the predicted edge currents remain open~\cite{Ishida1998,Anwar2016,Hicks2014,Scaffidi2014,Kirtley2007,Mackenzie2003a,Mackenzie2017}.
Meanwhile, the normal state of  \SRO{} attracts interest as one of the cleanest oxide Fermi liquids~\cite{Mackenzie1996b,Maeno1997,Bergemann2003,Stricker2014}. 
Its precise experimental characterization is equally important for understanding the unconventional superconducting ground state of \SRO~\cite{Maeno1994,Rice1995,Mackenzie2003a,Mackenzie2017,Ishida1998,Anwar2016,Hicks2014,Scaffidi2014,Kirtley2007,Raghu2010,Huo2013,Scaffidi2014,Komendova2017,Steppke2017}, as it is for benchmarking quantitative many-body calculations~\cite{Georges2013,Liebsch2000,Mravlje2011,Zhang2016,Mravlje2016,Kim2018}.

Transport, thermodynamic and optical data of \SRO{} display textbook Fermi-liquid behavior below a crossover temperature of $T_{\mathrm{FL}}\approx25$~K~\cite{Mackenzie1996b,Maeno1997,Bergemann2003,Stricker2014}.
Quantum oscillation and angle-resolved photoemission spectroscopy (ARPES) measurements~\cite{Oguchi1995,Mackenzie1996,Mackenzie1998,Bergemann2000,Damascelli2000,Iwasawa2005,Ingle2005,Iwasawa2010,Iwasawa2012,Zabolotnyy2013,Burganov2016} further reported a strong enhancement of the quasiparticle effective mass over the bare band mass. 
Theoretical progress has been made recently in revealing the important role of the intra-atomic Hund's coupling as a key source of correlation effects in \SRO~\cite{Mravlje2011,demedici_2011,Georges2013}. In this context, much attention was devoted to the intriguing properties of the unusual state above $T_{\mathrm{FL}}$, which displays metallic transport with no signs of resistivity saturation at the Mott-Ioffe-Regel limit~\cite{tyler_1998}.
Dynamical mean-field theory (DMFT)~\cite{georges_1996} calculations have proven successful in explaining several properties of this intriguing metallic state, as well as in elucidating the crossover from this unusual metallic state into the Fermi-liquid regime~\cite{Mravlje2011,Georges2013,Stricker2014,Mravlje2016,deng_2016,Zhang2016,Kim2018}. 
Within DMFT, the self-energies associated with each orbital component are assumed to be local. 
On the other hand, the low-temperature Fermi-liquid state is known to display strong magnetic fluctuations at specific wave-vectors, as revealed, e.g., by neutron scattering~\cite{Sidis1999,steffens_2018} and nuclear magnetic resonance spectroscopy (NMR)~\cite{ishida_2001,imai_1998}. 
These magnetic fluctuations were proposed early on to be an important source of correlations~\cite{Rice1995,lou_2003,kim_mazin_2017}. 
In this picture, it is natural to expect strong momentum dependence of the self-energy associated with these spin fluctuations.
Interestingly, a similar debate was raised long ago in the context of liquid $^3$He, with `paramagnon' theories emphasizing ferromagnetic spin-fluctuations and `quasi-localized' approaches \`a la Anderson-Brinkman emphasizing local correlations associated with the strong repulsive hard-core, leading to increasing Mott-like localization as the liquid is brought closer to solidification 
(for a review, see Ref.~\cite{vollhardt_1984}). 

In this work, we report on new insights into the nature of the Fermi-liquid state of \SRO.
Analyzing a comprehensive set of laser-based ARPES data with improved resolution and cleanliness, we reveal a strong angular (i.e., momentum) dependence of the self-energies associated with the quasiparticle bands. We demonstrate that this angular dependence originates in the variation of the orbital content of quasiparticle states as a function of angle, and can be understood quantitatively. 
Introducing a new framework for the analysis of ARPES data for multi-orbital systems, we extract the electronic self-energies associated with the three Ruthenium $t_{2g}$ orbitals with minimal theoretical input. 
We find that these orbital self-energies have strong frequency dependence, but surprisingly weak angular (i.e., momentum) 
dependence, and can thus be considered local to a very good approximation.  
Our results provide a direct experimental demonstration that the dominant effects of correlations in \SRO{}
are weakly momentum-dependent and can be understood from a local perspective, provided they 
are considered in relation to orbital degrees of freedom. 
One of the novel aspect of our work is to directly put the locality ansatz underlying DMFT to the experimental test. 
We also perform a direct comparison between DMFT calculations and our ARPES data, and find 
good agreement with the measured 
quasiparticle dispersions and angular dependence of the effective masses.

The experimentally determined real part of the self-energy displays
strong deviations from the low-energy Fermi-liquid behavior $\Sigma^\prime(\omega) \sim \omega (1-1/Z) + \cdots$ for binding energies $|\omega|$ larger than $\sim\SI{20}{meV}$. These deviations 
are reproduced by our DMFT calculations suggesting that the cause of these non-linearities are local electronic correlations. 
Our results thus call for a revision of earlier reports of strong electron-lattice coupling in \SRO~\cite{Aiura2004,Iwasawa2005,Ingle2005,Iwasawa2010,Kim2011,Iwasawa2013,Wang2017,Shuntaro2019}. We finally quantify the effective spin-orbit coupling (SOC) strength and confirm its enhancement due to correlations predicted theoretically~\cite{liu_prl_2008,Zhang2016,Kim2018}.

This article is organized as follows. In Sec.~\ref{sec:exp}, we briefly present the experimental method and 
report our main ARPES results for the Fermi surface and quasiparticle dispersions. In Sec.~\ref{subsec:theo}, we 
introduce the theoretical framework on which our data analysis is based. In Sec.~\ref{sec:fs}, we use our precise 
determination of the Fermi surface to reveal the correlation-induced enhancement of the effective SOC. 
In Sec.~\ref{sec:self} we proceed with a direct determination of the self-energies from the ARPES data. 
Sec.~\ref{sec:dmft} presents the DMFT calculations in comparison to experiments. Finally, our results are critically 
discussed and put in perspective in Secs.~\ref{sec:kinks}, \ref{sec:perspectives}.

\begin{figure*}[tb]
	\includegraphics[width=0.92 \textwidth]{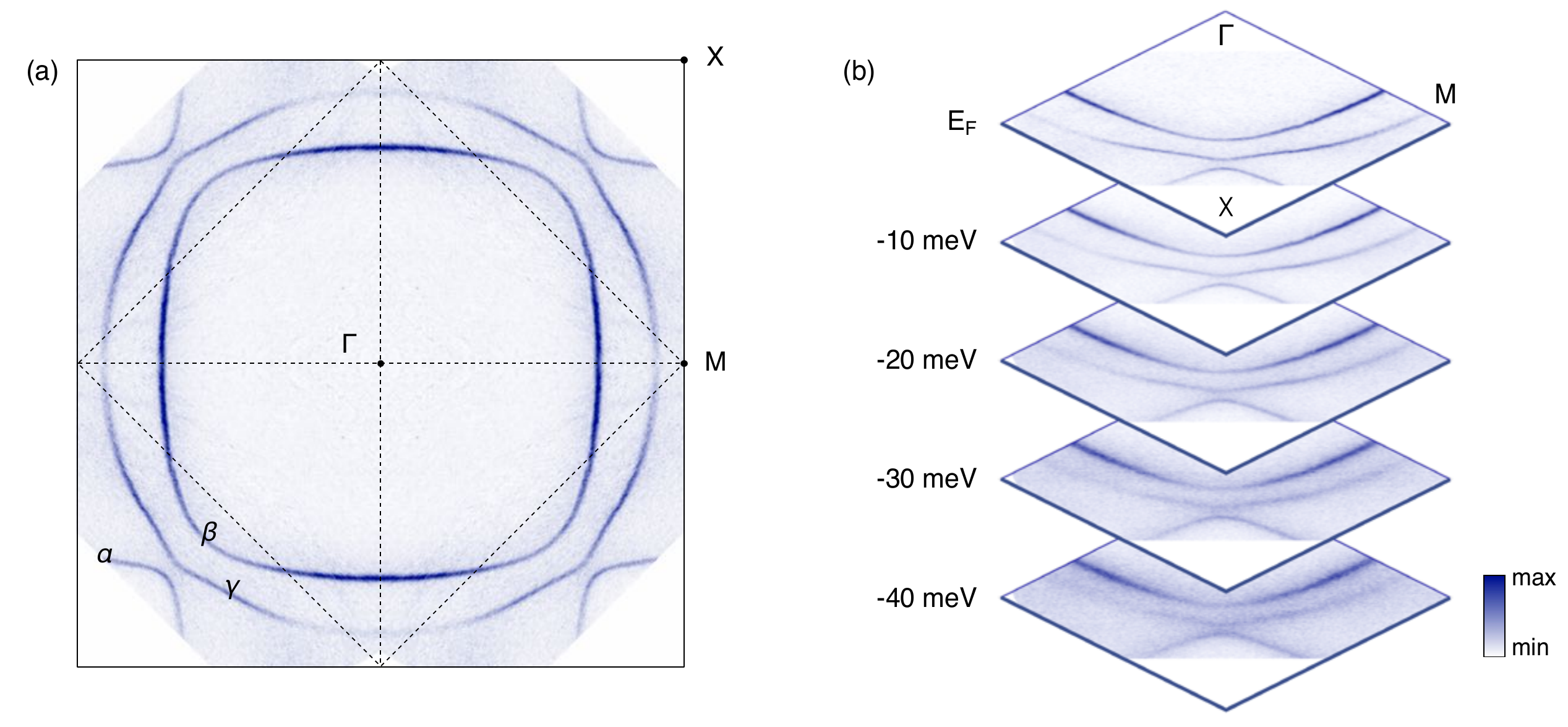} %
	\caption{
		(a) Fermi surface of \SRO. The data were acquired at 5~K on a CO passivated surface with a photon energy of 11~eV and $p$-polarization for measurements along the $\Gamma$X symmetry line. The sample tilt around the $\Gamma$X axis used to measure the full Fermi surface results in a mixed polarization for data away from this symmetry axis. The Brillouin zone of the reconstructed surface layer is indicated by diagonal dashed lines. Surface states and final state umklapp processes are suppressed to near the detection limit. A comparison with ARPES data from a pristine cleave is shown in appendix~\ref{sec:appA}. The data in (a) have been mirror-symmetrized for clarity. \added{Original measured data span slightly more than a quadrant of the Brillouin zone}.
		(b) Constant energy surfaces illustrating the progressive broadening of the quasiparticle states away from the Fermi level $E_F$.}
	\label{fig:f1}
\end{figure*}

\section{Experimental Results}
\label{sec:exp}

\subsection{Experimental methods}
\label{subsec:exp_meth}
The single crystals of \SRO{} used in our experiments were grown by the floating zone technique and showed a superconducting transition temperature of \mbox{$T_c=\SI{1.45}{K}$}. 
ARPES measurements were performed with an MBS electron spectrometer and a narrow bandwidth \SI{11}{eV}(\SI{113}{nm}) laser source from Lumeras that was operated at a repetition rate of \SI{50}{MHz} with \SI{30}{ps} pulse length of the \SI{1024}{nm} pump~\cite{He2016}. 
All experiments were performed at $T\approx5$~K using a cryogenic 6-axes sample goniometer, as described in Ref.~\cite{Hoesch2017}. 
A combined energy resolution of \SI{3}{meV} was determined from the width of the Fermi-Dirac distribution measured on a polycrystalline Au sample held at 4.2~K. The angular resolution was $\approx0.2^{\circ}$. 
In order to suppress the intensity of the surface layer states on pristine \SRO~\cite{Shen2001}, we exposed the cleaved surfaces to $\approx0.5$~\replaced{Langmuir}{L} \ce{CO} at a temperature of $\approx120$~K. Under these conditions, \ce{CO} preferentially fills surface defects and subsequently replaces apical oxygen ions to form a \ce{Ru-COO} carboxylate in which the \ce{C} end of a bent \ce{CO2} binds to \ce{Ru} ions of the reconstructed surface layer~\cite{Stoger2014}.

\subsection{Experimental Fermi surface and quasiparticle dispersions}
\label{subsec:exp_FS}
In Fig.~\ref{fig:f1} we show the Fermi surface and selected constant energy surfaces in the occupied states of \SRO. The rapid broadening of the excitations away from the Fermi level seen in the latter is typical for ruthenates and implies strong correlation effects on the quasiparticle properties. At the Fermi surface, one can readily identify the $\alpha$, $\beta$ and $\gamma$ sheets that were reported earlier~\cite{Mackenzie1996, Bergemann2000,Damascelli2000}. However, compared with previous ARPES studies we achieve a reduced line width and improved suppression of the surface layer states giving clean access to the bulk electronic structure. This is particularly evident along the Brillouin zone diagonal ($\Gamma$X) where we can clearly resolve all band splittings. 

\begin{figure*}[tb]
	\includegraphics[width=1 \textwidth]{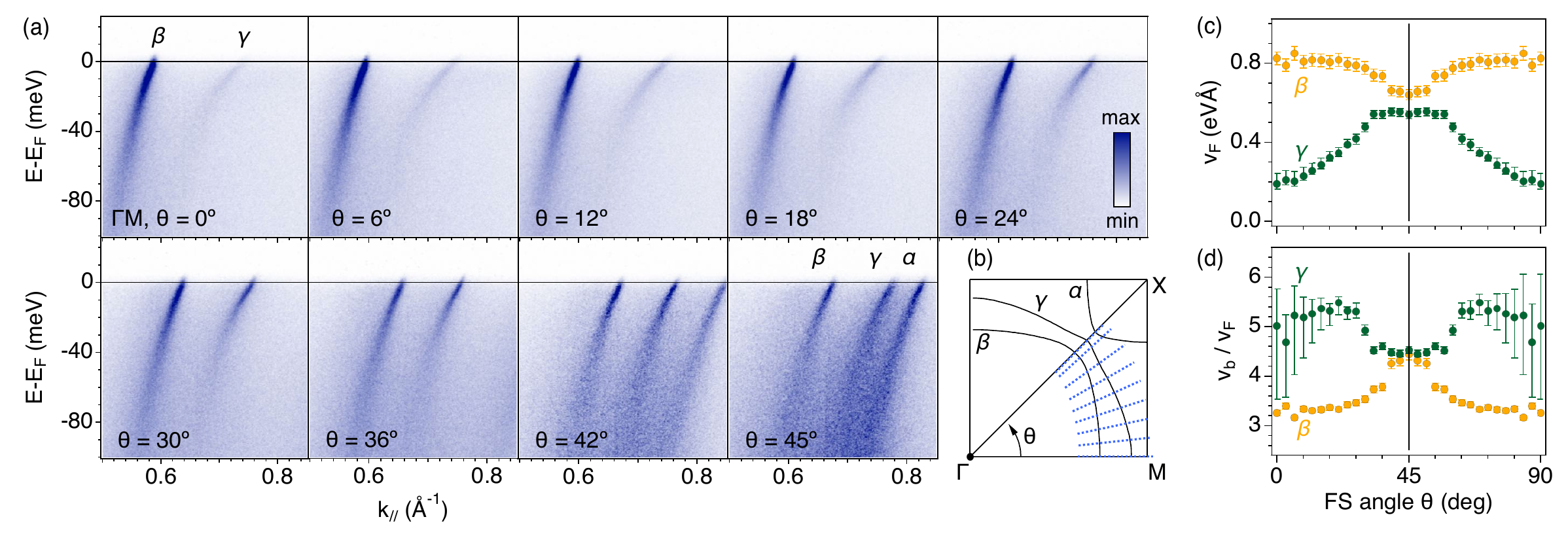} %
	\caption{(a) Quasiparticle dispersions measured with $p$-polarized light, for different azimuthal angles $\theta$ as defined in panel (b). 
	(c) Angular dependence of the quasiparticle velocity $v_{F}$ along the $\beta$ and $\gamma$ Fermi surface sheets. 
	(d) Angular dependence of the quasiparticle mass enhancement $v_{b}/v_{F}$. Here, $v_{b}$ is the bare velocity obtained from the single-particle Hamiltonian \Hzero=\Hdft+\Hsocdlam{} defined in Sec.~\ref{sec:fs} and $v_{F}$ is the quasiparticle Fermi velocity measured by ARPES. Error bars are obtained by propagating the experimental uncertainty on $v_{F}$. Thin lines are guides to the eye.}
	\label{fig:f1bis}
\end{figure*}

In the following, we will exploit this advance to quantify the effects of SOC in \SRO{} and to provide new insight into the renormalization of the quasiparticle excitations using minimal theoretical input only. To this end we acquired a set of 18 high resolution dispersion plots along radial $\bk$-space lines (as parameterized by the angle $\theta$ measured from the $\Gamma$M direction). The subset of data shown in Fig.~\ref{fig:f1bis}~(a) immediately reveals a rich behavior with a marked dependence of the low-energy dispersion on the Fermi surface angle $\theta$. Along the $\Gamma$M high-symmetry line our data reproduce the large difference in Fermi velocity $v_{F}^{\beta,\gamma}$ for the $\beta$ and $\gamma$ sheet, which is expected from the different cyclotron masses deduced from quantum oscillations~\cite{Mackenzie1996,Mackenzie1998,Bergemann2003} and was reported in earlier ARPES studies~\cite{Shen2007,Zabolotnyy2013}. Our systematic data, however, reveal that this difference gradually disappears towards the Brillouin zone diagonal $(\theta=45^{\circ})$, where all three bands disperse nearly parallel to one another. In Sec.~\ref{sec:fs} we will show that this equilibration of the Fermi velocity can be attributed to the strong effects of SOC around the zone diagonal.

To quantify the angle dependence of $v_{F}^{\beta,\gamma}$ from experiment, we determine the maxima $\kmaxnuom$ of the momentum distribution curves (MDCs) over the range of \SIrange{2}{6}{meV} below the Fermi level $E_F$ and fit these $\bk$-space loci with a second-order polynomial. We then define the Fermi velocity as the derivative of this polynomial at $E_F$. This procedure minimizes artifacts due to the finite energy resolution of the experiment.
As shown in Fig.~\ref{fig:f1bis}~(c), the Fermi velocities $v_{F}^{\beta,\gamma}$ obtained in this way show an opposite trend with azimuthal angle for the two Fermi sheets. For the $\beta$ band we observe a gentle decrease of $v_F$ as we approach the $\Gamma$X direction, whereas for $\gamma$ the velocity increases by more than a factor of two over the same range
\footnote{We note that radial $\bk$-space cuts are not exactly perpendicular to the Fermi surface. This can cause the velocities $v_F$ and $v_b$ given here to deviate by up to 10\% from the Fermi velocity. However, since we evaluate the experimental and theoretical dispersion along the same $\bk$-space cut, this effect cancels in Fig.~\ref{fig:f1bis}~(d) and does not affect the self-energy determination in Sec.~\ref{sec:self}}.
This provides a first indication for a strong momentum dependence of the self-energies $\Sigma^\prime_{\beta,\gamma}$, which we will analyze quantitatively in Sec.\ref{sec:self}. Here, we limit the discussion to the angle dependence of the mass enhancement $v_{b}/v_{F}$, which we calculate from the measured quasiparticle Fermi velocities of Fig.~\ref{fig:f1bis}~(c) and the corresponding \added{bare} velocities \added{$v_{b}$} of a reference Hamiltonian \Hzero{} defined in Sec.~\ref{sec:fs}.
As shown in Fig.~\ref{fig:f1bis}~(d), this confirms a substantial many-body effect on the anisotropy of the quasiparticle dispersion. Along $\Gamma$M, we find a strong differentiation with mass enhancements of $\approx5$ for the $\gamma$ sheet and $\approx3.2$ for $\beta$, whereas $v_{b}/v_{F}$ approaches a common value of $\approx4.4$ for both sheets along the Brillouin zone diagonal.

Before introducing the theoretical framework used to quantify the anisotropy of the self-energy and the effects of SOC, we compare our data quantitatively to bulk sensitive quantum oscillation measurements.
Using the experimental Fermi wave vectors $\bk_F$ and velocities determined from our data on a dense grid along the entire Fermi surface, we can compute the cyclotron masses measured by dHvA experiments, without relying on the approximation of circular Fermi surfaces and/or isotropic Fermi velocities used in earlier studies~\cite{Baumberger2006,Shen2007,Zabolotnyy2013,Wang2017}. Expressing the cyclotron mass $m^{*}$ as
\beq
m^{*} = \frac{\hbar^{2}}{2\pi}\frac{\partial A_{FS}}{\partial \epsilon}=\frac{\hbar^2}{2\pi}\int_{0}^{2\pi}\frac{k_F(\theta)}{\partial\epsilon/\partial k(\theta)}d\theta \, ,
\eeq
where $A_{FS}$ is the Fermi surface volume, and using the data shown in Fig.~\ref{fig:f1bis}~(c), we obtain $m^*_\gamma=17.3(2.0)$~$m_e$ and $m^*_\beta=6.1(1.0)$~$m_e$, in quantitative agreement with the values of $m^*_\gamma=16$~$m_e$ and $m^*_\beta=7$~$m_e$ found in dHvA experiments~\cite{Mackenzie1996,Mackenzie1998,Bergemann2003}. We thus conclude that the quasiparticle states probed by our experiments are representative of the bulk of \SRO
\footnote{We attribute the higher Fermi velocities reported in some earlier studies~\cite{Iwasawa2005,Iwasawa2010,Iwasawa2012,Burganov2016} to the lower energy resolution, which causes an extended range near the Fermi level where the dispersion extracted from fits to individual MDCs is artificially enhanced, rendering a precise determination of $v_{F}$ difficult. The much lower value of $v_{F}^{\beta}$ along $\Gamma$M reported in Ref.~\cite{Wang2017} corresponds to the surface $\beta$ band, as shown in appendix~\ref{sec:appA}.}.

\section{Theoretical Framework}
\label{subsec:theo}

\deleted{The goal of this work is to extract directly from the ARPES data some key information about quasiparticle properties. 
We are in particular interested in: 
(i) a precise determination of the Fermi surface, 
(ii) the quasiparticle velocities and their renormalization by electronic correlations, and
(iii) a direct determination of the electronic self-energy.}

In order to define the \replaced{electronic self-energy}{latter} and assess the effect of electronic correlations \added{in the spectral function measured by ARPES}, we need to 
specify a one-particle Hamiltonian \Hzero{} as a reference point. 
At this stage, we keep the presentation general. The particular choice of \Hzero{}
will be a focus of Sec.~\ref{sec:fs}.
The eigenstates $|\psiknu\rangle$ of \Hzero$\left(\bk\right)$ at a given quasi-momentum $\bk$ and
the corresponding eigenvalues $\epsknu$ define the `bare' band structure of the system, with respect to which the 
self-energy $\Sigma_{\nu\nup}(\omega,\bk)$ is defined \added{in the standard way} from the interacting Green's function
\begin{equation}
G^{-1}_{\nu\nup}(\omega,\bk)\,=\,\left[\omega+\mu-\epsknu\right]\delta_{\nu\nup} - \Sigma_{\nu\nup}(\omega,\bk)\, .
\end{equation}
In this expression $\nu$ and $\nu'$ label the bands and $\omega$ denotes the \deleted{binding} energy counted from $E_F$. \deleted{, which we set to zero throughout the rest of this work.} The interacting value of the chemical potential $\mu$ sets the total electron number. Since $\mu$ can be conventionally included in \Hzero, we shall omit it in the following.
\deleted{The electronic spectral function is related to the Green's function}
\deleted{It is this quantity which is most directly related to the ARPES signal: 
within the sudden approximation, in the absence of final-state interactions, the ARPES intensity is given by 
$\sum_{\nu\nu'}M_{\nu}M^*_{\nu'}A_{\nu\nu'}f(\omega)$
convolved with a function representing the experimental resolution and
assuming an average over the polarization. Here, $M$ is a matrix element and $f(\omega)$ is the Fermi function.}

\deleted{The Fermi surface of the system is the locus of zero-energy excitations, and hence corresponds to the 
momenta $\bk_F$ which are the solutions of}
The \added{Fermi surfaces and} dispersion relations of the quasiparticles are obtained as the solutions $\omega=0$ and $\omega=\omqp_\nu(\bk)$ of
\beq \label{eq:qp}
\mathrm{det} \left[\left(\omega-\epsknu\right)\, \delta_{\nu\nup} - \Sigma_{\nu\nup}^\prime(\omega,\bk) \right]\,=\,0\, .
\eeq
In the above equation $\Sigma^\prime$ denotes the real part of the self-energy. 
Its imaginary part $\Sigma^{\prime\prime}$ has been neglected, \textit{i.e.}, we assume that quasiparticles are coherent with a lifetime \deleted{much} longer than $1/\omqp$. 
\added{Our data indicate that this is indeed the case up to the highest energies analyzed here.
At very low frequency, the lifetime of quasiparticles cannot be reliably tested by ARPES, since the intrinsic quasiparticle width
is masked by contributions of the experimental resolution, impurity scattering and inhomogeneous broadening. However, the observation of strong quantum oscillations in the bulk provides direct evidence for well-defined quasiparticles in \SRO{} down to the lowest energies~\cite{Mackenzie1996,Bergemann2000}.
}

It is important to note that \added{the Green's function} $G$, \added{the self-energy }$\Sigma$ and \added{the spectral function} $A$ are in general non-diagonal matrices. \added{This has been overlooked thus far in self-energy analyses of ARPES data but is essential to determine the nature of many-body interactions in \SRO, as we will show in Sec.~\ref{sec:self}.}

\subsection{Localized orbitals and electronic structure}
\label{subsec:localorb}

Let us recall some of the important aspects of the electronic structure of \SRO{}. 
As shown in Sec.~\ref{subsec:exp_FS}, three bands, commonly denoted $\nu=\{\alpha,\beta,\gamma\}$,
cross the Fermi level. These bands correspond to states with $t_{2g}$ symmetry deriving from the hybridization between localized Ru-$4d$ ($d_{xy},d_{yz},d_{xz}$) orbitals and O-$2p$ states. Hence, we introduce a localized basis set of $t_{2g}$-like orbitals $|\chi_m\rangle$, with basis functions conveniently labeled as $m=\{xy,yz,xz\}$. In practice, we use maximally localized Wannier functions~\cite{MLWF1,MLWF2} constructed from the Kohn-Sham eigenbasis of a non-SOC density functional theory (DFT) calculation (see appendix~\ref{sec:appB1} for details). We term the corresponding Hamiltonian \Hdft{}.
It is important to note that the choice of a localized basis set is not unique and other ways of defining these orbitals are possible (see, e.g., Ref.~\cite{Lechermann_2006}).

In the following this set of orbitals plays two important roles. First, they are
atom-centered and provide a set of states localized in real-space $|\chi_m\left({\mathbf{R}}\right)\rangle$.
Secondly, the unitary transformation matrix to the band basis $|\psiknu\rangle$
\beq\label{eq:U}
U_{m\nu}\left(\bk\right) = \braket{\chikm|\psiknu}\, ,
\eeq
allows us to define an `orbital' character of each band $\nu$ as $|U_{m\nu}\left(\bk\right)|^2$.
In the localized-orbital basis the one-particle Hamiltonian is a non-diagonal matrix, which reads
\beq
\hat{H}^0_{mm^\prime}(\bk)\,=\,\sum_\nu U_{m\nu}\left(\bk\right)\, \epsknu\, U^*_{m^\prime\nu}\left(\bk\right)\, .
\eeq
The self-energy in the orbital basis is expressed as
\beq
\Sigma_{mm^\prime}(\omega,\bk)\,=\,\sum_{\nu\nu'} U_{m\nu}\left(\bk\right)\, \Sigma_{\nu\nup}(\omega,\bk)\, U^*_{m^\prime\nu^\prime }\left(\bk\right)\, ,
\eeq
and conversely in the band basis as
\beq\label{eq:signu_from_sigm}
\Sigma_{\nu\nup}(\omega,\bk)\,=\,\sum_{mm^\prime} U^*_{ m\nu}\left(\bk\right)\, \Sigma_{mm'}(\omega,\bk)\, U_{m^\prime \nu^\prime}\left(\bk\right)\, .
\eeq

\subsection{Spin-orbit coupling}
\label{subsec:so}

We treat SOC as an additional term to \Hdft{}, which is
independent of $\bk$ in the localized-orbital basis, but leads to a mixing of the individual orbitals. The single-particle SOC term for atomic $d$-orbitals projected to the $t_{2g}$-subspace reads~\footnote{By restricting the Hamiltonian to the $t_{2g}$-subspace we neglect the $e_{g}$-$t_{2g}$ coupling terms of $\hat{H}^{\mathrm{SOC}}_{\lambda}$. This approximation is valid as long as the $e_{g}$-$t_{2g}$ crystal-field splitting is large in comparison to $\lambda$, which is the case for \SRO{}.}
\beq \label{eq:hsoc}
\hat{H}^{\mathrm{SOC}}_{\lambda}\,=\, \frac{\lambda}{2} \sum_{mm^\prime}\sum_{\sigma\sigma^\prime} c^\dagger_{m\sigma} \left(\mathbf{l}_{mm^\prime} \cdot \text{\boldmath$\sigma$}_{\sigma\sigma^\prime}\right) c_{m^\prime\sigma^\prime}\, ,
\eeq
where $\mathbf{l}$ are the $t_{2g}$-projected angular momentum matrices, $\text{\boldmath$\sigma$}$ are Pauli matrices and $\lambda$ will be referred to in the following as the SOC coupling constant. 
As documented in appendix~\ref{sec:appB1}, the eigenenergies of a DFT+SOC calculation are well reproduced by \Hdft+\Hsoclam{} with $\lambda_{\mathrm{DFT}}=\SI{100}{meV}$.

\begin{figure*}[tb]
	\includegraphics[width=0.98 \textwidth]{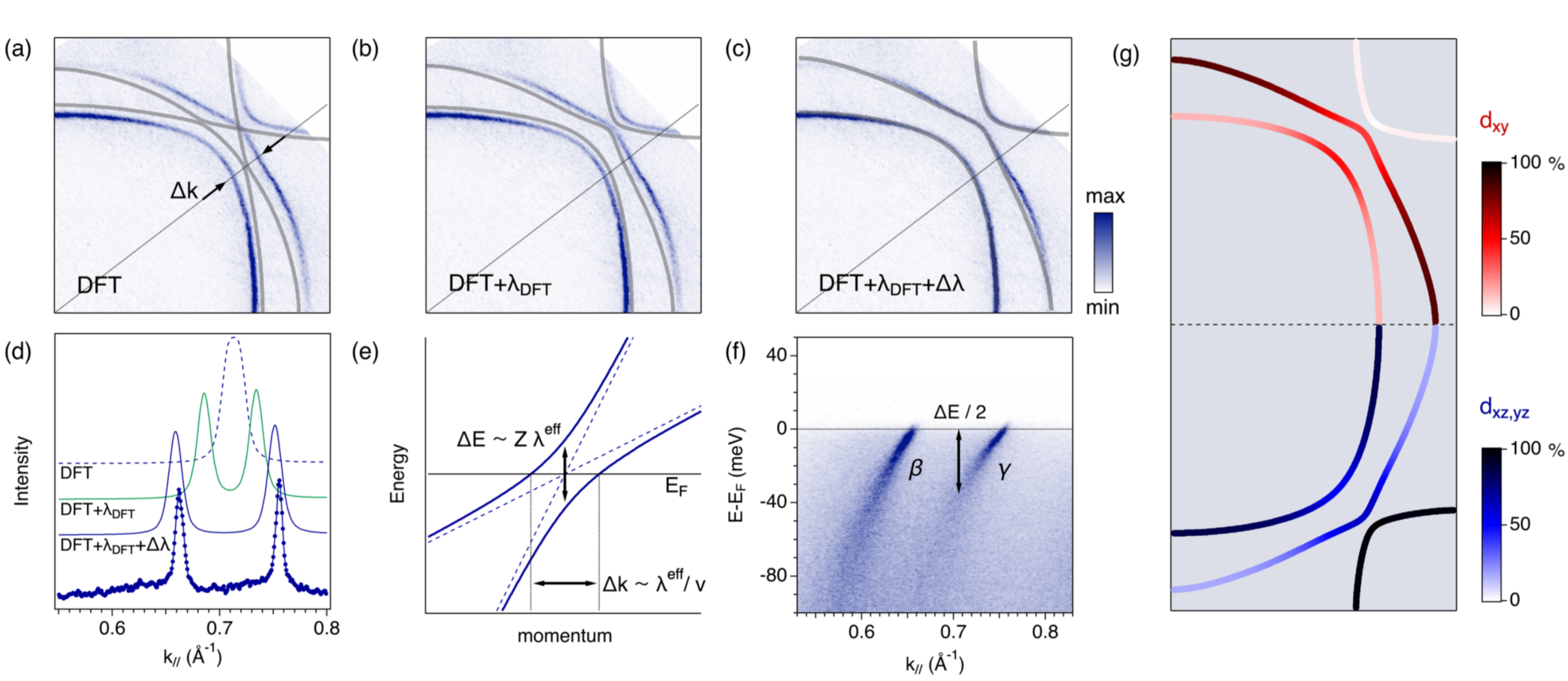} %
	\caption{Correlation enhanced effective SOC. (a) Quadrant of the experimental Fermi surface with a DFT calculation without SOC (\Hdft{}) at the experimental $k_z\approx0.4~\pi/c$ (grey lines). (b,c) Same as (a) with calculations including SOC (DFT+$\lambda_{\mathrm{DFT}}$) and enhanced SOC ($\lambda_{\mathrm{DFT}}+\Delta\lambda$), respectively. For details, see main text. (d) Comparison of the experimental MDC along the $\bk$-space cut indicated in (a) with the different calculations shown in (a-c). (e) Schematic illustration of the renormalization of a SOC induced degeneracy lifting. Here, $Z=\sqrt{Z_\nu Z_{\nu'}}$, where $\nu,\nu'$ labels the two bands and $v=\sqrt{v_{\nu}v_{\nu'}}$ where $v_{\nu}, v_{\nu'}$ are bare velocities in the absence of SOC~\cite{Kim2018} (see text). (f) Experimental quasiparticle dispersion along the $\bk$-space cut indicated in (a). (g) Orbital character of the DFT+$\lambda_{\mathrm{DFT}}+\Delta\lambda$ eigenstates along the Fermi surface.}
	\label{fig:f2}
\end{figure*}

\section{Enhanced effective spin-orbit coupling and single-particle Hamiltonian}
\label{sec:fs}

The importance of SOC for the low-energy physics of \SRO{} has been pointed out by several 
authors~\cite{ng_sigrist_2002,eremin_2002,Haverkort2008,Iwasawa2010,puetter_kee_2012,Veenstra2013,Scaffidi2014,Steppke2017,Zhang2016,Kim2018}. 
SOC lifts degeneracies found in its absence and causes a momentum dependent mixing of the orbital composition of quasiparticle states, which has non-trivial implications for superconductivity~\cite{Veenstra2013,Scaffidi2014,Steppke2017}. 
Signatures of SOC have been detected experimentally on the Fermi surface of \SRO{} in the form of a small protrusion of the $\gamma$ sheet along the zone diagonal~\cite{Haverkort2008,Iwasawa2010} and as a degeneracy lifting at the band bottom of the $\beta$ sheet~\cite{Veenstra2013}. These studies reported an overall good agreement between the experimental data and the effects of SOC calculated within DFT~\cite{Haverkort2008,Iwasawa2010,Veenstra2013}. This is in apparent contrast to more recent DMFT studies of \SRO, which predict large but frequency independent off-diagonal contributions to the local self-energy that can be interpreted as a contribution $\Delta\lambda$ to the effective coupling strength $\lambda^{\mathrm{eff}}=\lambda_{\mathrm{DFT}}+\Delta\lambda$~\cite{Zhang2016,Kim2018},
\replaced{This is also consistent with general perturbation-theory considerations~\cite{liu_prl_2008}, which show a Coulomb-enhancement of the level splitting in the $J$ basis, similar to a Coulomb-enhanced crystal-field splitting~\cite{Poteryaev2007}.}{consistent with general perturbation-theory considerations~\cite{liu_prl_2008}.}

In the absence of SOC, DFT yields a quasi-crossing between the $\beta$ and $\gamma$ Fermi surface sheets a few degrees away from the zone diagonal, as displayed on Fig.~\ref{fig:f2}~(a). Near such a point we expect the degeneracy to be lifted by SOC, leading to a momentum splitting $\Delta k = \lambda^{\mathrm{eff}}/v$ and to an energy splitting 
of $\Delta E = Z \lambda^{\mathrm{eff}}$ \cite{Kim2018}, as depicted schematically in Fig.~\ref{fig:f2}~(e).
In these expressions, $v\equiv\sqrt{v_\beta v_\gamma}$, with $v_\beta$ and $v_\gamma$ the bare band velocities in the 
absence of SOC and correlations, and $Z\equiv \sqrt{Z_\beta Z_\gamma}$ involves the quasiparticle residues $Z_{\nu}$ associated 
with each band (also in the absence of SOC). 

It is clear from these expressions that a quantitative determination of \leff{} is not possible from experiment alone. 
Earlier studies on \SRO~\cite{Veenstra2013} and iron-based superconductors~\cite{Borisenko2015}, have interpreted the energy splitting $\Delta E$ at avoided crossings as a direct measure of the SOC strength $\lambda^{\mathrm{eff}}$. However, in interacting systems $\Delta E$ is not a robust measure of SOC since correlations can both enhance $\Delta E$ by enhancing \leff{} and reduce it via the renormalization factor $Z$.
We thus quantify the enhancement of SOC from the momentum splitting $\Delta k$, which is not renormalized by
the quasiparticle residue $Z$.
The experimental splitting at the avoided crossing between the $\beta$ and $\gamma$ Fermi surface sheets indicated in Fig.~\ref{fig:f2}~(a) is $\Delta k^{\mathrm{QP}}=0.094(9)$~\AA$^{-1}$ whereas DFT predicts $\Delta k^{\mathrm{DFT+\lambda_{\mathrm{DFT}}}}=0.046$~\AA$^{-1}$.
We thus obtain an effective SOC strength $\lambda^{\mathrm{eff}}=\lambda_{\mathrm{DFT}}\Delta k^{\mathrm{QP}}/\Delta k^{\mathrm{DFT+\lambda_{\mathrm{DFT}}}}=205(20)$~meV, in quantitative agreement with the predictions in Refs.~\cite{Zhang2016,Kim2018}.
We note that despite this large enhancement of the effective SOC, the energy splitting remains smaller than $\lambda_{\mathrm{DFT}}$ as \replaced{shown in Fig.~\ref{fig:f2}~(f).}{illustrated in Fig.~\ref{fig:f2}~(e).}
When deviations from linearity in band dispersions are small, the splitting $\Delta E$ is symmetric around the
$E_F$ and can thus be determined from the occupied states probed in experiment. Direct inspection of the data in Fig.~\ref{fig:f2}~(f) yields $\Delta E\approx\SI{70}{meV}$, which is about $2/3$ of $\lambda_{\mathrm{DFT}}$ and thus clearly not a good measure of SOC.

The experimental splitting is slightly larger than that expected from the expression $\Delta E = Z \lambda^{\mathrm{eff}}$ and our theoretical determination of $Z_\beta$ and $Z_\gamma$ at the Fermi surface \added{(which will be described in Sec.~\ref{sec:dmft})}. This can be attributed to the energy dependence of $Z$, which, in \SRO, is not negligible over the energy scale of SOC.
Note that the magnitude of the SOC-induced splitting of the bands at the $\Gamma$ point reported in Ref.~\cite{Veenstra2013} can also be explained by the competing effects of enhancement by correlations and reduction by the quasiparticle weight as shown in Ref.~\cite{Kim2018}.
We also point out that the equilibration of quasiparticle velocities
close to the diagonal, apparent from Figs.~\ref{fig:f1bis}~(a,c) and \ref{fig:f2}~(f) is indeed the behavior expected close to an avoided crossing~\cite{Kim2018}.

Including the enhanced SOC determined from this non-crossing gap leads to a much improved theoretical description of the entire Fermi surface
\footnote{Ref.~\cite{Rozbicki2011b} noted that an enhance effective SOC also improves the description of de Haas van Alphen data.}.
As shown in Fig.~\ref{fig:f2}~(b), our high-resolution experimental Fermi surface deviates systematically from a DFT calculation with SOC. Most notably, \Hdft+\Hsocdft{} underestimates the size of the $\gamma$ sheet and overestimates the $\beta$ sheet. 
Intriguingly, this is almost completely corrected in \Hdft+\Hsocdlam{}, with $\lambda_{\mathrm{DFT}}+\Delta\lambda=\SI{200}{meV}$, as demonstrated in Fig.~\ref{fig:f2}~(c).
Indeed, a close inspection shows that the remaining discrepancies between experiment and \Hdft+\Hsocdlam{} break the crystal symmetry, suggesting that they are dominated by experimental artifacts. A likely source for these image distortions is imperfections in the electron optics arising from variations of the work function around the electron emission spot on the sample.
Such distortions can presently not be fully eliminated in low-energy photoemission from cleaved single crystals.

Importantly, the change in Fermi surface sheet volume with the inclusion of $\Delta\lambda$ is not driven by a change in the crystal-field splitting between the $xy$ and $xz,yz$ orbitals (see appendix~\ref{sec:appB}). The volume change occurs solely because of a further increase in the orbital mixing induced by the enhanced SOC. As shown in Fig.~\ref{fig:f2}~(g), this mixing is not limited to the vicinity of the avoided crossing but extends along the entire Fermi surface \added{(see Fig.~\ref{fig:fB1} in appendix~\ref{sec:appB1} for the orbital character without SOC)}. For $\lambda_{\mathrm{DFT}}+\Delta\lambda$ we find a minimal $d_{xy}$ and $d_{xz,yz}$ mixing for the $\gamma$ and $\beta$ bands of $20/80$\% along the $\Gamma$M direction with a monotonic increase to $\approx50$\% along the Brillouin zone diagonal $\Gamma$X.
We note that this mixing varies with the perpendicular momentum $k_z$. However, around the experimental value of $k_z \approx 0.4~\pi/c$ the variation is weak
\footnote{Using the free electron final state approximation, we obtain  $k_z = (2m_e\hbar^{-2}(h\nu+U-\Phi) - k_{\parallel}^2)^{1/2}\approx 0.4(0.12) \pi/c$ assuming an inner potential relative to $E_F$ of $U-\Phi=8.5(1.0)$~eV.}.
The analysis presented here and in Secs.~\ref{subsec:localorb} and \ref{sec:dmft} is thus robust with respect to a typical uncertainty in $k_z$.
These findings suggest that a natural reference single-particle Hamiltonian is \Hzero=\Hdft+\Hsocdlam{}. 
This choice ensures that the Fermi surface of \Hzero{} is very close to that of the interacting system. 
From Eq.~\ref{eq:qp}, this implies that the self-energy matrix approximately vanishes at zero binding energy: 
$\Sigma'_{\nu\nu'}(\omega=0,\bk)\simeq 0$. We choose \Hzero{} in this manner in all the following. Hence, from now on 
$|\psiknu\rangle$ and $\epsknu$ refer to the eigenstates and band structure of \Hzero=\Hdft+\Hsocdlam. 
We point out that although \Hzero{} is a single-particle Hamiltonian, the effective enhancement $\Delta\lambda$ of SOC included in \Hzero{} is a correlation effect beyond DFT.

\begin{figure*}[htb]
	\includegraphics[width=0.95 \textwidth]{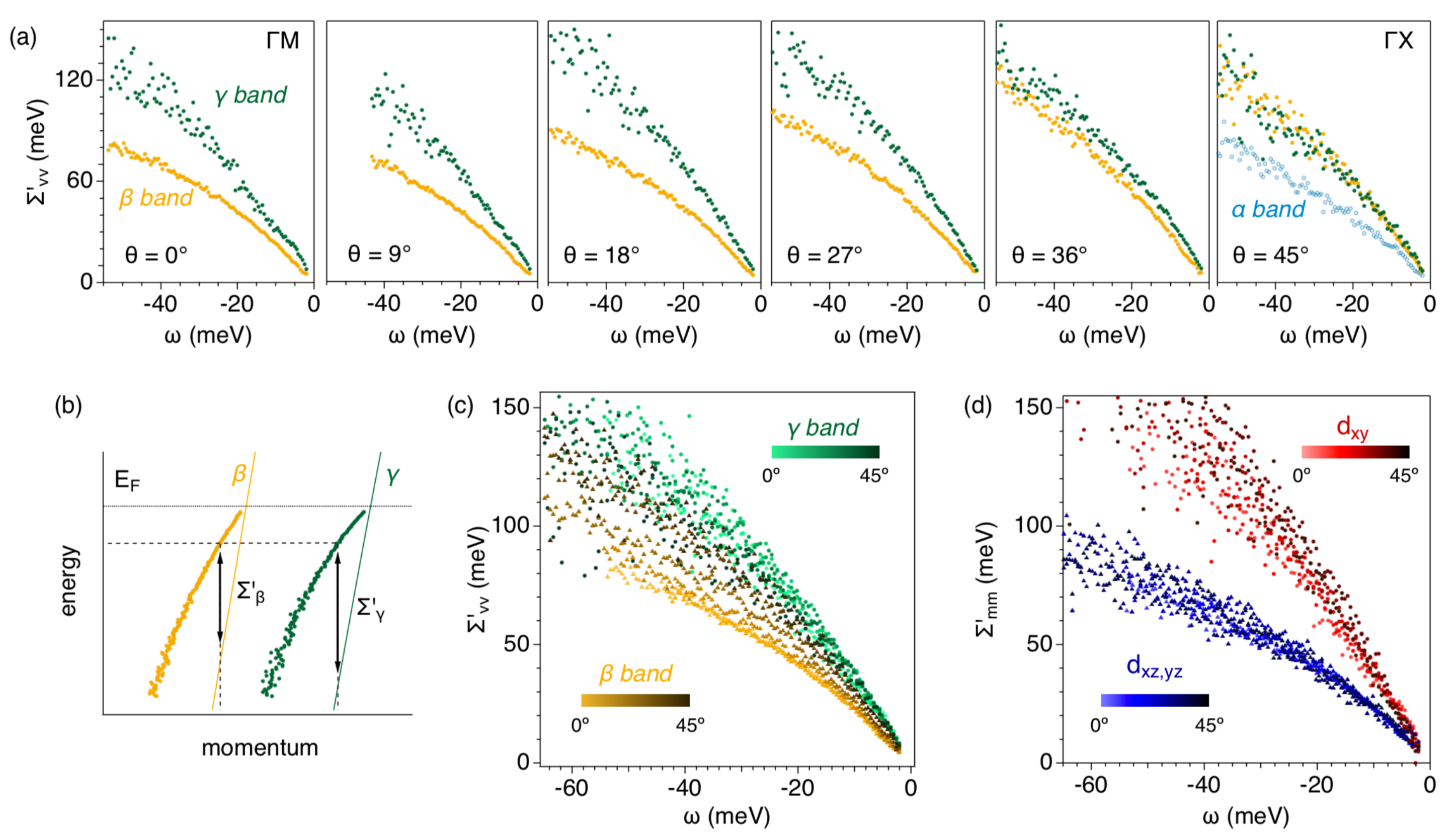} %
	\caption{Self-energies extracted in the band and orbital bases. (a) Real part of the self-energy $\Sigma'_{\nu\nu}$ in the band basis (solid symbols) in $\num{9}^{\circ}$ steps of the Fermi surface angle $\theta$. (b) Illustration of the relation between $\Sigma'_{\nu\nu}$, the bare bands given by \Hzero (thin lines) and the quasiparticle peak positions (solid symbols). (c) Compilation of $\Sigma'_{\nu\nu}$ from panel (a). (d) Real part of the self-energy $\Sigma'_{mm}$ in the orbital basis.}
	\label{fig:f4}
\end{figure*}

\section{Experimental determination of self-energies}
\label{sec:self}

\subsection{Self-energies in the quasiparticle/band basis}
\label{subsec:self_qp}

Working in the band basis, i.e., with the eigenstates of \Hzero{}, the maximum of 
the ARPES intensity for a given binding energy $\omega$ (maximum of the MDCs) 
corresponds to the momenta $\bk$ which satisfy (following Eq.~\ref{eq:qp}):  
$\omega-\epsknu-\Sigma^\prime_{\nu\nu}(\omega,\bk)=0$.  
Hence, for each binding energy, each azimuthal cut, and each sheet of the quasiparticle dispersions, we fit the MDCs and determine the momentum $\kmaxnuom$ at their maximum.
Using the value of $\varepsilon_\nu\left(\kmaxnuom \right)$ at this momentum yields the following quantity
\beq
\label{eq:qp_v}
\omega-\varepsilon_\nu\left(\kmaxnuom \right)= \Sigma^\prime_{\nu\nu}\left(\omega,\kmaxnuom\right)
\,\equiv\,\Sigma'_\nu(\omega,\theta)\, .
\eeq

This equation corresponds to the simple construction illustrated graphically in Fig.~\ref{fig:f4}~(b), and 
it is a standard way of extracting a self-energy from ARPES, as used in previous works on several materials~\cite{Hengsberger1999,Lanzara2001,Iwasawa2012,Iwasawa2013,Tamai2013}.
We note that this procedure assumes that the off-diagonal components 
$\Sigma'_{\nu\neq\nup}(\omega,\bk)$ can be neglected for states close to the Fermi surface (i.e., for small $\omega$ and $\bk$ close to a Fermi crossing). This assumption can be validated, as shown in appendix~\ref{sec:appB3}. 
When performing this analysis, we only include the $\alpha$ sheet for $\theta=45^{\circ}$.
Whenever the constraint $\Sigma'_{\nu}(\omega\rightarrow 0)\rightarrow 0$ on the self-energy is not precisely obeyed, a small shift is applied to set it to zero. We chose this procedure to correct for the minor differences
between the experimental and the reference \Hzero{} Fermi wave-vectors because we attribute these differences predominantly to experimental artifacts.
 
The determined self-energies for each band $\nu=\alpha,\beta,\gamma$ and the different values of $\theta$ are depicted in Fig.~\ref{fig:f4}~(a,c). For the $\beta$ and $\gamma$ sheets they show a substantial dependence on the azimuthal angle. 
Around $\Gamma$M we find that $\Sigma'_{\gamma}$ exceeds $\Sigma'_{\beta}$ by almost a factor of two 
(at $\omega = \SI{-50}{meV}$), whereas they essentially coincide along the zone diagonal ($\Gamma$X). 
This change evolves as a function of $\theta$ and occurs 
via a simultaneous increase in $\Sigma'_{\beta}$ and a decrease in $\Sigma'_{\gamma}$ for all energies 
as $\theta$ is increased from $0^\circ$ ($\Gamma$M) to $45^\circ$ ($\Gamma$X). 
In order to better visualize this angular dependence, a compilation of $\Sigma'_{\nu}(\omega)$ for different values of $\theta$ is displayed in Fig.~\ref{fig:f4}~(c). 

\subsection{Accounting for the angular dependence: local self-energies in the orbital basis}
\label{sec:self_orb}

In this section, we introduce a different procedure for extracting self-energies from ARPES, 
by working in the orbital basis $|\chikm\rangle$. We do this by making two key assumptions: 

\begin{enumerate}

\item We assume that the off-diagonal components are negligible, i.e., $\Sigma'_{m\neq m^\prime}\simeq 0$.
Let us note that in \SRO{} even a $\bk$-independent self-energy has non-zero off-diagonal elements if \Hdft+\Hsocdft{}
is considered. Using DMFT, these off-diagonal elements have been shown to be very weakly dependent 
on frequency in this material~\citep{Kim2018}, leading to the notion of a static correlation enhancement of the effective SOC ($\Delta\lambda$).
In the present work, these off-diagonal frequency-independent components are already incorporated into \Hzero{} (see Sec.~\ref{sec:fs}), 
and thus the frequency-dependent part of the self-energy is (approximately) orbital diagonal by virtue of the tetragonal crystal structure.

\item We assume that the diagonal components of the self-energy in the orbital basis depend on the momentum $\bk$ only through the azimuthal angle $\theta_\bk$: 
$\Sigma'_{mm}(\omega,\bk)\simeq \Sigma'_m(\omega,\theta_\bk)$.
We neglect the dependence on the momentum which is parallel to the angular cut.

\end{enumerate}

Under these assumptions, the equation determining the quasiparticle dispersions reads
\beq\label{eq:qp_orb_loc}
\det\left[(\omega-\Sigma'_m(\omega,\theta_\bk))\delta_{mm'}-\hat{H}^0_{mm'}(\bk) \right] \,=\,0 \, .
\eeq
In this equation, we have neglected the lifetime effects associated with the imaginary part $\Sigma''_m$.
In order to extract the functions $\Sigma'_m(\omega,\theta_\bk)$ directly from the ARPES data, we first determine the peak positions $\bk_{\mathrm{max}}^\nu(\omega,\theta)$ for MDCs at a given angle $\theta$ and binding energy $\omega$. 
We then compute (for the same $\omega$ and $\theta$) the matrix $A_{mm'}\equiv\omega\delta_{mm'}-\hat{H}^0_{mm'}\left(\bk_{\mathrm{max}}^{\alpha}(\omega,\theta_{\bk})\right)$
and similarly $B_{mm'}$, $G_{mm'}$ for the $\beta$ and $\gamma$ band MDCs, $\bk_{\mathrm{max}}^\beta(\omega,\theta)$ and
$\bk_{\mathrm{max}}^\gamma(\omega,\theta)$, respectively.
In terms of these matrices, the quasiparticle equations (\ref{eq:qp_orb_loc}) read
\begin{align}\label{eq:sigma_m_eq}
&\det [ A_{mm'}-\Sigma'_m \delta_{mm'}] = \det [ B_{mm'}-\Sigma'_m \delta_{mm'}] = \nonumber \\
&=\det [ G_{mm'}-\Sigma'_m \delta_{mm'}] = 0 \, .
\end{align}

However, when taking symmetry into account, the self-energy has only two independent components: $\Sigma'_{xy}$ and $\Sigma'_{xz}=\Sigma'_{yz}$.
Hence, we only need two of the above equations to solve for the two unknown components of the self-energy. This means that we can also extract a self-energy in the directions where only two bands ($\beta$ and $\gamma$) are present in the considered energy range of $\omega \lesssim \SI{100}{meV}$, e.g., along $\Gamma$M.
The resulting functions $\Sigma'_m(\omega,\theta_\bk)$ determined at several angles $\theta$ 
are displayed in Fig.~\ref{fig:f4}~(d). 
It is immediately apparent that, in contrast to $\Sigma'_{\nu\nu}$, the self-energies in the orbital basis 
do not show a strong angular (momentum) dependence, but rather collapse into two sets of points, one for the $xy$ orbital and one for the $xz/yz$ orbitals.
Thus, we reach the remarkable conclusion that the angular dependence of the self-energy in the orbital 
basis is negligible, within the	range of binding energies investigated here: $\Sigma'_m(\omega,\theta_\bk)\simeq\Sigma'_m(\omega)$.
This implies that a good approximation of the full momentum and energy dependence of the self-energy in the 
band (quasiparticle) basis is given by
\beq\label{eq:sig_local_ansatz}
\Sigma_{\nu\nup}(\omega,\bk)\,=\,\sum_{m} U^*_{m \nu}\left(\bk\right)\, \Sigma_{m}(\omega)\, U_{m\nup}\left(\bk\right)\, .
\eeq
The physical content of this expression is that the angular (momentum) dependence
of the quasiparticle self-energies emphasized above is actually due to
the matrix elements $U_{m\nu}\left(\bk\right)$ defined in Eq.~\ref{eq:U}.
In \SRO{} the angular dependence of these matrix elements is mainly due to the SOC, as seen from
the variation of the orbital content of quasiparticles in Fig.~\ref{fig:f2}~(g).
In appendix~\ref{sec:appB4} we show the back-transform of $\Sigma'_{m}\left(\omega,\theta_\bk = 18^\circ\right)$ into $\Sigma'_{\nu}\left(\omega,\theta_\bk=0,18,45^\circ\right)$. The good agreement with $\Sigma'_{\nu\nu}$ directly extracted from experiment further justifies the above expression and also confirms the validity of the approximations made throughout this section.

Finally, we stress that expression~(\ref{eq:sig_local_ansatz}) precisely coincides with the {\it ansatz} made by DMFT: 
within this theory, the self-energy is approximated as a local \mbox{($\bk$-independent)} object {\it when expressed in a basis of localized orbitals}, while it acquires momentum dependence when transformed to the band basis.

\begin{figure*}[tb]
	\includegraphics[width=0.98 \textwidth]{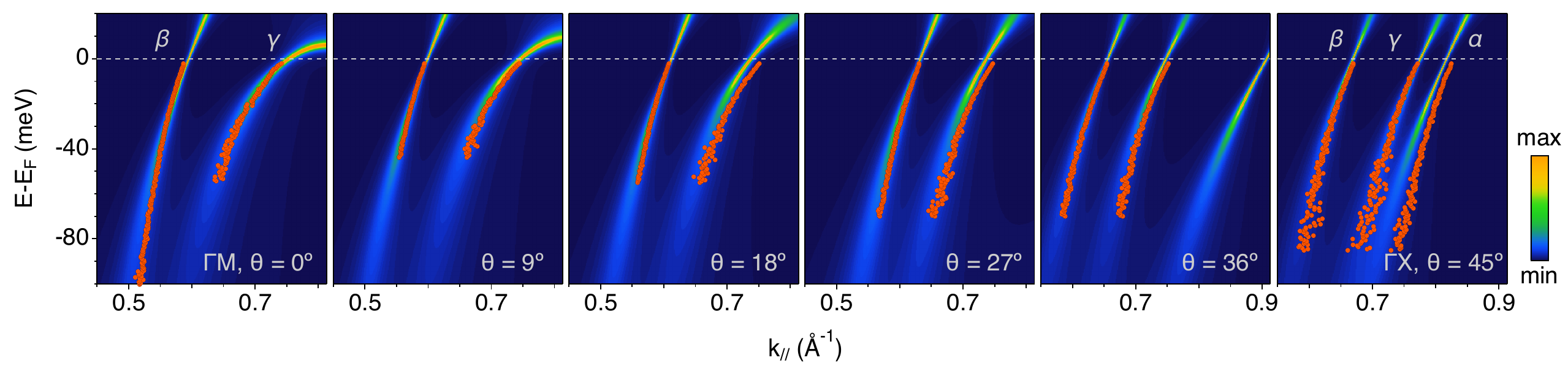} %
	\caption{Comparison of experimental quasiparticle dispersions (markers) with DMFT spectral functions (color plots) calculated for different Fermi surface angles $\theta$. }
	\label{fig:f5}
\end{figure*}

\begin{figure}[tb]
	\includegraphics[width=0.45 \textwidth]{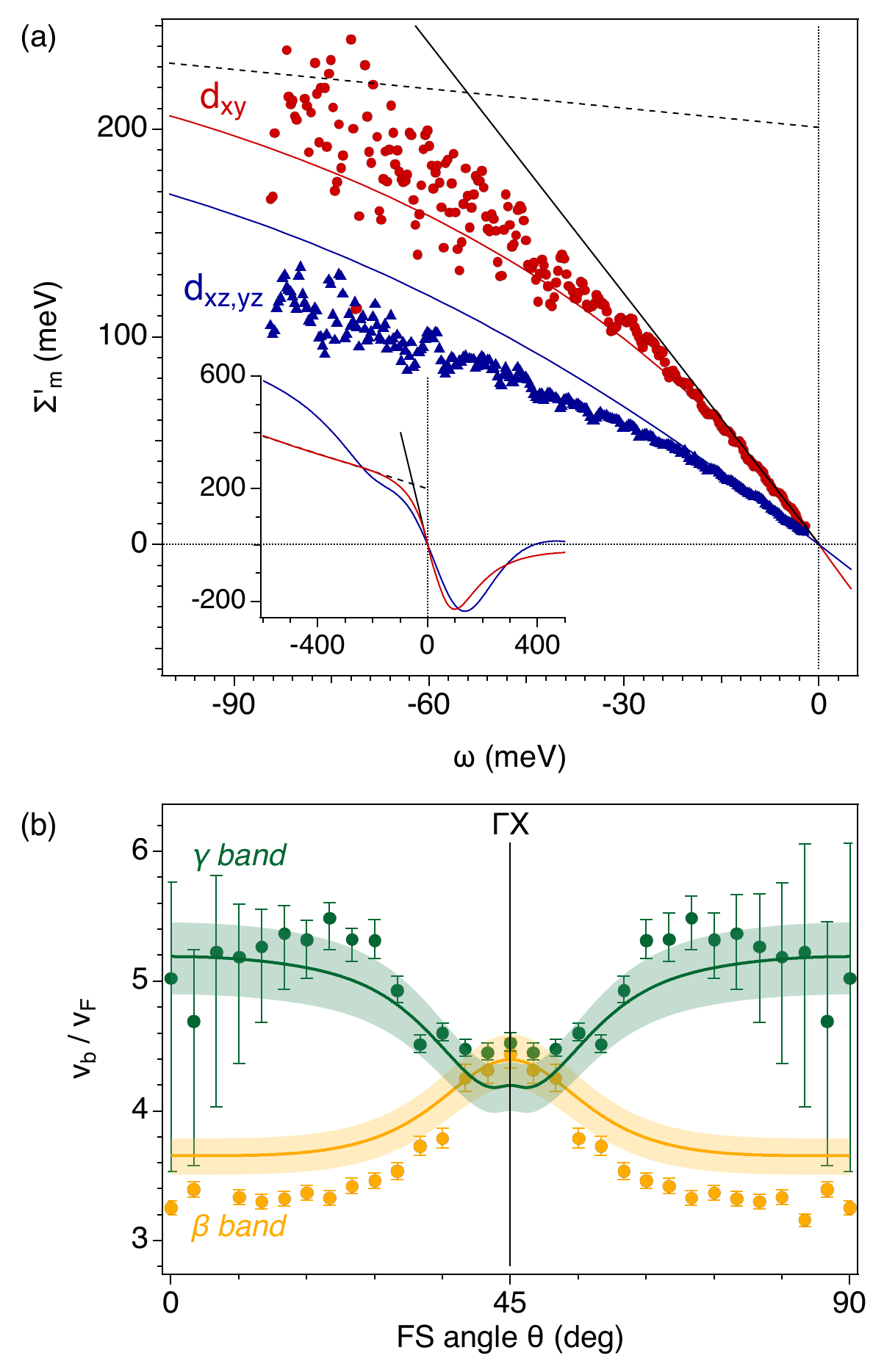} %
	\caption{(a) Average of the self-energies $\Sigma'_{xz/yz}$, $\Sigma'_{xy}$ shown in Fig.~\ref{fig:f4}~(d) compared with DMFT self-energies calculated at \SI{29}{K}. The self-energies are shifted such that $\Sigma'_{m}(\omega = 0) = 0$. The inset shows the DMFT self-energies over a larger energy range. Linear fits at low and high energy of $\Sigma'_{xy}$ from DMFT are shown as solid and dashed black lines, respectively. (b) Angle dependence of the mass enhancement $v_{b}/v_{F}$ from ARPES (markers) and DMFT (solid line). The range indicated by the shaded areas corresponds to mass enhancements calculated from the numerical data by using different methods (see appendix~\ref{sec:appB}). Error bars on the experimental data are obtained from propagating the estimated uncertainty of the Fermi velocities shown in Fig.~\ref{fig:f1bis}~(c).}
	\label{fig:f6}
\end{figure}

\section{Comparison to Dynamical Mean-Field Theory}
\label{sec:dmft}

In this section we perform an explicit comparison of the measured quasiparticle dispersions and self-energies to DMFT results.
The latter are based on the Hamiltonian \Hdft{}, to which the Hubbard-Kanamori interaction with on-site 
interaction $U = \SI{2.3}{eV}$ and Hund's coupling $J=\SI{0.4}{eV}$~\cite{Mravlje2011} is added. 
For the details of the DMFT calculation and especially the treatment of SOC in this framework, we refer the reader to appendix~\ref{sec:appB}. There, we also comment on some of the limitations and shortcomings of the current state of the art for DFT+DMFT calculations in this context.   
Fig.~\ref{fig:f5}~(a) shows the experimental quasiparticle dispersion extracted from our ARPES data (circles) on top of the DMFT spectral function $A\left(\omega,\bk\right)$ displayed as a color-intensity map. 
Clearly, the theoretical results are in near quantitative agreement with the data: both the strong renormalization of the Fermi velocity and the angular-dependent curvature of the quasiparticle bands are very well reproduced \added{by the purely local, and thus momentum-independent DMFT self-energies}. 
\added{This validates the assumption of no momentum dependence along the radial $k$-space direction of the self-energy made in Sec.~\ref{sec:self_orb} for the $k/\omega$ range studied here.}
The small deviations in Fermi wave vectors discernible in Fig.~\ref{fig:f5} are consistent with Fig.~\ref{fig:f2}(c) and the overall experimental precision of the Fermi surface determination.

In Fig.~\ref{fig:f6}~(a), we compare the experimental self-energies for each orbital with the DMFT results. 
The overall agreement is notable. At low-energy, the self-energies are linear in frequency and the agreement is excellent. 
The slope of the self-energies in this regime controls the angular-dependence of the effective mass renormalisation. 
Using the local ansatz (\ref{eq:sig_local_ansatz}) into the quasiparticle dispersion equation, and performing an expansion 
around $E_F$, we obtain
\beq
\frac{v_b}{v_F^\nu(\theta)}\,=\,\sum_m \frac{1}{Z_m}\, |U_{m\nu}(\theta)|^2\,\,\,,\,\,\,
\frac{1}{Z_m} \equiv 1-\left.\frac{\partial\Sigma^\prime_m}{\partial\omega}\right|_{\omega=0} \, .
\eeq
In Fig.~\ref{fig:f6}~(b), we show  $v_b/v_F^\nu(\theta)$ for the $\beta$ and $\gamma$ bands using the DMFT values $Z_{xy}=0.18\pm0.01$ and $Z_{xz/yz}=0.30\pm0.01$ obtained at \SI{29}{K} (appendix~\ref{sec:appB2}). The overall angular dependence and the absolute value of the $\gamma$ band mass enhancement is very well captured by DMFT, while the $\beta$ band is a bit overestimated. Close to the zone-diagonal ($\theta=45^\circ$), the two mass enhancements are approximately equal, due to the strong orbital mixing induced by the SOC. 

Turning to larger binding energies, we see that the theoretical $\Sigma'_{xy}$ is in \replaced{good}{remarkable} agreement with the experimental data over the full energy range of \SIrange{2}{80}{meV} covered in our experiments. Both the theoretical and experimental self-energies deviate significantly from the linear regime down to low energies ($\sim\SI{20}{meV}$), causing curved quasiparticle bands with progressively steeper dispersion as the energy increases (Fig.~\ref{fig:f5}).
In contrast, the agreement between theory and experiment for the $xz/yz$ self-energy is somewhat less impressive at binding energies larger than $\sim\SI{ 30}{meV}$. Our DMFT self-energy $\Sigma'_{xz/yz}$ overestimates the strength of correlations in this regime (by $20-25\%$), with a theoretical slope larger than the experimental one. Correspondingly, the quasiparticle dispersion is slightly steeper in this regime than the theoretical result, as can be also seen in Fig.~\ref{fig:f5}. 

There may be several reasons for this discrepancy. Even while staying in the framework of a local self-energy, we note that the present DMFT calculation is performed 
with an on-site value of $U$ which is the same for all orbitals. Earlier cRPA calculations have suggested 
that this on-site interaction is slightly larger for the $xy$ orbital ($U_{xy} = \SI{2.5}{eV}$ and $U_{xz/yz} = \SI{2.2}{eV}$)~\cite{Mravlje2011} and recent work has advocated the relevance of this for DFT+DMFT calculations of \SRO{}~\cite{Zhang2016}.
Another possible explanation is that this discrepancy is actually a hint of some momentum dependent contribution to the self-energy, especially dependence on momentum perpendicular to the Fermi surface. We note in this respect 
that the discrepancy is larger for the $\alpha,\beta$ sheets which have dominant $xz/yz$ character. 
These orbitals have, in the absence of SOC, a strong one-dimensional character, for which momentum 
dependence is definitely expected and DMFT is less appropriate. Furthermore, these FS sheets are also the ones associated with nesting and spin-density wave correlations, which are expected to lead to an additional momentum-dependence of the self-energy. 
We further discuss possible contributions of spin fluctuations in Sec.~\ref{sec:perspectives}.

\section{Kinks}
\label{sec:kinks}

The self-energies $\Sigma'\left(\omega\right)$ shown in Figs.~\ref{fig:f4} and \ref{fig:f6} display a fairly smooth curvature, rather than pronounced `kinks'. 
Over a larger range, however, $\Sigma'_{xy}$ from DMFT does show an energy scale marking the crossover from the strongly renormalized low-energy regime to weakly renormalized excitations. This is illustrated in the inset to Fig.~\ref{fig:f6}~(a). Such purely electronic kinks were reported in DMFT calculations of a generic system with Mott-Hubbard sidebands~\cite{Byczuk2007} and have been abundantly documented \added{in the theoretical literature} since then~\cite{Raas2009,Mravlje2011,Held2013,Deng2013,Zitko2013,Stricker2014}. In \SRO{} they are associated with the crossover from the Fermi-liquid behavior into a more incoherent regime~\cite{Mravlje2011,Georges2013}. 
The near quantitative agreement of the frequency dependence of the experimental self-energies $\Sigma'_{m}\left(\omega\right)$ and our single-site DMFT calculation provides strong evidence for the existence of such electronic kinks in \SRO.
In addition, it implies that the local DMFT treatment of electronic correlations is capturing the dominant effects.

Focusing on the low-energy regime \added{of our experimental data}, we find deviations from the linear form $\Sigma'\left(\omega\right)=\omega(1-1/Z)$ characteristic of a Fermi liquid for $|\omega|> \SI{20}{meV}$, irrespective of the basis. However, this is only an upper limit for the Fermi-liquid energy scale in \SRO. Despite the improved resolution of our experiments, we cannot exclude an even lower crossover energy to non-Fermi-liquid like excitations. 
We note that \deleted{this is consistent with} the crossover temperature of $T_{\mathrm{FL}}\approx\SI{25}{K}$ reported from transport  and thermodynamic experiments~\cite{Mackenzie1996b,Maeno1997,Bergemann2003} \added{indeed suggests a cross-over energy scale that is significantly below \SI{20}{meV}}.

The overall behavior of $\Sigma'\left(\omega\right)$ including the energy range where we find strong changes in the slope agrees with previous photoemission experiments, which were interpreted as evidence for electron-phonon coupling~\cite{Aiura2004,Iwasawa2005,Ingle2005,Iwasawa2010,Iwasawa2013}. 
Such an interpretation, however, relies on a linear Fermi-liquid regime of electronic correlations over the entire phonon bandwidth of $\approx\SI{90}{meV}$~\cite{Braden2007}, which is inconsistent with our DMFT calculations. Moreover, attributing the entire curvature of $\Sigma'_m$ in our data to electron-phonon coupling would result in unrealistic coupling constants far into the polaronic regime, which is hard to reconcile with the transport properties of \SRO~\cite{Mackenzie1996b,Maeno1997,Bergemann2003}.
We also note that a recent scanning tunneling microscope (STM) study reported very strong kinks in the $\beta$ and $\gamma$ sheets of \SRO~\cite{Wang2017}, which is inconsistent with our data. We discuss the reason for this discrepancy in appendix~\ref{sec:appA}.

\section{Discussion and Perspectives}
\label{sec:perspectives}
In this article, we have reported on high-resolution ARPES measurements which allow for a determination of the Fermi surface 
and quasiparticle dispersions of \SRO{} with unprecedented accuracy. 
Our data reveal an enhancement (by a factor of about two) of the splitting between Fermi surface sheets along the zone diagonal, 
in comparison to the DFT value. 
This can be interpreted as a correlation-induced enhancement of the effective SOC, 
an effect predicted theoretically~\cite{liu_prl_2008,Zhang2016,Kim2018} and demonstrated experimentally here for this material, for the first time.

Thanks to the \replaced{improved cleanliness of our data}{high resolution}, we have been able to determine the electronic self-energies directly from \replaced{experiment}{the ARPES data},
using both a standard procedure applied in the band (quasiparticle) basis as well as a novel procedure, introduced in the present article, in the orbital basis.  
Combining these two approaches, we have demonstrated that the large angular (momentum) dependence of the 
quasiparticle self-energies and dispersions can be mostly attributed to the fact that quasiparticle states have 
an orbital content which is strongly angular dependent, due to the SOC. 
Hence, assuming self-energies which are frequency-dependent but essentially independent of angle (momentum), when considered in the orbital basis, is a very good approximation. 
This provides a direct experimental validation of the DMFT {\it ansatz}.\deleted{, and indeed, the comparison between the ARPES 
data and DMFT calculations is found to be remarkably accurate.}

\added{The key importance of atomic-like orbitals in correlated insulators is well established~\cite{kugel_khomskii_1982,tokura_nagaosa_orbital_2000}. The present work demonstrates that orbitals retain a considerable physical relevance even in a metal in the low-temperature Fermi-liquid regime. \deleted[id=2nd]{Thinking in terms of quasiparticles associated with a specific Fermi surface sheet is insufficient to unravel the hidden simplicity in the nature of correlations.} \added[id=2nd]{Although the band and orbital basis used here are equivalent, our analysis shows that the underlying simplicity in the nature of correlations emerges only when working in the latter and taking into account the orbital origin of quasiparticles.} Beyond \SRO{}, this is an observation of general relevance to multiband metals with strong correlations such as the iron-based superconductors~\cite{Miao2016,Sprau75}.}

Notwithstanding its success, the excellent agreement of the DMFT results with ARPES data does raise puzzling questions. 
\SRO{} is known to be host to strong magnetic fluctuations~\cite{Sidis1999,steffens_2018,ishida_2001,imai_1998}, 
with a strong peak in its spin response 
$\chi({\mathbf{Q}})$ close to the spin-density wave (SDW)  
vector $\mathbf{Q}\sim (2\pi/3,2\pi/3)$, as well as quasi-ferromagnetic fluctuations which are broader in momentum around 
$\mathbf{Q}=0$. Indeed, tiny amounts of substitutional impurities induce long-range magnetic order in this material, of 
either SDW of ferromagnetic type~\cite{braden_impurities_2002,ortmann_impurities_2013}. 
Hence, it is a prominent open question to understand how these long-wavelength fluctuations affect 
the physics of quasiparticles in the Fermi-liquid state. Single-site DMFT does not capture this feedback,
and the excellent agreement with the overall quasiparticle physics must imply that these effects have a
comparatively smaller magnitude than the dominant local effect of correlations (on-site $U$ and especially Hund's $J$)
captured by DMFT. A closely related question is how much momentum dependence is present in the low-energy (Landau) interactions between quasiparticles. These effects are expected to be fundamental for subsequent instabilities of the 
Fermi liquid, into either the superconducting state in pristine samples or magnetic ordering in samples with 
impurities. Making progress on this issue is also key to the understanding of the superconducting state of \SRO{}, for 
which the precise nature of the pairing mechanism as well as symmetry of the order parameter are still 
outstanding open questions~\cite{Mackenzie2017}.

\begin{acknowledgments}

The experimental work has been supported by the European Research Council (ERC), the Scottish Funding Council, the UK EPSRC and the Swiss National Science Foundation (SNSF). 
Theoretical work was supported by the ERC grant ERC-319286-QMAC and by the SNSF (NCCR MARVEL). 
The Flatiron Institute is a division of the Simons Foundation. 
AG and MZ gratefully acknowledge useful discussions with Gabriel Kotliar, Andrew J.~Millis and Jernej Mravlje.
\end{acknowledgments}

\appendix

\section{Bulk and surface electronic structure of \SRO}
\label{sec:appA}

\begin{figure}[hb]
	\includegraphics[width=0.44 \textwidth]{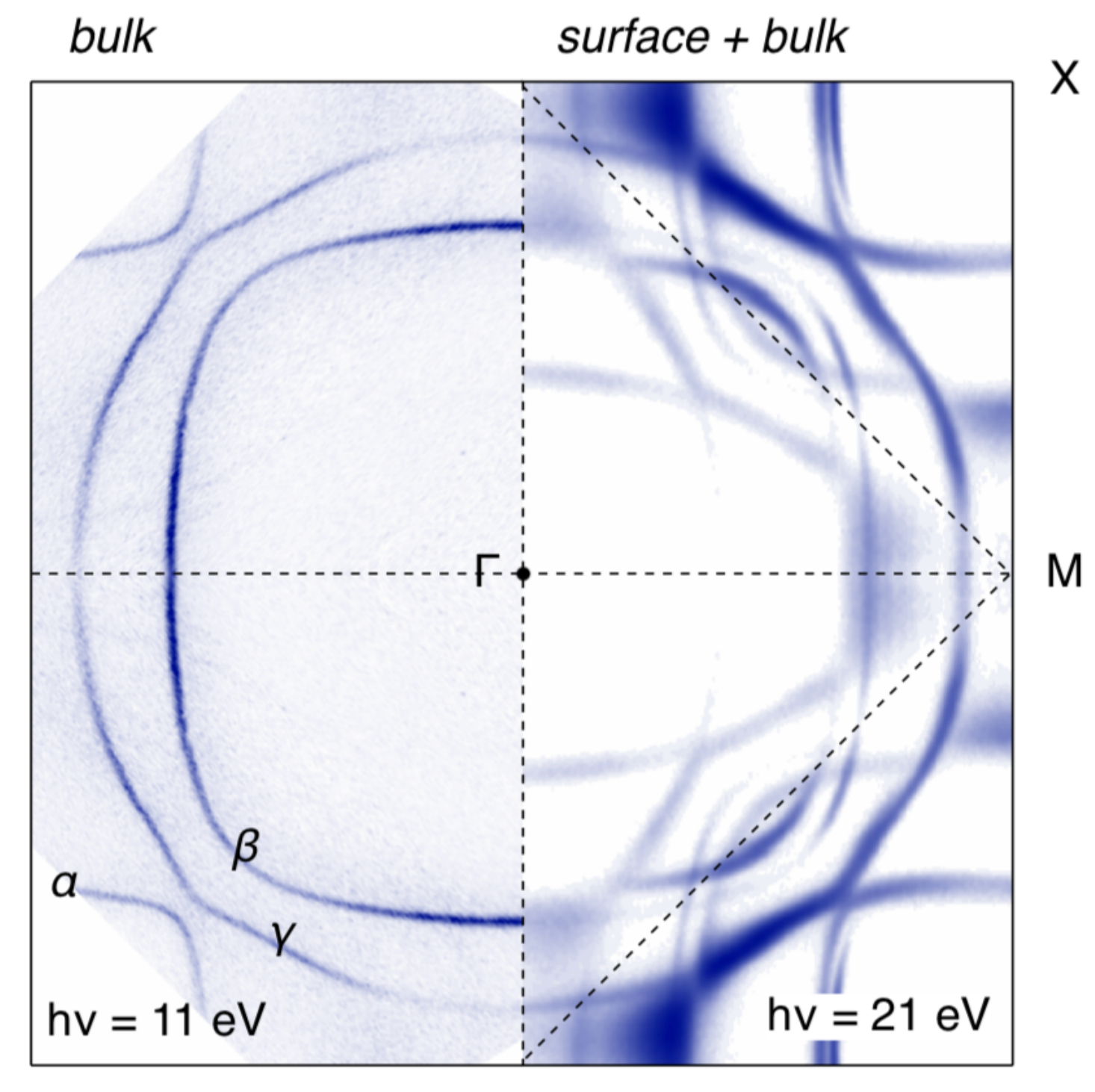} %
	\caption{Bulk and surface electronic structure of \SRO. Left half: Fermi surface map of a CO passivated surface shown in the main text. Right half: Fermi surface acquired on a pristine cleave at \SI{21}{eV} photon energy. Intense surface states are evident in addition to the bulk bands observed in the left panel. }
	\label{fig:fA1}
\end{figure}

\begin{figure}[ht]
	\includegraphics[width=0.49 \textwidth]{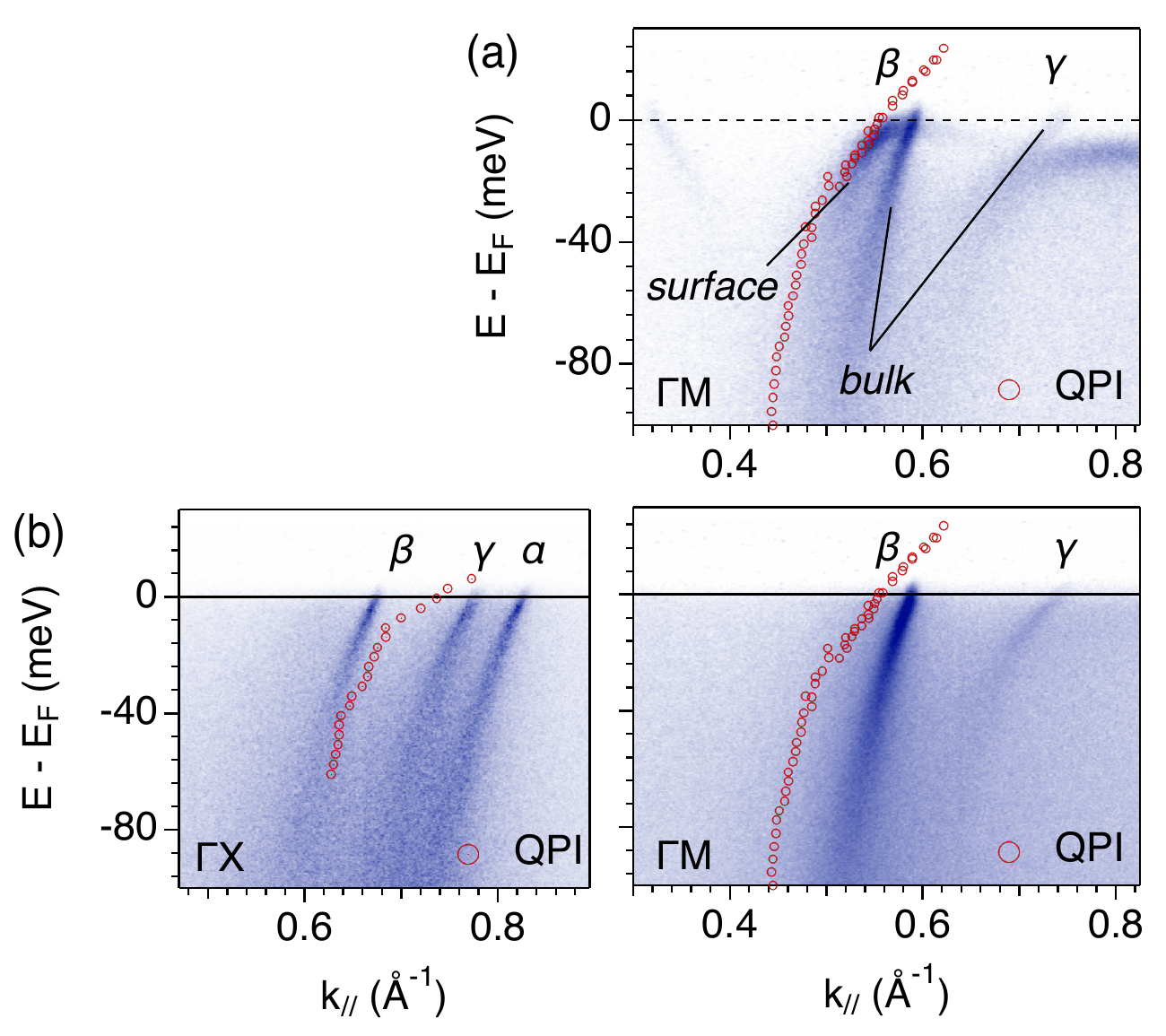} %
	\caption{Comparison with the STM data from Ref.~\cite{Wang2017}. (a) $\Gamma$M high-symmetry cut on a pristine cleave. The bulk $\beta$ and $\gamma$ and the surface $\beta$ bands are labeled. (b) Laser-ARPES data from CO passivated surface showing the bulk band dispersion along $\Gamma$X and $\Gamma$M. The dispersion obtained from quasiparticle interference in Ref.~\cite{Wang2017} is overlaid with red markers.}
	\label{fig:fA2}
\end{figure}

In Fig.~\ref{fig:fA1} we compare the data presented in the main text with data from a pristine cleave taken with $h\nu=\SI{21}{eV}$ at the SIS beamline of the Swiss Light Source. This comparison confirms the identification of bulk and surface bands by Shen~\etal{}~\cite{Shen2001}. In particular, we find that the larger $\beta$ sheet has bulk character. This band assignment is used by the vast majority of subsequent ARPES publications~\cite{Shen2007,Aiura2004,Wang2004,Iwasawa2005,Ingle2005,Kidd2005,Iwasawa2010,Zabolotnyy2013}, except for Ref.~\cite{Kim2011}, which reports a dispersion with much lower Fermi velocity and a strong kink at \SI{15}{meV} for the smaller $\beta$ sheet that we identify as a surface band.

Wang~\etal{}~\cite{Wang2017} have recently probed the low-energy electronic structure of \SRO{} by STM. Analyzing quasiparticle interference patterns along the $\Gamma$X and $\Gamma$M high-symmetry directions, they obtained band dispersions with low Fermi velocities and strong kinks at \SI{10}{meV} and \SI{37}{meV}.
In Fig.~\ref{fig:fA2} we compare the band dispersion reported by Wang~\etal{} with our ARPES data. Along both high-symmetry directions, we find a clear discrepancy with our data, which are in quantitative agreement with bulk de Haas van Alphen measurements, as demonstrated in the main text. On the other hand, we find a striking similarity between the STM data along $\Gamma$M and the band commonly identified as the surface $\beta$ band~\cite{Shen2001,Ingle2005}. We thus conclude that the experiments reported in Ref.~\cite{Wang2017} probed the surface states of \SRO. This is fully consistent with the \added{strong low-energy kinks and overall
enhanced low-energy renormalization seen by ARPES in the surface bands~\cite{Kondo2016,Shuntaro2019}.}

\section{Computational details}\label{sec:appB}
\subsection{DFT and model Hamiltonian}\label{sec:appB1}

These orbitals are centered on the Ru atoms and have $t_{2g}$ symmetry, but are indeed linear combinations of Ru-$d$ and O-$p$ states. We do not add Wannier functions centered on the oxygen atoms, because the resulting three orbital Wannier model already accurately reproduces the three bands crossing the Fermi energy, as demonstrated in Fig. 9. Also note that the Wannier function construction allows to disentangle the $\gamma$ band from the bands with dominantly O-$p$ character below \SI{-2}{eV}.

\begin{figure}[t]
	\centering
	\includegraphics[width=0.43 \textwidth]{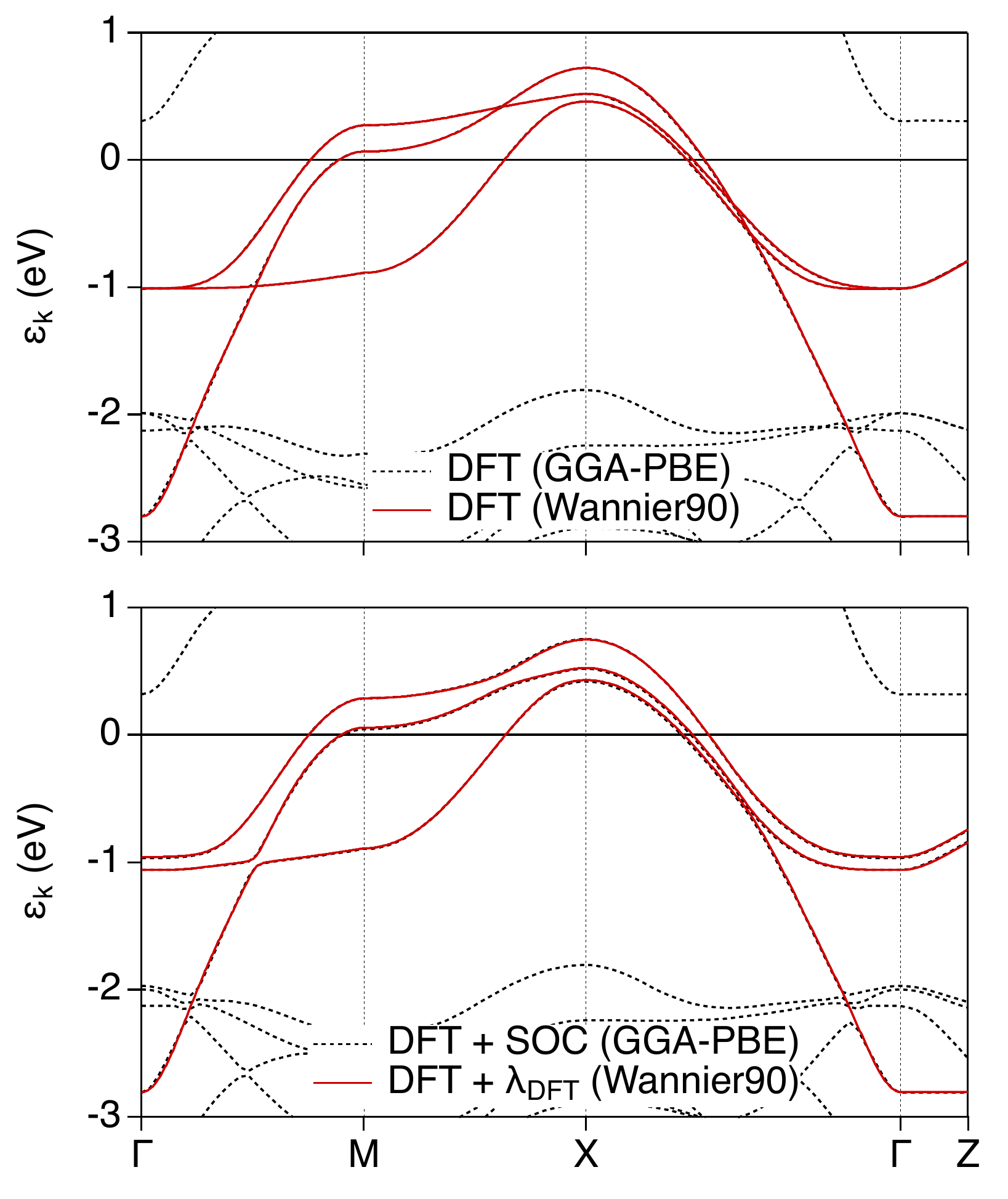}
	\caption{DFT band structure along the high-symmetry path $\Gamma$MX$\Gamma$Z compared to the eigenstates of our maximally-localized Wannier Hamiltonian \Hdft{} for the three $t_{2g}$ bands. Top: DFT (GGA-PBE) and eigenstates of \Hdft{}. Bottom: DFT+SOC (GGA-PBE) and eigenstates of \Hdft+\Hsocdft{} with a local SOC term (Eq.~\ref{eq:hsoc}) and a coupling strength of $\lambda_{\mathrm{DFT}} = \SI{100}{meV}$.}
	\label{fig:fB1}
\end{figure}

\begin{figure}[t]
	\centering
	\includegraphics[width=0.41 \textwidth]{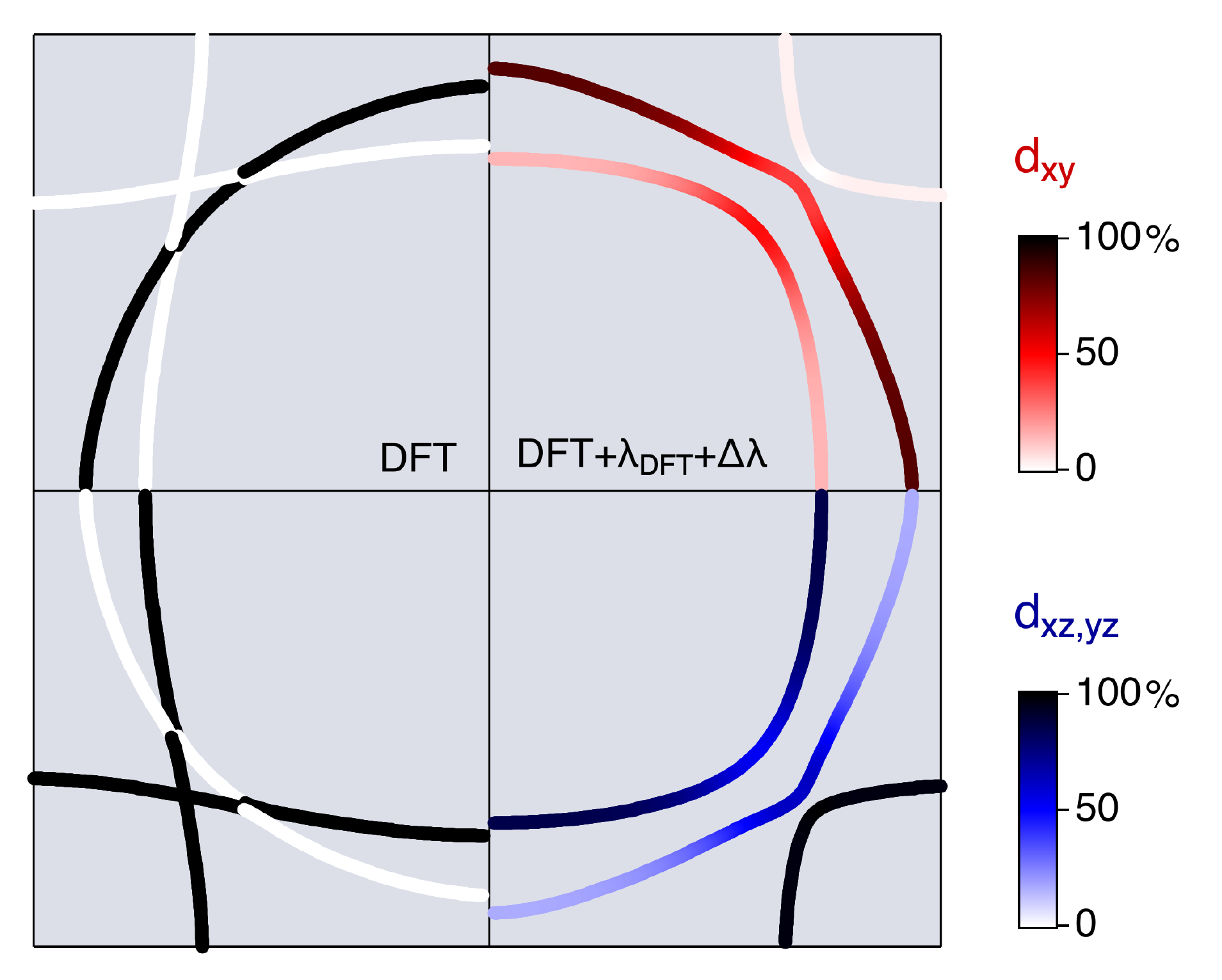}
	\caption{Orbital character of the DFT FS without SOC at $k_z = 0.4~\pi/c$ (left). The orbital character of the DFT+$\lambda_{\mathrm{DFT}}+\Delta\lambda$ eigenstates at the same $k_z$ is reproduced on the right from Fig.~\ref{fig:f2}.}
	\label{fig:fB1_bis}
\end{figure}

We generate our theoretical model Hamiltonian \added{\Hdft} with a maximally-localized Wannier function~\cite{MLWF1,MLWF2} \replaced{construction of $t_{2g}$-like orbitals for the three bands crossing the Fermi surface}{for the three Ru-$4d$ $t_2g$ orbitals}. These \added{Wannier orbitals} are \replaced{obtained}{constructed} on a $10\times 10\times 10$ $\bk$ grid based on a non-SOC DFT calculation using WIEN2k~\cite{Blaha2001} with the GGA-PBE functional~\cite{PBE}, wien2wannier~\cite{wien2wannier} and Wannier90~\cite{wannier90}. The DFT calculation is performed with lattice parameters from Ref.~\cite{Vogt1995} (measured at \SI{100}{K}) and converged with twice as many $\bk$-points in each dimension.

The eigenenergies of the resulting Wannier Hamiltonian, \Hdft{}, accurately reproduce the DFT band structure (Fig.~\ref{fig:fB1}~top). 
\added{Note that in the absence of SOC, the eigenstates retain pure orbital character, as shown in Fig.~\ref{fig:fB1_bis}.}
To take SOC into account, we add the local single-particle term \Hsoclam{}, as given in Eq.~\ref{eq:hsoc}, with coupling constant $\lambda$. In the bottom panel of Fig.~\ref{fig:fB1} we show that the eigenenergies of \Hdft+\Hsoclam{}
are in nearly perfect agreement with the DFT+SOC band structure at a value of $\lambda_{\mathrm{DFT}}=\SI{100}{meV}$.

\begin{figure}[t]
	\centering
	\includegraphics[width=0.4 \textwidth]{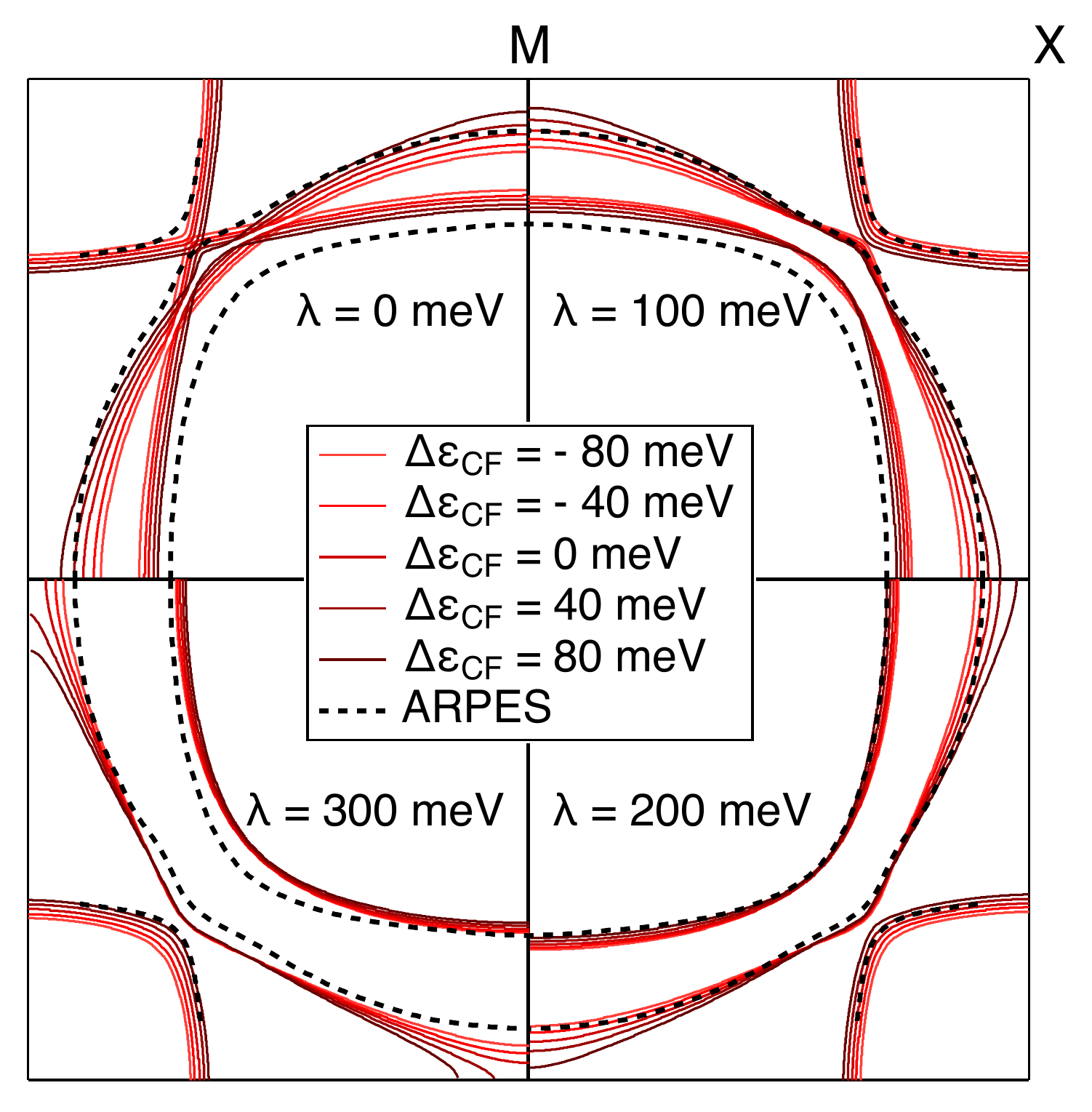} %
	\caption{Fermi surface of \SRO{} for $\lambda=\SI{0}{meV}$ (top left), $\lambda=\SI{100}{meV}$ (top right), $\lambda=\SI{200}{meV}$ (bottom right) and $\lambda=\SI{300}{meV}$ (bottom left) compared to ARPES (dashed black line). $\lambda=\SI{100}{meV}$ corresponds to a DFT+SOC calculation and $\lambda=\SI{200}{meV}$ to an effective SOC enhanced by electronic correlations (see main text). 
		The different shades of red indicate additional crystal-field splittings \Dcf{} added to \epscf{} = \SI{85}{meV} of \Hdft.}
	\label{fig:fB2}
\end{figure}

Our model Hamiltonian provides the reference point to which we define a self-energy, but it is also a perfect playground to study the change in the Fermi surface under the influence of SOC and the crystal-field splitting between the $xy$ and $xz/yz$ orbitals. In the following, we will confirm that the best agreement with the experimental Fermi surface is found with an effective SOC of $\lambda^{\mathrm{eff}} = \lambda_{\mathrm{DFT}}+\Delta\lambda=\SI{200}{meV}$, but at the same time keeping the DFT crystal-field splitting of \mbox{\epscf{} $ = \epsilon_{xz/yz} - \epsilon_{xy}  = \SI{85}{meV}$} unchanged. We compare in Fig.~\ref{fig:fB2} the experimental Fermi surface (dashed lines) to the one of \Hdft+\Hsocdft. The Fermi surfaces for additionally introduced crystal-field splittings \Dcf{} between \num{-80} and \SI{80}{meV} are shown with solid lines in different shades of red. 
In contrast to the Fermi surface without SOC ($\lambda=\SI{0}{meV}$, top left panel), the Fermi surfaces with the DFT SOC of $\lambda=\SI{100}{meV}$ (top right panel) resembles the overall structure of the experimental Fermi surface. However, the areas of the $\alpha$ and $\beta$ sheets are too large and the $\gamma$ sheet is too small. Importantly, the agreement cannot be improved by adding \Dcf{}. For example, along $\Gamma$M a \Dcf{} of \SI{-40}{meV} would move the Fermi surface closer to the experiment, but, on the other hand, along $\Gamma$X a \Dcf{} of \SI{80}{meV} would provide the best agreement. The situation is different if we consider an enhanced SOC of $\lambda=\SI{200}{meV}$ (bottom right panel). Then, we find a nearly perfect agreement with experiment without any additional crystal-field splitting (\Dcf$=\SI{0}{meV}$). At an even higher SOC of $\lambda=\SI{300}{meV}$ (bottom left panel) we see again major discrepancies, but with an opposite trend: The $\alpha$ and $\beta$ sheets are now too small and the $\gamma$ sheet is too large. Like in the case of $\lambda=\SI{100}{meV}$ this can not be cured by an adjustment of \epscf.

\begin{figure}[t]
	\centering
	\includegraphics[width=0.38 \textwidth]{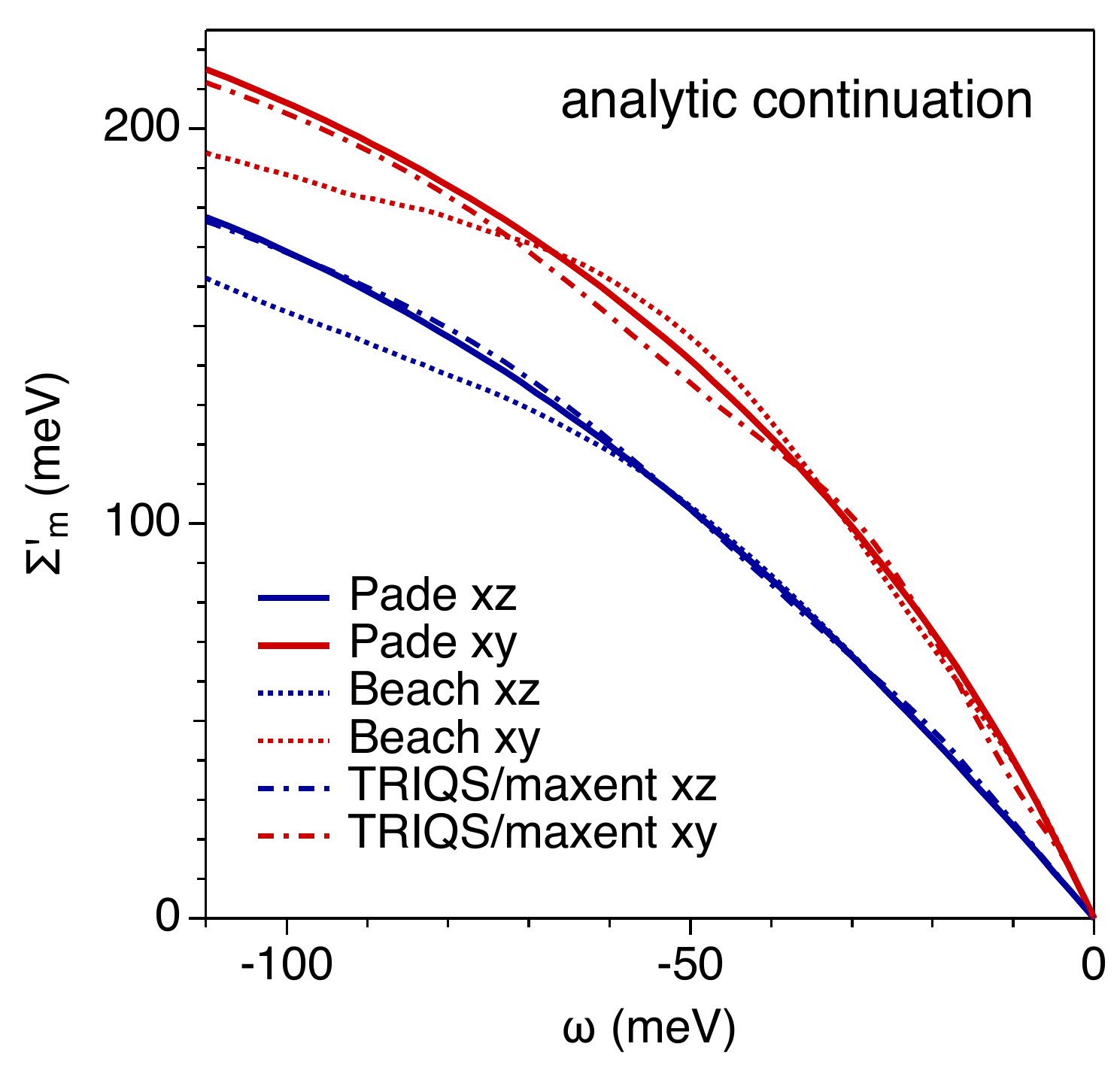} %
	\caption{Real part of DFT+DMFT self-energy in the considered energy range obtained with three different analytic continuation methods: Pad\'e approximants (using TRIQS~\cite{TRIQS}), Stochastic continuation (after Beach~\cite{MaxEntBeach}) and Maximum Entropy (using TRIQS/maxent~\cite{TRIQS/MAXENT}). The difference between the analytic continuation methods is smaller than the experimental error.}
	\label{fig:fB3}
\end{figure}

\subsection{DMFT}\label{sec:appB2}
We perform single-site DMFT calculation with the TRIQS/DFTTools~\cite{TRIQS/DFTTOOLS} package for \Hdft{} and Hubbard-Kanamori
interactions with a screened Coulomb repulsion $U = \SI{2.3}{eV}$ and a Hund's coupling
$J=\SI{0.4}{eV}$ based on previous works~\cite{Mravlje2011,Kim2018}.
The impurity problem is solved on the imaginary-time axis with the TRIQS/CTHYB~\cite{TRIQS/CTHYB} solver at a temperature of \SI{29}{K}. 
The employed open-source software tools are based on the TRIQS library~\cite{TRIQS}.
We assume an orbital-independent double counting, and hence it can be absorbed 
into an effective chemical potential, which is adjusted such that the filling
is equal to four electrons. For the analytic continuation of the self-energy to the real-frequency axis we employ three different methods: 
Pad\'e approximants (using TRIQS~\cite{TRIQS}), Stochastic continuation (after Beach~\cite{MaxEntBeach}) and Maximum Entropy (using TRIQS/maxent~\cite{TRIQS/MAXENT}). In the relevant energy range from \num{-100} to \SI{0}{meV}
the difference in the resulting self-energies (Fig.~\ref{fig:fB3}) is \replaced{below the experimental uncertainty}{smaller than the experimental resolution} .
The averaged quasiparticle renormalizations (of the three continuations) are: $Z_{xy}=0.18\pm0.01$ and $Z_{xz} = Z_{yz} = 0.30\pm0.01$. For all other results presented in the main text the Pad\'e solution has been used.

Our calculations at a temperature of \SI{29}{K} use \Hdft{}, as the sign problem prohibits reaching such low temperatures with SOC included.
Nevertheless, calculations with SOC were successfully carried out at a temperature of \SI{290}{K} using CT-INT~\cite{Zhang2016} and at \SI{230}{K} using CT-HYB with a simplified two-dimensional tight-binding model~\cite{Kim2018}. These works
pointed out that electronic correlations in \SRO{} lead to an enhanced SOC. To be more precise, Kim~\etal{}~\cite{Kim2018} observed that electronic correlations in this material are described by a self-energy with diagonal elements close to the ones without SOC plus, to a good approximation, frequency-independent off-diagonal elements, which can be absorbed in a static effective SOC strength of $\lambda^{\mathrm{eff}}=\lambda_{\mathrm{DFT}}+\Delta\lambda\simeq\SI{200}{meV}$ -- this is the approach followed in the present article.

\begin{figure*}[t]
	\centering
	\includegraphics[width=0.98 \textwidth]{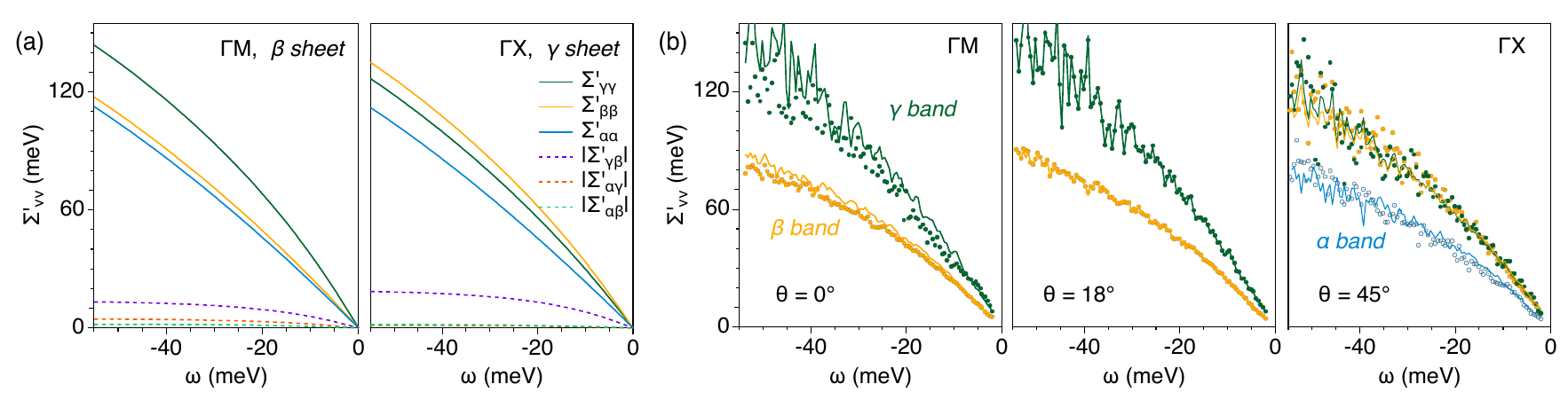} %
	\caption{Reconstructed self-energies in the quasiparticle basis. (a) Full matrix $\Sigma'_{\nu\nup}(\omega,\bk)$ calculated with Eq.~\ref{eq:sig_local_ansatz} using the DMFT self-energy (shown in Fig.~\ref{fig:f6}) at two selected $\bk$ points: on the $\beta$ sheet for $\theta = 0^{\circ}$ (left) and on the $\gamma$ sheet for $\theta = 45^{\circ}$ (right). (b) Directly extracted $\Sigma'_{\nu\nu}\left(\omega,\kmaxnuom\right)$ (Sec.~\ref{subsec:self_qp}) compared to the $\Sigma'_{m}(\omega,\theta = 18^{\circ})$ (Sec.~\ref{sec:self_orb}) transformed to $\Sigma'_{\nu\nu}\left(\omega,\kmaxnuom\right)$ at 0, 18, and 45$^\circ$.}
	\label{fig:fB4}
\end{figure*}

In addition to the enhancement of SOC it was observed that 
\replaced{low-energy many-body effect}{DFT+DMFT} also lead to an enhancement of the crystal-field splitting~\cite{Zhang2016, Kim2018}. 
In our DFT+DMFT calculation without SOC this results in \added{a orbital-dependent splitting in the real part of the self-energies} (\Dcf{} $=\SI{60}{meV}$), which would move the $\gamma$ sheet \replaced{closer}{nearly} to \added{the} van Hove \replaced{singularity}{filling} and consequently worsen the agreement with the experimental Fermi surface \added{along the $\Gamma$M direction} (see bottom right panel of Fig.~\ref{fig:fB2}). Different roots of this \added{small} discrepancy are possible, ranging from orbital-dependent double counting corrections to, in general, DFT being not \replaced{perfect}{ideal} as `non-interacting' reference point for DMFT. Zhang~\etal{}~\cite{Zhang2016} showed that by considering the anisotropy of the Coulomb tensor the additional crystal-field splitting is suppressed and consequently the disagreement between theory and experiment can be cured. \added{We point out that in comparision to the present work a large enhancement of the crystal-field splitting was observed in Ref.~\cite{Zhang2016}, presumably due to the larger interactions employed.}

Based on these considerations we calculate the correlated spectral function $A\left(\omega,\bk\right)$ (shown in Fig.~\ref{fig:f5}~(a) of the main text) using the Hamiltonian with enhanced SOC (\Hdft+\Hsocdlam{}) in combination with the frequency-dependent part of the non-SOC (diagonal) self-energy, but neglect the additional static part introduced by DMFT.

\section{Off-diagonal elements of $\Sigma'_{\nu\nup}$}\label{sec:appB3}

In Sec.~\ref{subsec:self_qp} we extracted $\Sigma'_{\nu\nu}$ under the assumption that the off-diagonal elements can be neglected. To obtain insights about the size of the off-diagonal elements we use Eq.~\ref{eq:sig_local_ansatz} to calculate the full matrix $\Sigma'_{\nu\nup}(\omega,\bk)$ in the band basis from the DMFT self-energy in the orbital basis $\Sigma'_m(\omega)$, as shown in Fig.~\ref{fig:f6}~(a). This allows us to obtain the full self-energy matrix $\Sigma'_{\nu\nup}(\omega,\bk)$ at one specific combination of $\bk$ and $\omega$. Note that for the results presented in Fig.~\ref{fig:f4} and Fig.~\ref{fig:fB4}~(b) this is not the case, because the extracted self-energies for each band correspond to different $\bk^\nu_{\text{max}}$, which are further defined by the experimental MDCs.

In Fig.~\ref{fig:fB4}~(a) we show the result for two selected $\bk$ points: on the $\beta$
sheet for $\theta = 0$ and on the $\gamma$ sheet for $\theta = 45^\circ$. 
For these $\bk$ points the largest off-diagonal element is $\Sigma'_{\gamma\beta}$, which is 
about $10-20\%$ of the size of the diagonal elements. A scan performed for the whole $k_z = 0.4~\pi/c$
plane further confirms that $|\Sigma'_{\nu\neq\nup}|$ is smaller than \SI{20}{meV}. 

However, when neglect the off-diagonal elements it is also important to have a large enough energy separation of the bands. This can be understood by considering a simplified case of two bands ($\nu,\nu'$) and rewriting Eq.~\ref{eq:qp}, which determines the quasiparticle dispersion $\omqp_\nu(\bk)$, as
\beq
\label{eq:sigma_off_det}
\omega-\varepsilon_{\nu}\left(\bk\right)-\Sigma^\prime_{\nu\nu}(\omega, \bk) - \frac{\Sigma'_{\nu\nu'}(\omega, \bk)\,  \Sigma'_{\nu'\nu}(\omega, \bk)}{\omega-\varepsilon_{\nu'}\left(\bk\right)-\Sigma^\prime_{\nu'\nu'}(\omega, \bk)} = 0 \, .
\eeq
Setting the last term to zero, i.e., using the procedure described in Sec.~\ref{subsec:self_qp} to extract $\Sigma^\prime_{\nu\nu}$, is justified at \mbox{$\omega = \omqp_\nu(\bk)$} as long as
\begin{align}
\label{eq:sigma_off_cond}
\Sigma^\prime_{\nu\nu}(\omega, \bk) \gg \frac{\Sigma'_{\nu\nu'}(\omega, \bk)\,  \Sigma'_{\nu'\nu}(\omega, \bk)}{\omega-\varepsilon_{\nu'}\left(\bk\right)-\Sigma^\prime_{\nu'\nu'}(\omega, \bk)} \,
\end{align}
In this condition the already small off-diagonal elements enter quadratically, but also the denominator is not a small quantity, because the energy separation of the bare bands ($\varepsilon_{\nu}\left(\bk\right) - \varepsilon_{\nu'}\left(\bk\right)$) is larger than the difference of the diagonal self-energies.

By using the generalized version of Eq.~\ref{eq:sigma_off_cond} for all three bands, we find that the right-hand side of this equation is indeed less than $1.2\%$ of $\Sigma^\prime_{\nu\nu}(\omega, \kmaxnuom)$ for all experimentally determined $\kmaxnuom$. This means that for \SRO{} treating each band separately when extracting $\Sigma^\prime_{\nu\nu}$ is well justified in the investigated energy range.

\section{Reconstruction of $\Sigma'_{\nu\nu}$}\label{sec:appB4}

In order to further test the validity of the local ansatz (Eq.~\ref{eq:sig_local_ansatz}) and establish the overall consistency of the two procedures used to extract the self-energy in Sec.~\ref{sec:self}, we perform the following `reconstruction procedure'. We use the $\Sigma'_{m}(\omega,\theta_\bk)$ (from Sec.~\ref{sec:self_orb}) at one angle, e.g., $\theta=18^{\circ}$, and transform it into $\Sigma'_{\nu\nu}\left(\omega,\kmaxnuom\right)$ for other measured angles, using Eq.~\ref{eq:sig_local_ansatz}.
The good agreement between the self-energy reconstructed in this manner (thin lines in Fig.~\ref{fig:fB4}~(b)) and its direct determination following the procedure of Sec.~\ref{subsec:self_qp} (dots) confirms the validity of the approximations used throughout Sec.~\ref{sec:self}. It also shows that the origin of the strong momentum dependence of $\Sigma'_{\nu\nu}$ is almost entirely due to the momentum dependence of the orbital content of quasiparticle states, i.e., of $U_{m\nu}\left(\bk\right) = \langle\chikm|\psiknu\rangle$. In \SRO{} the momentum dependence of these matrix elements is mainly due to the SOC.


\begin{thebibliography}{101}%
\makeatletter
\providecommand \@ifxundefined [1]{%
 \@ifx{#1\undefined}
}%
\providecommand \@ifnum [1]{%
 \ifnum #1\expandafter \@firstoftwo
 \else \expandafter \@secondoftwo
 \fi
}%
\providecommand \@ifx [1]{%
 \ifx #1\expandafter \@firstoftwo
 \else \expandafter \@secondoftwo
 \fi
}%
\providecommand \natexlab [1]{#1}%
\providecommand \enquote  [1]{``#1''}%
\providecommand \bibnamefont  [1]{#1}%
\providecommand \bibfnamefont [1]{#1}%
\providecommand \citenamefont [1]{#1}%
\providecommand \href@noop [0]{\@secondoftwo}%
\providecommand \href [0]{\begingroup \@sanitize@url \@href}%
\providecommand \@href[1]{\@@startlink{#1}\@@href}%
\providecommand \@@href[1]{\endgroup#1\@@endlink}%
\providecommand \@sanitize@url [0]{\catcode `\\12\catcode `\$12\catcode
  `\&12\catcode `\#12\catcode `\^12\catcode `\_12\catcode `\%12\relax}%
\providecommand \@@startlink[1]{}%
\providecommand \@@endlink[0]{}%
\providecommand \url  [0]{\begingroup\@sanitize@url \@url }%
\providecommand \@url [1]{\endgroup\@href {#1}{\urlprefix }}%
\providecommand \urlprefix  [0]{URL }%
\providecommand \Eprint [0]{\href }%
\providecommand \doibase [0]{http://dx.doi.org/}%
\providecommand \selectlanguage [0]{\@gobble}%
\providecommand \bibinfo  [0]{\@secondoftwo}%
\providecommand \bibfield  [0]{\@secondoftwo}%
\providecommand \translation [1]{[#1]}%
\providecommand \BibitemOpen [0]{}%
\providecommand \bibitemStop [0]{}%
\providecommand \bibitemNoStop [0]{.\EOS\space}%
\providecommand \EOS [0]{\spacefactor3000\relax}%
\providecommand \BibitemShut  [1]{\csname bibitem#1\endcsname}%
\let\auto@bib@innerbib\@empty
\bibitem [{\citenamefont {Maeno}\ \emph {et~al.}(1994)\citenamefont {Maeno},
  \citenamefont {Hashimoto}, \citenamefont {Yoshida}, \citenamefont
  {Nishizaki}, \citenamefont {Fujita}, \citenamefont {Bednorz},\ and\
  \citenamefont {Lichtenberg}}]{Maeno1994}%
  \BibitemOpen
  \bibfield  {author} {\bibinfo {author} {\bibfnamefont {Y.}~\bibnamefont
  {Maeno}}, \bibinfo {author} {\bibfnamefont {H.}~\bibnamefont {Hashimoto}},
  \bibinfo {author} {\bibfnamefont {K.}~\bibnamefont {Yoshida}}, \bibinfo
  {author} {\bibfnamefont {S.}~\bibnamefont {Nishizaki}}, \bibinfo {author}
  {\bibfnamefont {T.}~\bibnamefont {Fujita}}, \bibinfo {author} {\bibfnamefont
  {J.~G.}\ \bibnamefont {Bednorz}}, \ and\ \bibinfo {author} {\bibfnamefont
  {F.}~\bibnamefont {Lichtenberg}},\ }\bibfield  {title} {\enquote {\bibinfo
  {title} {Superconductivity in a layered perovskite without copper},}\ }\href
  {https://doi.org/10.1038/372532a0} {\bibfield  {journal} {\bibinfo  {journal}
  {Nature}\ }\textbf {\bibinfo {volume} {372}},\ \bibinfo {pages} {532}
  (\bibinfo {year} {1994})}\BibitemShut {NoStop}%
\bibitem [{\citenamefont {Rice}\ and\ \citenamefont
  {Sigrist}(1995)}]{Rice1995}%
  \BibitemOpen
  \bibfield  {author} {\bibinfo {author} {\bibfnamefont {T.~M.}\ \bibnamefont
  {Rice}}\ and\ \bibinfo {author} {\bibfnamefont {M.}~\bibnamefont {Sigrist}},\
  }\bibfield  {title} {\enquote {\bibinfo {title} {{Sr$_2$RuO$_4$ : an
  electronic analogue of $^3$He?}}}\ }\href
  {http://stacks.iop.org/0953-8984/7/i=47/a=002} {\bibfield  {journal}
  {\bibinfo  {journal} {J. Phys.: Condens. Matter}\ }\textbf {\bibinfo {volume}
  {7}},\ \bibinfo {pages} {L643} (\bibinfo {year} {1995})}\BibitemShut
  {NoStop}%
\bibitem [{\citenamefont {Mackenzie}\ and\ \citenamefont
  {Maeno}(2003)}]{Mackenzie2003a}%
  \BibitemOpen
  \bibfield  {author} {\bibinfo {author} {\bibfnamefont {A.~P.}\ \bibnamefont
  {Mackenzie}}\ and\ \bibinfo {author} {\bibfnamefont {Y.}~\bibnamefont
  {Maeno}},\ }\bibfield  {title} {\enquote {\bibinfo {title} {{The
  superconductivity of Sr$_2$RuO$_4$ and the physics of spin-triplet
  pairing}},}\ }\href {\doibase 10.1103/RevModPhys.75.657} {\bibfield
  {journal} {\bibinfo  {journal} {Rev. Mod. Phys.}\ }\textbf {\bibinfo {volume}
  {75}},\ \bibinfo {pages} {657--712} (\bibinfo {year} {2003})}\BibitemShut
  {NoStop}%
\bibitem [{\citenamefont {Mackenzie}\ \emph {et~al.}(2017)\citenamefont
  {Mackenzie}, \citenamefont {Scaffidi}, \citenamefont {Hicks},\ and\
  \citenamefont {Maeno}}]{Mackenzie2017}%
  \BibitemOpen
  \bibfield  {author} {\bibinfo {author} {\bibfnamefont {Andrew~P.}\
  \bibnamefont {Mackenzie}}, \bibinfo {author} {\bibfnamefont {Thomas}\
  \bibnamefont {Scaffidi}}, \bibinfo {author} {\bibfnamefont {Clifford~W.}\
  \bibnamefont {Hicks}}, \ and\ \bibinfo {author} {\bibfnamefont {Yoshiteru}\
  \bibnamefont {Maeno}},\ }\bibfield  {title} {\enquote {\bibinfo {title}
  {{Even odder after twenty-three years: the superconducting order parameter
  puzzle of Sr$_2$RuO$_4$}},}\ }\href {\doibase 10.1038/s41535-017-0045-4}
  {\bibfield  {journal} {\bibinfo  {journal} {npj Quantum Materials}\ }\textbf
  {\bibinfo {volume} {2}},\ \bibinfo {pages} {40} (\bibinfo {year}
  {2017})}\BibitemShut {NoStop}%
\bibitem [{\citenamefont {Ishida}\ \emph {et~al.}(1998)\citenamefont {Ishida},
  \citenamefont {Mukuda}, \citenamefont {Kitaoka}, \citenamefont {Asayama},
  \citenamefont {Mao}, \citenamefont {Mori},\ and\ \citenamefont
  {Maeno}}]{Ishida1998}%
  \BibitemOpen
  \bibfield  {author} {\bibinfo {author} {\bibfnamefont {K.}~\bibnamefont
  {Ishida}}, \bibinfo {author} {\bibfnamefont {H.}~\bibnamefont {Mukuda}},
  \bibinfo {author} {\bibfnamefont {Y.}~\bibnamefont {Kitaoka}}, \bibinfo
  {author} {\bibfnamefont {K.}~\bibnamefont {Asayama}}, \bibinfo {author}
  {\bibfnamefont {Z.~Q.}\ \bibnamefont {Mao}}, \bibinfo {author} {\bibfnamefont
  {Y.}~\bibnamefont {Mori}}, \ and\ \bibinfo {author} {\bibfnamefont
  {Y.}~\bibnamefont {Maeno}},\ }\bibfield  {title} {\enquote {\bibinfo {title}
  {{Spin-triplet superconductivity in Sr$_2$RuO$_4$ identified by $^{17}$O
  Knight shift}},}\ }\href {https://www.nature.com/articles/25315} {\bibfield
  {journal} {\bibinfo  {journal} {Nature}\ }\textbf {\bibinfo {volume} {396}},\
  \bibinfo {pages} {658--660} (\bibinfo {year} {1998})}\BibitemShut {NoStop}%
\bibitem [{\citenamefont {Anwar}\ \emph {et~al.}(2016)\citenamefont {Anwar},
  \citenamefont {Lee}, \citenamefont {Ishiguro}, \citenamefont {Sugimoto},
  \citenamefont {Tano}, \citenamefont {Kang}, \citenamefont {Shin},
  \citenamefont {Yonezawa}, \citenamefont {Manske}, \citenamefont {Takayanagi},
  \citenamefont {Noh},\ and\ \citenamefont {Maeno}}]{Anwar2016}%
  \BibitemOpen
  \bibfield  {author} {\bibinfo {author} {\bibfnamefont {M.~S.}\ \bibnamefont
  {Anwar}}, \bibinfo {author} {\bibfnamefont {S.~R.}\ \bibnamefont {Lee}},
  \bibinfo {author} {\bibfnamefont {R.}~\bibnamefont {Ishiguro}}, \bibinfo
  {author} {\bibfnamefont {Y.}~\bibnamefont {Sugimoto}}, \bibinfo {author}
  {\bibfnamefont {Y.}~\bibnamefont {Tano}}, \bibinfo {author} {\bibfnamefont
  {S.~J.}\ \bibnamefont {Kang}}, \bibinfo {author} {\bibfnamefont {Y.~J.}\
  \bibnamefont {Shin}}, \bibinfo {author} {\bibfnamefont {S.}~\bibnamefont
  {Yonezawa}}, \bibinfo {author} {\bibfnamefont {D.}~\bibnamefont {Manske}},
  \bibinfo {author} {\bibfnamefont {H.}~\bibnamefont {Takayanagi}}, \bibinfo
  {author} {\bibfnamefont {T.~W.}\ \bibnamefont {Noh}}, \ and\ \bibinfo
  {author} {\bibfnamefont {Y.}~\bibnamefont {Maeno}},\ }\bibfield  {title}
  {\enquote {\bibinfo {title} {{Direct penetration of spin-triplet
  superconductivity into a ferromagnet in Au/SrRuO$_3$/Sr$_2$RuO$_4$
  junctions}},}\ }\href {\doibase 10.1038/ncomms13220} {\bibfield  {journal}
  {\bibinfo  {journal} {Nat. Commun.}\ }\textbf {\bibinfo {volume} {7}},\
  \bibinfo {pages} {1--7} (\bibinfo {year} {2016})}\BibitemShut {NoStop}%
\bibitem [{\citenamefont {Hicks}\ \emph {et~al.}(2014)\citenamefont {Hicks},
  \citenamefont {Brodsky}, \citenamefont {Yelland}, \citenamefont {Gibbs},
  \citenamefont {Bruin}, \citenamefont {Barber}, \citenamefont {Edkins},
  \citenamefont {Nishimura}, \citenamefont {Yonezawa}, \citenamefont {Maeno},\
  and\ \citenamefont {Mackenzie}}]{Hicks2014}%
  \BibitemOpen
  \bibfield  {author} {\bibinfo {author} {\bibfnamefont {Clifford~W.}\
  \bibnamefont {Hicks}}, \bibinfo {author} {\bibfnamefont {Daniel~O.}\
  \bibnamefont {Brodsky}}, \bibinfo {author} {\bibfnamefont {Edward~A.}\
  \bibnamefont {Yelland}}, \bibinfo {author} {\bibfnamefont {Alexandra~S.}\
  \bibnamefont {Gibbs}}, \bibinfo {author} {\bibfnamefont {Jan A.~N.}\
  \bibnamefont {Bruin}}, \bibinfo {author} {\bibfnamefont {Mark~E.}\
  \bibnamefont {Barber}}, \bibinfo {author} {\bibfnamefont {Stephen~D.}\
  \bibnamefont {Edkins}}, \bibinfo {author} {\bibfnamefont {Keigo}\
  \bibnamefont {Nishimura}}, \bibinfo {author} {\bibfnamefont {Shingo}\
  \bibnamefont {Yonezawa}}, \bibinfo {author} {\bibfnamefont {Yoshiteru}\
  \bibnamefont {Maeno}}, \ and\ \bibinfo {author} {\bibfnamefont {Andrew~P.}\
  \bibnamefont {Mackenzie}},\ }\bibfield  {title} {\enquote {\bibinfo {title}
  {{Strong Increase of Tc of Sr$_2$RuO$_4$ Under Both Tensile and Compressive
  Strain}},}\ }\href {\doibase 10.1126/science.1248292} {\bibfield  {journal}
  {\bibinfo  {journal} {Science}\ }\textbf {\bibinfo {volume} {344}},\ \bibinfo
  {pages} {283--285} (\bibinfo {year} {2014})}\BibitemShut {NoStop}%
\bibitem [{\citenamefont {Scaffidi}\ \emph {et~al.}(2014)\citenamefont
  {Scaffidi}, \citenamefont {Romers},\ and\ \citenamefont
  {Simon}}]{Scaffidi2014}%
  \BibitemOpen
  \bibfield  {author} {\bibinfo {author} {\bibfnamefont {T.}~\bibnamefont
  {Scaffidi}}, \bibinfo {author} {\bibfnamefont {J.~C.}\ \bibnamefont
  {Romers}}, \ and\ \bibinfo {author} {\bibfnamefont {S.~H.}\ \bibnamefont
  {Simon}},\ }\bibfield  {title} {\enquote {\bibinfo {title} {{Pairing symmetry
  and dominant band in Sr$_2$RuO$_4$}},}\ }\href {\doibase
  10.1103/PhysRevB.89.220510} {\bibfield  {journal} {\bibinfo  {journal} {Phys.
  Rev. B}\ }\textbf {\bibinfo {volume} {89}},\ \bibinfo {pages} {220510}
  (\bibinfo {year} {2014})}\BibitemShut {NoStop}%
\bibitem [{\citenamefont {Kirtley}\ \emph {et~al.}(2007)\citenamefont
  {Kirtley}, \citenamefont {Kallin}, \citenamefont {Hicks}, \citenamefont
  {Kim}, \citenamefont {Liu}, \citenamefont {Moler}, \citenamefont {Maeno},\
  and\ \citenamefont {Nelson}}]{Kirtley2007}%
  \BibitemOpen
  \bibfield  {author} {\bibinfo {author} {\bibfnamefont {J.~R.}\ \bibnamefont
  {Kirtley}}, \bibinfo {author} {\bibfnamefont {C.}~\bibnamefont {Kallin}},
  \bibinfo {author} {\bibfnamefont {C.~W.}\ \bibnamefont {Hicks}}, \bibinfo
  {author} {\bibfnamefont {E.-A.}\ \bibnamefont {Kim}}, \bibinfo {author}
  {\bibfnamefont {Y.}~\bibnamefont {Liu}}, \bibinfo {author} {\bibfnamefont
  {K.~A.}\ \bibnamefont {Moler}}, \bibinfo {author} {\bibfnamefont
  {Y.}~\bibnamefont {Maeno}}, \ and\ \bibinfo {author} {\bibfnamefont {K.~D.}\
  \bibnamefont {Nelson}},\ }\bibfield  {title} {\enquote {\bibinfo {title}
  {{Upper limit on spontaneous supercurrents in Sr$_2$RuO$_4$}},}\ }\href
  {\doibase 10.1103/PhysRevB.76.014526} {\bibfield  {journal} {\bibinfo
  {journal} {Phys. Rev. B}\ }\textbf {\bibinfo {volume} {76}},\ \bibinfo
  {pages} {014526} (\bibinfo {year} {2007})}\BibitemShut {NoStop}%
\bibitem [{\citenamefont {Mackenzie}\ \emph
  {et~al.}(1996{\natexlab{a}})\citenamefont {Mackenzie}, \citenamefont
  {Julian}, \citenamefont {Diver}, \citenamefont {Lonzarich}, \citenamefont
  {Hussey}, \citenamefont {Maeno}, \citenamefont {Nishizaki},\ and\
  \citenamefont {Fujita}}]{Mackenzie1996b}%
  \BibitemOpen
  \bibfield  {author} {\bibinfo {author} {\bibfnamefont {A~P}\ \bibnamefont
  {Mackenzie}}, \bibinfo {author} {\bibfnamefont {S~R}\ \bibnamefont {Julian}},
  \bibinfo {author} {\bibfnamefont {A~J}\ \bibnamefont {Diver}}, \bibinfo
  {author} {\bibfnamefont {G~G}\ \bibnamefont {Lonzarich}}, \bibinfo {author}
  {\bibfnamefont {N~E}\ \bibnamefont {Hussey}}, \bibinfo {author}
  {\bibfnamefont {Y}~\bibnamefont {Maeno}}, \bibinfo {author} {\bibfnamefont
  {S}~\bibnamefont {Nishizaki}}, \ and\ \bibinfo {author} {\bibfnamefont
  {T}~\bibnamefont {Fujita}},\ }\bibfield  {title} {\enquote {\bibinfo {title}
  {{Calculation of thermodynamic and transport properties of Sr$_2$RuO$_4$ at
  low temperatures using known Fermi surface parameters}},}\ }\href
  {https://www.sciencedirect.com/science/article/pii/0921453495007709}
  {\bibfield  {journal} {\bibinfo  {journal} {Phys. C}\ }\textbf {\bibinfo
  {volume} {263}},\ \bibinfo {pages} {510--515} (\bibinfo {year}
  {1996}{\natexlab{a}})}\BibitemShut {NoStop}%
\bibitem [{\citenamefont {Maeno}\ \emph {et~al.}(1997)\citenamefont {Maeno},
  \citenamefont {Yoshida}, \citenamefont {Hashimoto}, \citenamefont
  {Nishizaki}, \citenamefont {Ikeda}, \citenamefont {Nohara}, \citenamefont
  {Fujita}, \citenamefont {Mackenzie}, \citenamefont {Hussey}, \citenamefont
  {Bednorz},\ and\ \citenamefont {Lichtenberg}}]{Maeno1997}%
  \BibitemOpen
  \bibfield  {author} {\bibinfo {author} {\bibfnamefont {Yoshiteru}\
  \bibnamefont {Maeno}}, \bibinfo {author} {\bibfnamefont {Koji}\ \bibnamefont
  {Yoshida}}, \bibinfo {author} {\bibfnamefont {Hiroaki}\ \bibnamefont
  {Hashimoto}}, \bibinfo {author} {\bibfnamefont {Shuji}\ \bibnamefont
  {Nishizaki}}, \bibinfo {author} {\bibfnamefont {Shin-Ichi}\ \bibnamefont
  {Ikeda}}, \bibinfo {author} {\bibfnamefont {Minoru}\ \bibnamefont {Nohara}},
  \bibinfo {author} {\bibfnamefont {Toshizo}\ \bibnamefont {Fujita}}, \bibinfo
  {author} {\bibfnamefont {Andrew~P.}\ \bibnamefont {Mackenzie}}, \bibinfo
  {author} {\bibfnamefont {Nigel~E.}\ \bibnamefont {Hussey}}, \bibinfo {author}
  {\bibfnamefont {J.~Georg}\ \bibnamefont {Bednorz}}, \ and\ \bibinfo {author}
  {\bibfnamefont {Frank}\ \bibnamefont {Lichtenberg}},\ }\bibfield  {title}
  {\enquote {\bibinfo {title} {{Two-Dimensional Fermi Liquid Behavior of the
  Superconductor Sr$_2$RuO$_4$}},}\ }\href {\doibase 10.1143/JPSJ.66.1405}
  {\bibfield  {journal} {\bibinfo  {journal} {J. Phys. Soc. Japan}\ }\textbf
  {\bibinfo {volume} {66}},\ \bibinfo {pages} {1405--1408} (\bibinfo {year}
  {1997})}\BibitemShut {NoStop}%
\bibitem [{\citenamefont {Bergemann}\ \emph {et~al.}(2003)\citenamefont
  {Bergemann}, \citenamefont {Mackenzie}, \citenamefont {Julian}, \citenamefont
  {Forsythe},\ and\ \citenamefont {Ohmichi}}]{Bergemann2003}%
  \BibitemOpen
  \bibfield  {author} {\bibinfo {author} {\bibfnamefont {C.}~\bibnamefont
  {Bergemann}}, \bibinfo {author} {\bibfnamefont {A.~P.}\ \bibnamefont
  {Mackenzie}}, \bibinfo {author} {\bibfnamefont {S.~R.}\ \bibnamefont
  {Julian}}, \bibinfo {author} {\bibfnamefont {D.}~\bibnamefont {Forsythe}}, \
  and\ \bibinfo {author} {\bibfnamefont {E.}~\bibnamefont {Ohmichi}},\
  }\bibfield  {title} {\enquote {\bibinfo {title} {{Quasi-two-dimensional Fermi
  liquid properties of the unconventional superconductor Sr$_2$RuO$_4$}},}\
  }\href {\doibase 10.1080/00018730310001621737} {\bibfield  {journal}
  {\bibinfo  {journal} {Adv. Phys.}\ }\textbf {\bibinfo {volume} {52}},\
  \bibinfo {pages} {639--725} (\bibinfo {year} {2003})}\BibitemShut {NoStop}%
\bibitem [{\citenamefont {Stricker}\ \emph {et~al.}(2014)\citenamefont
  {Stricker}, \citenamefont {Mravlje}, \citenamefont {Berthod}, \citenamefont
  {Fittipaldi}, \citenamefont {Vecchione}, \citenamefont {Georges},\ and\
  \citenamefont {{van der Marel}}}]{Stricker2014}%
  \BibitemOpen
  \bibfield  {author} {\bibinfo {author} {\bibfnamefont {D.}~\bibnamefont
  {Stricker}}, \bibinfo {author} {\bibfnamefont {J.}~\bibnamefont {Mravlje}},
  \bibinfo {author} {\bibfnamefont {C.}~\bibnamefont {Berthod}}, \bibinfo
  {author} {\bibfnamefont {R.}~\bibnamefont {Fittipaldi}}, \bibinfo {author}
  {\bibfnamefont {A.}~\bibnamefont {Vecchione}}, \bibinfo {author}
  {\bibfnamefont {A.}~\bibnamefont {Georges}}, \ and\ \bibinfo {author}
  {\bibfnamefont {D.}~\bibnamefont {{van der Marel}}},\ }\bibfield  {title}
  {\enquote {\bibinfo {title} {{Optical response of Sr$_2$RuO$_4$ reveals
  universal Fermi-liquid scaling and quasiparticles beyond landau theory}},}\
  }\href {\doibase 10.1103/PhysRevLett.113.087404} {\bibfield  {journal}
  {\bibinfo  {journal} {Phys. Rev. Lett.}\ }\textbf {\bibinfo {volume} {113}},\
  \bibinfo {pages} {087404} (\bibinfo {year} {2014})}\BibitemShut {NoStop}%
\bibitem [{\citenamefont {Raghu}\ \emph {et~al.}(2010)\citenamefont {Raghu},
  \citenamefont {Kapitulnik},\ and\ \citenamefont {Kivelson}}]{Raghu2010}%
  \BibitemOpen
  \bibfield  {author} {\bibinfo {author} {\bibfnamefont {S.}~\bibnamefont
  {Raghu}}, \bibinfo {author} {\bibfnamefont {A.}~\bibnamefont {Kapitulnik}}, \
  and\ \bibinfo {author} {\bibfnamefont {S.~A.}\ \bibnamefont {Kivelson}},\
  }\bibfield  {title} {\enquote {\bibinfo {title} {{Hidden
  Quasi-One-Dimensional Superconductivity in Sr$_2$RuO$_4$}},}\ }\href
  {http://link.aps.org/doi/10.1103/PhysRevLett.105.136401} {\bibfield
  {journal} {\bibinfo  {journal} {Phys. Rev. Lett.}\ }\textbf {\bibinfo
  {volume} {105}},\ \bibinfo {pages} {136401} (\bibinfo {year}
  {2010})}\BibitemShut {NoStop}%
\bibitem [{\citenamefont {Huo}\ \emph {et~al.}(2013)\citenamefont {Huo},
  \citenamefont {Rice},\ and\ \citenamefont {Zhang}}]{Huo2013}%
  \BibitemOpen
  \bibfield  {author} {\bibinfo {author} {\bibfnamefont {Jia~Wei}\ \bibnamefont
  {Huo}}, \bibinfo {author} {\bibfnamefont {T.~M.}\ \bibnamefont {Rice}}, \
  and\ \bibinfo {author} {\bibfnamefont {Fu~Chun}\ \bibnamefont {Zhang}},\
  }\bibfield  {title} {\enquote {\bibinfo {title} {{Spin density wave
  fluctuations and p-wave pairing in Sr$_2$RuO$_4$}},}\ }\href {\doibase
  10.1103/PhysRevLett.110.167003} {\bibfield  {journal} {\bibinfo  {journal}
  {Phys. Rev. Lett.}\ }\textbf {\bibinfo {volume} {110}},\ \bibinfo {pages}
  {167003} (\bibinfo {year} {2013})}\BibitemShut {NoStop}%
\bibitem [{\citenamefont {Komendov\'a}\ and\ \citenamefont
  {Black-Schaffer}(2017)}]{Komendova2017}%
  \BibitemOpen
  \bibfield  {author} {\bibinfo {author} {\bibfnamefont {L.}~\bibnamefont
  {Komendov\'a}}\ and\ \bibinfo {author} {\bibfnamefont {A.~M.}\ \bibnamefont
  {Black-Schaffer}},\ }\bibfield  {title} {\enquote {\bibinfo {title}
  {{Odd-Frequency Superconductivity in Sr$_2$RuO$_4$ Measured by Kerr
  Rotation}},}\ }\href {\doibase 10.1103/PhysRevLett.119.087001} {\bibfield
  {journal} {\bibinfo  {journal} {Phys. Rev. Lett.}\ }\textbf {\bibinfo
  {volume} {119}},\ \bibinfo {pages} {087001} (\bibinfo {year}
  {2017})}\BibitemShut {NoStop}%
\bibitem [{\citenamefont {Steppke}\ \emph {et~al.}(2017)\citenamefont
  {Steppke}, \citenamefont {Zhao}, \citenamefont {Barber}, \citenamefont
  {Scaffidi}, \citenamefont {Jerzembeck}, \citenamefont {Rosner}, \citenamefont
  {Gibbs}, \citenamefont {Maeno}, \citenamefont {Simon}, \citenamefont
  {Mackenzie},\ and\ \citenamefont {Hicks}}]{Steppke2017}%
  \BibitemOpen
  \bibfield  {author} {\bibinfo {author} {\bibfnamefont {Alexander}\
  \bibnamefont {Steppke}}, \bibinfo {author} {\bibfnamefont {Lishan}\
  \bibnamefont {Zhao}}, \bibinfo {author} {\bibfnamefont {Mark~E.}\
  \bibnamefont {Barber}}, \bibinfo {author} {\bibfnamefont {Thomas}\
  \bibnamefont {Scaffidi}}, \bibinfo {author} {\bibfnamefont {Fabian}\
  \bibnamefont {Jerzembeck}}, \bibinfo {author} {\bibfnamefont {Helge}\
  \bibnamefont {Rosner}}, \bibinfo {author} {\bibfnamefont {Alexandra~S.}\
  \bibnamefont {Gibbs}}, \bibinfo {author} {\bibfnamefont {Yoshiteru}\
  \bibnamefont {Maeno}}, \bibinfo {author} {\bibfnamefont {Steven~H.}\
  \bibnamefont {Simon}}, \bibinfo {author} {\bibfnamefont {Andrew~P.}\
  \bibnamefont {Mackenzie}}, \ and\ \bibinfo {author} {\bibfnamefont
  {Clifford~W.}\ \bibnamefont {Hicks}},\ }\bibfield  {title} {\enquote
  {\bibinfo {title} {{Strong peak in Tc of Sr$_2$RuO$_4$ under uniaxial
  pressure}},}\ }\href
  {http://science.sciencemag.org/content/355/6321/eaaf9398} {\bibfield
  {journal} {\bibinfo  {journal} {Science}\ }\textbf {\bibinfo {volume} {355}}
  (\bibinfo {year} {2017})}\BibitemShut {NoStop}%
\bibitem [{\citenamefont {Georges}\ \emph {et~al.}(2013)\citenamefont
  {Georges}, \citenamefont {de' Medici},\ and\ \citenamefont
  {Mravlje}}]{Georges2013}%
  \BibitemOpen
  \bibfield  {author} {\bibinfo {author} {\bibfnamefont {Antoine}\ \bibnamefont
  {Georges}}, \bibinfo {author} {\bibfnamefont {Luca}\ \bibnamefont {de'
  Medici}}, \ and\ \bibinfo {author} {\bibfnamefont {Jernej}\ \bibnamefont
  {Mravlje}},\ }\bibfield  {title} {\enquote {\bibinfo {title} {{Strong
  Correlations from Hund's Coupling}},}\ }\href {\doibase
  10.1146/annurev-conmatphys-020911-125045} {\bibfield  {journal} {\bibinfo
  {journal} {Annu. Rev. Condens. Matter Phys.}\ }\textbf {\bibinfo {volume}
  {4}},\ \bibinfo {pages} {137--178} (\bibinfo {year} {2013})}\BibitemShut
  {NoStop}%
\bibitem [{\citenamefont {Liebsch}\ and\ \citenamefont
  {Lichtenstein}(2000)}]{Liebsch2000}%
  \BibitemOpen
  \bibfield  {author} {\bibinfo {author} {\bibfnamefont {A.}~\bibnamefont
  {Liebsch}}\ and\ \bibinfo {author} {\bibfnamefont {A.}~\bibnamefont
  {Lichtenstein}},\ }\bibfield  {title} {\enquote {\bibinfo {title}
  {{Photoemission Quasiparticle Spectra of Sr$_2$RuO$_4$}},}\ }\href
  {http://link.aps.org/abstract/PRL/v84/p1591} {\bibfield  {journal} {\bibinfo
  {journal} {Phys. Rev. Lett.}\ }\textbf {\bibinfo {volume} {84}},\ \bibinfo
  {pages} {1591} (\bibinfo {year} {2000})}\BibitemShut {NoStop}%
\bibitem [{\citenamefont {Mravlje}\ \emph {et~al.}(2011)\citenamefont
  {Mravlje}, \citenamefont {Aichhorn}, \citenamefont {Miyake}, \citenamefont
  {Haule}, \citenamefont {Kotliar},\ and\ \citenamefont
  {Georges}}]{Mravlje2011}%
  \BibitemOpen
  \bibfield  {author} {\bibinfo {author} {\bibfnamefont {Jernej}\ \bibnamefont
  {Mravlje}}, \bibinfo {author} {\bibfnamefont {Markus}\ \bibnamefont
  {Aichhorn}}, \bibinfo {author} {\bibfnamefont {Takashi}\ \bibnamefont
  {Miyake}}, \bibinfo {author} {\bibfnamefont {Kristjan}\ \bibnamefont
  {Haule}}, \bibinfo {author} {\bibfnamefont {Gabriel}\ \bibnamefont
  {Kotliar}}, \ and\ \bibinfo {author} {\bibfnamefont {Antoine}\ \bibnamefont
  {Georges}},\ }\bibfield  {title} {\enquote {\bibinfo {title}
  {{Coherence-Incoherence Crossover and the Mass-Renormalization Puzzles in
  Sr$_2$RuO$_4$}},}\ }\href {\doibase 10.1103/PhysRevLett.106.096401}
  {\bibfield  {journal} {\bibinfo  {journal} {Phys. Rev. Lett.}\ }\textbf
  {\bibinfo {volume} {106}},\ \bibinfo {pages} {096401} (\bibinfo {year}
  {2011})}\BibitemShut {NoStop}%
\bibitem [{\citenamefont {Zhang}\ \emph {et~al.}(2016)\citenamefont {Zhang},
  \citenamefont {Gorelov}, \citenamefont {Sarvestani},\ and\ \citenamefont
  {Pavarini}}]{Zhang2016}%
  \BibitemOpen
  \bibfield  {author} {\bibinfo {author} {\bibfnamefont {Guoren}\ \bibnamefont
  {Zhang}}, \bibinfo {author} {\bibfnamefont {Evgeny}\ \bibnamefont {Gorelov}},
  \bibinfo {author} {\bibfnamefont {Esmaeel}\ \bibnamefont {Sarvestani}}, \
  and\ \bibinfo {author} {\bibfnamefont {Eva}\ \bibnamefont {Pavarini}},\
  }\bibfield  {title} {\enquote {\bibinfo {title} {{Fermi Surface of
  Sr$_2$RuO$_4$: Spin-Orbit and Anisotropic Coulomb Interaction Effects}},}\
  }\href {\doibase 10.1103/PhysRevLett.116.106402} {\bibfield  {journal}
  {\bibinfo  {journal} {Phys. Rev. Lett.}\ }\textbf {\bibinfo {volume} {116}},\
  \bibinfo {pages} {106402} (\bibinfo {year} {2016})}\BibitemShut {NoStop}%
\bibitem [{\citenamefont {Mravlje}\ and\ \citenamefont
  {Georges}(2016)}]{Mravlje2016}%
  \BibitemOpen
  \bibfield  {author} {\bibinfo {author} {\bibfnamefont {Jernej}\ \bibnamefont
  {Mravlje}}\ and\ \bibinfo {author} {\bibfnamefont {Antoine}\ \bibnamefont
  {Georges}},\ }\bibfield  {title} {\enquote {\bibinfo {title} {{Thermopower
  and Entropy: Lessons from Sr$_2$RuO$_4$}},}\ }\href {\doibase
  10.1103/PhysRevLett.117.036401} {\bibfield  {journal} {\bibinfo  {journal}
  {Phys. Rev. Lett.}\ }\textbf {\bibinfo {volume} {117}},\ \bibinfo {pages}
  {036401} (\bibinfo {year} {2016})}\BibitemShut {NoStop}%
\bibitem [{\citenamefont {Kim}\ \emph {et~al.}(2018)\citenamefont {Kim},
  \citenamefont {Mravlje}, \citenamefont {Ferrero}, \citenamefont {Parcollet},\
  and\ \citenamefont {Georges}}]{Kim2018}%
  \BibitemOpen
  \bibfield  {author} {\bibinfo {author} {\bibfnamefont {Minjae}\ \bibnamefont
  {Kim}}, \bibinfo {author} {\bibfnamefont {Jernej}\ \bibnamefont {Mravlje}},
  \bibinfo {author} {\bibfnamefont {Michel}\ \bibnamefont {Ferrero}}, \bibinfo
  {author} {\bibfnamefont {Olivier}\ \bibnamefont {Parcollet}}, \ and\ \bibinfo
  {author} {\bibfnamefont {Antoine}\ \bibnamefont {Georges}},\ }\bibfield
  {title} {\enquote {\bibinfo {title} {{Spin-Orbit Coupling and Electronic
  Correlations in Sr$_2$RuO$_4$}},}\ }\href {\doibase
  10.1103/PhysRevLett.120.126401} {\bibfield  {journal} {\bibinfo  {journal}
  {Phys. Rev. Lett.}\ }\textbf {\bibinfo {volume} {120}},\ \bibinfo {pages}
  {126401} (\bibinfo {year} {2018})}\BibitemShut {NoStop}%
\bibitem [{\citenamefont {Oguchi}(1995)}]{Oguchi1995}%
  \BibitemOpen
  \bibfield  {author} {\bibinfo {author} {\bibfnamefont {Tamio}\ \bibnamefont
  {Oguchi}},\ }\bibfield  {title} {\enquote {\bibinfo {title} {{Electronic band
  structure of the superconductor Sr$_2$RuO$_4$}},}\ }\href
  {http://link.aps.org/abstract/PRB/v51/p1385} {\bibfield  {journal} {\bibinfo
  {journal} {Phys. Rev. B}\ }\textbf {\bibinfo {volume} {51}},\ \bibinfo
  {pages} {1385} (\bibinfo {year} {1995})}\BibitemShut {NoStop}%
\bibitem [{\citenamefont {Mackenzie}\ \emph
  {et~al.}(1996{\natexlab{b}})\citenamefont {Mackenzie}, \citenamefont
  {Julian}, \citenamefont {Diver}, \citenamefont {McMullan}, \citenamefont
  {Ray}, \citenamefont {Lonzarich}, \citenamefont {Maeno}, \citenamefont
  {Nishizaki},\ and\ \citenamefont {Fujita}}]{Mackenzie1996}%
  \BibitemOpen
  \bibfield  {author} {\bibinfo {author} {\bibfnamefont {A.~P.}\ \bibnamefont
  {Mackenzie}}, \bibinfo {author} {\bibfnamefont {S.~R.}\ \bibnamefont
  {Julian}}, \bibinfo {author} {\bibfnamefont {A.~J.}\ \bibnamefont {Diver}},
  \bibinfo {author} {\bibfnamefont {G.~J.}\ \bibnamefont {McMullan}}, \bibinfo
  {author} {\bibfnamefont {M.~P.}\ \bibnamefont {Ray}}, \bibinfo {author}
  {\bibfnamefont {G.~G.}\ \bibnamefont {Lonzarich}}, \bibinfo {author}
  {\bibfnamefont {Y.}~\bibnamefont {Maeno}}, \bibinfo {author} {\bibfnamefont
  {S.}~\bibnamefont {Nishizaki}}, \ and\ \bibinfo {author} {\bibfnamefont
  {T.}~\bibnamefont {Fujita}},\ }\bibfield  {title} {\enquote {\bibinfo {title}
  {{Quantum Oscillations in the Layered Perovskite Superconductor
  Sr$_2$RuO$_4$}},}\ }\href {\doibase 10.1103/PhysRevLett.76.3786} {\bibfield
  {journal} {\bibinfo  {journal} {Phys. Rev. Lett.}\ }\textbf {\bibinfo
  {volume} {76}},\ \bibinfo {pages} {3786--3789} (\bibinfo {year}
  {1996}{\natexlab{b}})}\BibitemShut {NoStop}%
\bibitem [{\citenamefont {Mackenzie}\ \emph {et~al.}(1998)\citenamefont
  {Mackenzie}, \citenamefont {Ikeda}, \citenamefont {Maeno}, \citenamefont
  {Fujita}, \citenamefont {Julian},\ and\ \citenamefont
  {Lonzarich}}]{Mackenzie1998}%
  \BibitemOpen
  \bibfield  {author} {\bibinfo {author} {\bibfnamefont {Andrew~P.}\
  \bibnamefont {Mackenzie}}, \bibinfo {author} {\bibfnamefont {Shin}\
  \bibnamefont {Ikeda}}, \bibinfo {author} {\bibfnamefont {Yoshiteru}\
  \bibnamefont {Maeno}}, \bibinfo {author} {\bibfnamefont {Toshizo}\
  \bibnamefont {Fujita}}, \bibinfo {author} {\bibfnamefont {Stephen~R.}\
  \bibnamefont {Julian}}, \ and\ \bibinfo {author} {\bibfnamefont {Gilbert~G.}\
  \bibnamefont {Lonzarich}},\ }\bibfield  {title} {\enquote {\bibinfo {title}
  {{The Fermi Surface Topography of Sr$_2$RuO$_4$}},}\ }\href {\doibase
  10.1143/JPSJ.67.385} {\bibfield  {journal} {\bibinfo  {journal} {J. Phys.
  Soc. Japan}\ }\textbf {\bibinfo {volume} {67}},\ \bibinfo {pages} {385--388}
  (\bibinfo {year} {1998})}\BibitemShut {NoStop}%
\bibitem [{\citenamefont {Bergemann}\ \emph {et~al.}(2000)\citenamefont
  {Bergemann}, \citenamefont {Julian}, \citenamefont {Mackenzie}, \citenamefont
  {NishiZaki},\ and\ \citenamefont {Maeno}}]{Bergemann2000}%
  \BibitemOpen
  \bibfield  {author} {\bibinfo {author} {\bibfnamefont {C.}~\bibnamefont
  {Bergemann}}, \bibinfo {author} {\bibfnamefont {S.}~\bibnamefont {Julian}},
  \bibinfo {author} {\bibfnamefont {A.}~\bibnamefont {Mackenzie}}, \bibinfo
  {author} {\bibfnamefont {S.}~\bibnamefont {NishiZaki}}, \ and\ \bibinfo
  {author} {\bibfnamefont {Y.}~\bibnamefont {Maeno}},\ }\bibfield  {title}
  {\enquote {\bibinfo {title} {{Detailed Topography of the Fermi Surface of
  Sr$_2$RuO$_4$}},}\ }\href {\doibase 10.1103/PhysRevLett.84.2662} {\bibfield
  {journal} {\bibinfo  {journal} {Phys. Rev. Lett.}\ }\textbf {\bibinfo
  {volume} {84}},\ \bibinfo {pages} {2662--2665} (\bibinfo {year}
  {2000})}\BibitemShut {NoStop}%
\bibitem [{\citenamefont {Damascelli}\ \emph {et~al.}(2000)\citenamefont
  {Damascelli}, \citenamefont {Lu}, \citenamefont {Shen}, \citenamefont
  {Armitage}, \citenamefont {Ronning}, \citenamefont {Feng}, \citenamefont
  {Kim}, \citenamefont {Shen}, \citenamefont {Kimura}, \citenamefont {Tokura},
  \citenamefont {Mao},\ and\ \citenamefont {Maeno}}]{Damascelli2000}%
  \BibitemOpen
  \bibfield  {author} {\bibinfo {author} {\bibfnamefont {A.}~\bibnamefont
  {Damascelli}}, \bibinfo {author} {\bibfnamefont {D.}~\bibnamefont {Lu}},
  \bibinfo {author} {\bibfnamefont {K.}~\bibnamefont {Shen}}, \bibinfo {author}
  {\bibfnamefont {N.}~\bibnamefont {Armitage}}, \bibinfo {author}
  {\bibfnamefont {F.}~\bibnamefont {Ronning}}, \bibinfo {author} {\bibfnamefont
  {D.}~\bibnamefont {Feng}}, \bibinfo {author} {\bibfnamefont {C.}~\bibnamefont
  {Kim}}, \bibinfo {author} {\bibfnamefont {Z.-X.}\ \bibnamefont {Shen}},
  \bibinfo {author} {\bibfnamefont {T.}~\bibnamefont {Kimura}}, \bibinfo
  {author} {\bibfnamefont {Y.}~\bibnamefont {Tokura}}, \bibinfo {author}
  {\bibfnamefont {Z.}~\bibnamefont {Mao}}, \ and\ \bibinfo {author}
  {\bibfnamefont {Y.}~\bibnamefont {Maeno}},\ }\bibfield  {title} {\enquote
  {\bibinfo {title} {{Fermi Surface, Surface States, and Surface Reconstruction
  in Sr$_2$RuO$_4$}},}\ }\href {\doibase 10.1103/PhysRevLett.85.5194}
  {\bibfield  {journal} {\bibinfo  {journal} {Phys. Rev. Lett.}\ }\textbf
  {\bibinfo {volume} {85}},\ \bibinfo {pages} {5194--5197} (\bibinfo {year}
  {2000})}\BibitemShut {NoStop}%
\bibitem [{\citenamefont {Iwasawa}\ \emph {et~al.}(2005)\citenamefont
  {Iwasawa}, \citenamefont {Aiura}, \citenamefont {Saitoh}, \citenamefont
  {Hase}, \citenamefont {Ikeda}, \citenamefont {Yoshida}, \citenamefont
  {Bando}, \citenamefont {Higashiguchi}, \citenamefont {Miura}, \citenamefont
  {Cui}, \citenamefont {Shimada}, \citenamefont {Namatame},\ and\ \citenamefont
  {Taniguchi}}]{Iwasawa2005}%
  \BibitemOpen
  \bibfield  {author} {\bibinfo {author} {\bibfnamefont {H.}~\bibnamefont
  {Iwasawa}}, \bibinfo {author} {\bibfnamefont {Y.}~\bibnamefont {Aiura}},
  \bibinfo {author} {\bibfnamefont {T.}~\bibnamefont {Saitoh}}, \bibinfo
  {author} {\bibfnamefont {I.}~\bibnamefont {Hase}}, \bibinfo {author}
  {\bibfnamefont {S.~I.}\ \bibnamefont {Ikeda}}, \bibinfo {author}
  {\bibfnamefont {Y.}~\bibnamefont {Yoshida}}, \bibinfo {author} {\bibfnamefont
  {H.}~\bibnamefont {Bando}}, \bibinfo {author} {\bibfnamefont
  {M.}~\bibnamefont {Higashiguchi}}, \bibinfo {author} {\bibfnamefont
  {Y.}~\bibnamefont {Miura}}, \bibinfo {author} {\bibfnamefont {X.~Y.}\
  \bibnamefont {Cui}}, \bibinfo {author} {\bibfnamefont {K.}~\bibnamefont
  {Shimada}}, \bibinfo {author} {\bibfnamefont {H.}~\bibnamefont {Namatame}}, \
  and\ \bibinfo {author} {\bibfnamefont {M.}~\bibnamefont {Taniguchi}},\
  }\bibfield  {title} {\enquote {\bibinfo {title} {{Orbital selectivity of the
  kink in the dispersion of Sr$_2$RuO$_4$}},}\ }\href
  {http://link.aps.org/abstract/PRB/v72/e104514} {\bibfield  {journal}
  {\bibinfo  {journal} {Phys. Rev. B}\ }\textbf {\bibinfo {volume} {72}},\
  \bibinfo {pages} {104514--104515} (\bibinfo {year} {2005})}\BibitemShut
  {NoStop}%
\bibitem [{\citenamefont {Ingle}\ \emph {et~al.}(2005)\citenamefont {Ingle},
  \citenamefont {Shen}, \citenamefont {Baumberger}, \citenamefont {Meevasana},
  \citenamefont {Lu}, \citenamefont {Shen}, \citenamefont {Damascelli},
  \citenamefont {Nakatsuji}, \citenamefont {Mao}, \citenamefont {Maeno},
  \citenamefont {Kimura},\ and\ \citenamefont {Tokura}}]{Ingle2005}%
  \BibitemOpen
  \bibfield  {author} {\bibinfo {author} {\bibfnamefont {N.~J.~C.}\
  \bibnamefont {Ingle}}, \bibinfo {author} {\bibfnamefont {K.~M.}\ \bibnamefont
  {Shen}}, \bibinfo {author} {\bibfnamefont {F.}~\bibnamefont {Baumberger}},
  \bibinfo {author} {\bibfnamefont {W.}~\bibnamefont {Meevasana}}, \bibinfo
  {author} {\bibfnamefont {D.~H.}\ \bibnamefont {Lu}}, \bibinfo {author}
  {\bibfnamefont {Z.~X.}\ \bibnamefont {Shen}}, \bibinfo {author}
  {\bibfnamefont {A.}~\bibnamefont {Damascelli}}, \bibinfo {author}
  {\bibfnamefont {S.}~\bibnamefont {Nakatsuji}}, \bibinfo {author}
  {\bibfnamefont {Z.~Q.}\ \bibnamefont {Mao}}, \bibinfo {author} {\bibfnamefont
  {Y.}~\bibnamefont {Maeno}}, \bibinfo {author} {\bibfnamefont
  {T.}~\bibnamefont {Kimura}}, \ and\ \bibinfo {author} {\bibfnamefont
  {Y.}~\bibnamefont {Tokura}},\ }\bibfield  {title} {\enquote {\bibinfo {title}
  {{Quantitative analysis of Sr$_2$RuO$_4$ angle-resolved photoemission
  spectra: Many-body interactions in a model Fermi liquid}},}\ }\href
  {https://journals.aps.org/prb/abstract/10.1103/PhysRevB.72.205114} {\bibfield
   {journal} {\bibinfo  {journal} {Phys. Rev. B}\ }\textbf {\bibinfo {volume}
  {72}},\ \bibinfo {pages} {205114} (\bibinfo {year} {2005})}\BibitemShut
  {NoStop}%
\bibitem [{\citenamefont {Iwasawa}\ \emph {et~al.}(2010)\citenamefont
  {Iwasawa}, \citenamefont {Yoshida}, \citenamefont {Hase}, \citenamefont
  {Koikegami}, \citenamefont {Hayashi}, \citenamefont {Jiang}, \citenamefont
  {Shimada}, \citenamefont {Namatame}, \citenamefont {Taniguchi},\ and\
  \citenamefont {Aiura}}]{Iwasawa2010}%
  \BibitemOpen
  \bibfield  {author} {\bibinfo {author} {\bibfnamefont {H.}~\bibnamefont
  {Iwasawa}}, \bibinfo {author} {\bibfnamefont {Y.}~\bibnamefont {Yoshida}},
  \bibinfo {author} {\bibfnamefont {I.}~\bibnamefont {Hase}}, \bibinfo {author}
  {\bibfnamefont {S.}~\bibnamefont {Koikegami}}, \bibinfo {author}
  {\bibfnamefont {H.}~\bibnamefont {Hayashi}}, \bibinfo {author} {\bibfnamefont
  {J.}~\bibnamefont {Jiang}}, \bibinfo {author} {\bibfnamefont
  {K.}~\bibnamefont {Shimada}}, \bibinfo {author} {\bibfnamefont
  {H.}~\bibnamefont {Namatame}}, \bibinfo {author} {\bibfnamefont
  {M.}~\bibnamefont {Taniguchi}}, \ and\ \bibinfo {author} {\bibfnamefont
  {Y.}~\bibnamefont {Aiura}},\ }\bibfield  {title} {\enquote {\bibinfo {title}
  {{Interplay among Coulomb Interaction, Spin-Orbit Interaction, and Multiple
  Electron-Boson Interactions in Sr$_2$RuO$_4$}},}\ }\href
  {http://link.aps.org/doi/10.1103/PhysRevLett.105.226406} {\bibfield
  {journal} {\bibinfo  {journal} {Phys. Rev. Lett.}\ }\textbf {\bibinfo
  {volume} {105}},\ \bibinfo {pages} {226406} (\bibinfo {year}
  {2010})}\BibitemShut {NoStop}%
\bibitem [{\citenamefont {Iwasawa}\ \emph {et~al.}(2012)\citenamefont
  {Iwasawa}, \citenamefont {Yoshida}, \citenamefont {Hase}, \citenamefont
  {Shimada}, \citenamefont {Namatame}, \citenamefont {Taniguchi},\ and\
  \citenamefont {Aiura}}]{Iwasawa2012}%
  \BibitemOpen
  \bibfield  {author} {\bibinfo {author} {\bibfnamefont {H.}~\bibnamefont
  {Iwasawa}}, \bibinfo {author} {\bibfnamefont {Y.}~\bibnamefont {Yoshida}},
  \bibinfo {author} {\bibfnamefont {I.}~\bibnamefont {Hase}}, \bibinfo {author}
  {\bibfnamefont {K.}~\bibnamefont {Shimada}}, \bibinfo {author} {\bibfnamefont
  {H.}~\bibnamefont {Namatame}}, \bibinfo {author} {\bibfnamefont
  {M.}~\bibnamefont {Taniguchi}}, \ and\ \bibinfo {author} {\bibfnamefont
  {Y.}~\bibnamefont {Aiura}},\ }\bibfield  {title} {\enquote {\bibinfo {title}
  {{High-energy anomaly in the band dispersion of the ruthenate
  superconductor}},}\ }\href {\doibase 10.1103/PhysRevLett.109.066404}
  {\bibfield  {journal} {\bibinfo  {journal} {Phys. Rev. Lett.}\ }\textbf
  {\bibinfo {volume} {109}},\ \bibinfo {pages} {1} (\bibinfo {year}
  {2012})}\BibitemShut {NoStop}%
\bibitem [{\citenamefont {Zabolotnyy}\ \emph {et~al.}(2013)\citenamefont
  {Zabolotnyy}, \citenamefont {Evtushinsky}, \citenamefont {Kordyuk},
  \citenamefont {Kim}, \citenamefont {Carleschi}, \citenamefont {Doyle},
  \citenamefont {Fittipaldi}, \citenamefont {Cuoco}, \citenamefont
  {Vecchione},\ and\ \citenamefont {Borisenko}}]{Zabolotnyy2013}%
  \BibitemOpen
  \bibfield  {author} {\bibinfo {author} {\bibfnamefont {V.~B.}\ \bibnamefont
  {Zabolotnyy}}, \bibinfo {author} {\bibfnamefont {D.~V.}\ \bibnamefont
  {Evtushinsky}}, \bibinfo {author} {\bibfnamefont {A.~A.}\ \bibnamefont
  {Kordyuk}}, \bibinfo {author} {\bibfnamefont {T.~K.}\ \bibnamefont {Kim}},
  \bibinfo {author} {\bibfnamefont {E.}~\bibnamefont {Carleschi}}, \bibinfo
  {author} {\bibfnamefont {B.~P.}\ \bibnamefont {Doyle}}, \bibinfo {author}
  {\bibfnamefont {R.}~\bibnamefont {Fittipaldi}}, \bibinfo {author}
  {\bibfnamefont {M.}~\bibnamefont {Cuoco}}, \bibinfo {author} {\bibfnamefont
  {A.}~\bibnamefont {Vecchione}}, \ and\ \bibinfo {author} {\bibfnamefont
  {S.~V.}\ \bibnamefont {Borisenko}},\ }\bibfield  {title} {\enquote {\bibinfo
  {title} {{Renormalized band structure of Sr$_2$RuO$_4$: A quasiparticle
  tight-binding approach}},}\ }\href {\doibase 10.1016/j.elspec.2013.10.003}
  {\bibfield  {journal} {\bibinfo  {journal} {J. Electron Spectros. Relat.
  Phenomena}\ }\textbf {\bibinfo {volume} {191}},\ \bibinfo {pages} {48}
  (\bibinfo {year} {2013})}\BibitemShut {NoStop}%
\bibitem [{\citenamefont {Burganov}\ \emph {et~al.}(2016)\citenamefont
  {Burganov}, \citenamefont {Adamo}, \citenamefont {Mulder}, \citenamefont
  {Uchida}, \citenamefont {King}, \citenamefont {Harter}, \citenamefont {Shai},
  \citenamefont {Gibbs}, \citenamefont {Mackenzie}, \citenamefont {Uecker},
  \citenamefont {Bruetzam}, \citenamefont {Beasley}, \citenamefont {Fennie},
  \citenamefont {Schlom},\ and\ \citenamefont {Shen}}]{Burganov2016}%
  \BibitemOpen
  \bibfield  {author} {\bibinfo {author} {\bibfnamefont {B.}~\bibnamefont
  {Burganov}}, \bibinfo {author} {\bibfnamefont {C.}~\bibnamefont {Adamo}},
  \bibinfo {author} {\bibfnamefont {A.}~\bibnamefont {Mulder}}, \bibinfo
  {author} {\bibfnamefont {M.}~\bibnamefont {Uchida}}, \bibinfo {author}
  {\bibfnamefont {P.~D.C.}\ \bibnamefont {King}}, \bibinfo {author}
  {\bibfnamefont {J.~W.}\ \bibnamefont {Harter}}, \bibinfo {author}
  {\bibfnamefont {D.~E.}\ \bibnamefont {Shai}}, \bibinfo {author}
  {\bibfnamefont {A.~S.}\ \bibnamefont {Gibbs}}, \bibinfo {author}
  {\bibfnamefont {A.~P.}\ \bibnamefont {Mackenzie}}, \bibinfo {author}
  {\bibfnamefont {R.}~\bibnamefont {Uecker}}, \bibinfo {author} {\bibfnamefont
  {M.}~\bibnamefont {Bruetzam}}, \bibinfo {author} {\bibfnamefont {M.~R.}\
  \bibnamefont {Beasley}}, \bibinfo {author} {\bibfnamefont {C.~J.}\
  \bibnamefont {Fennie}}, \bibinfo {author} {\bibfnamefont {D.~G.}\
  \bibnamefont {Schlom}}, \ and\ \bibinfo {author} {\bibfnamefont {K.~M.}\
  \bibnamefont {Shen}},\ }\bibfield  {title} {\enquote {\bibinfo {title}
  {{Strain Control of Fermiology and Many-Body Interactions in Two-Dimensional
  Ruthenates}},}\ }\href {\doibase 10.1103/PhysRevLett.116.197003} {\bibfield
  {journal} {\bibinfo  {journal} {Phys. Rev. Lett.}\ }\textbf {\bibinfo
  {volume} {116}},\ \bibinfo {pages} {1} (\bibinfo {year} {2016})}\BibitemShut
  {NoStop}%
\bibitem [{\citenamefont {de' Medici}\ \emph {et~al.}(2011)\citenamefont {de'
  Medici}, \citenamefont {Mravlje},\ and\ \citenamefont
  {Georges}}]{demedici_2011}%
  \BibitemOpen
  \bibfield  {author} {\bibinfo {author} {\bibfnamefont {Luca}\ \bibnamefont
  {de' Medici}}, \bibinfo {author} {\bibfnamefont {Jernej}\ \bibnamefont
  {Mravlje}}, \ and\ \bibinfo {author} {\bibfnamefont {Antoine}\ \bibnamefont
  {Georges}},\ }\bibfield  {title} {\enquote {\bibinfo {title} {{Janus-Faced
  Influence of Hund's Rule Coupling in Strongly Correlated Materials}},}\
  }\href {\doibase 10.1103/PhysRevLett.107.256401} {\bibfield  {journal}
  {\bibinfo  {journal} {Phys. Rev. Lett.}\ }\textbf {\bibinfo {volume} {107}},\
  \bibinfo {pages} {256401} (\bibinfo {year} {2011})}\BibitemShut {NoStop}%
\bibitem [{\citenamefont {Tyler}\ \emph {et~al.}(1998)\citenamefont {Tyler},
  \citenamefont {Mackenzie}, \citenamefont {NishiZaki},\ and\ \citenamefont
  {Maeno}}]{tyler_1998}%
  \BibitemOpen
  \bibfield  {author} {\bibinfo {author} {\bibfnamefont {A.~W.}\ \bibnamefont
  {Tyler}}, \bibinfo {author} {\bibfnamefont {A.~P.}\ \bibnamefont
  {Mackenzie}}, \bibinfo {author} {\bibfnamefont {S.}~\bibnamefont
  {NishiZaki}}, \ and\ \bibinfo {author} {\bibfnamefont {Y.}~\bibnamefont
  {Maeno}},\ }\bibfield  {title} {\enquote {\bibinfo {title} {{High-temperature
  resistivity of Sr$_2$RuO$_4$: Bad metallic transport in a good metal}},}\
  }\href {\doibase 10.1103/PhysRevB.58.R10107} {\bibfield  {journal} {\bibinfo
  {journal} {Phys. Rev. B}\ }\textbf {\bibinfo {volume} {58}},\ \bibinfo
  {pages} {R10107} (\bibinfo {year} {1998})}\BibitemShut {NoStop}%
\bibitem [{\citenamefont {Georges}\ \emph {et~al.}(1996)\citenamefont
  {Georges}, \citenamefont {Kotliar}, \citenamefont {Krauth},\ and\
  \citenamefont {Rozenberg}}]{georges_1996}%
  \BibitemOpen
  \bibfield  {author} {\bibinfo {author} {\bibfnamefont {Antoine}\ \bibnamefont
  {Georges}}, \bibinfo {author} {\bibfnamefont {Gabriel}\ \bibnamefont
  {Kotliar}}, \bibinfo {author} {\bibfnamefont {Werner}\ \bibnamefont
  {Krauth}}, \ and\ \bibinfo {author} {\bibfnamefont {Marcelo~J.}\ \bibnamefont
  {Rozenberg}},\ }\bibfield  {title} {\enquote {\bibinfo {title} {{Dynamical
  mean-field theory of strongly correlated fermion systems and the limit of
  infinite dimensions}},}\ }\href {\doibase 10.1103/RevModPhys.68.13}
  {\bibfield  {journal} {\bibinfo  {journal} {Rev. Mod. Phys.}\ }\textbf
  {\bibinfo {volume} {68}},\ \bibinfo {pages} {13} (\bibinfo {year}
  {1996})}\BibitemShut {NoStop}%
\bibitem [{\citenamefont {Deng}\ \emph {et~al.}(2016)\citenamefont {Deng},
  \citenamefont {Haule},\ and\ \citenamefont {Kotliar}}]{deng_2016}%
  \BibitemOpen
  \bibfield  {author} {\bibinfo {author} {\bibfnamefont {Xiaoyu}\ \bibnamefont
  {Deng}}, \bibinfo {author} {\bibfnamefont {Kristjan}\ \bibnamefont {Haule}},
  \ and\ \bibinfo {author} {\bibfnamefont {Gabriel}\ \bibnamefont {Kotliar}},\
  }\bibfield  {title} {\enquote {\bibinfo {title} {Transport properties of
  metallic ruthenates: A $\mathrm{DFT}+\mathrm{DMFT}$ investigation},}\ }\href
  {\doibase 10.1103/PhysRevLett.116.256401} {\bibfield  {journal} {\bibinfo
  {journal} {Phys. Rev. Lett.}\ }\textbf {\bibinfo {volume} {116}},\ \bibinfo
  {pages} {256401} (\bibinfo {year} {2016})}\BibitemShut {NoStop}%
\bibitem [{\citenamefont {Sidis}\ \emph {et~al.}(1999)\citenamefont {Sidis},
  \citenamefont {Braden}, \citenamefont {Bourges}, \citenamefont {Hennion},
  \citenamefont {NishiZaki}, \citenamefont {Maeno},\ and\ \citenamefont
  {Mori}}]{Sidis1999}%
  \BibitemOpen
  \bibfield  {author} {\bibinfo {author} {\bibfnamefont {Y.}~\bibnamefont
  {Sidis}}, \bibinfo {author} {\bibfnamefont {M.}~\bibnamefont {Braden}},
  \bibinfo {author} {\bibfnamefont {P.}~\bibnamefont {Bourges}}, \bibinfo
  {author} {\bibfnamefont {B.}~\bibnamefont {Hennion}}, \bibinfo {author}
  {\bibfnamefont {S.}~\bibnamefont {NishiZaki}}, \bibinfo {author}
  {\bibfnamefont {Y.}~\bibnamefont {Maeno}}, \ and\ \bibinfo {author}
  {\bibfnamefont {Y.}~\bibnamefont {Mori}},\ }\bibfield  {title} {\enquote
  {\bibinfo {title} {{Evidence for Incommensurate Spin Fluctuations in
  Sr$_2$RuO$_4$}},}\ }\href {http://link.aps.org/abstract/PRL/v83/p3320}
  {\bibfield  {journal} {\bibinfo  {journal} {Phys. Rev. Lett.}\ }\textbf
  {\bibinfo {volume} {83}},\ \bibinfo {pages} {3320} (\bibinfo {year}
  {1999})}\BibitemShut {NoStop}%
\bibitem [{\citenamefont {Steffens}\ \emph {et~al.}(2019)\citenamefont
  {Steffens}, \citenamefont {Sidis}, \citenamefont {Kulda}, \citenamefont
  {Mao}, \citenamefont {Maeno}, \citenamefont {Mazin},\ and\ \citenamefont
  {Braden}}]{steffens_2018}%
  \BibitemOpen
  \bibfield  {author} {\bibinfo {author} {\bibfnamefont {P.}~\bibnamefont
  {Steffens}}, \bibinfo {author} {\bibfnamefont {Y.}~\bibnamefont {Sidis}},
  \bibinfo {author} {\bibfnamefont {J.}~\bibnamefont {Kulda}}, \bibinfo
  {author} {\bibfnamefont {Z.~Q.}\ \bibnamefont {Mao}}, \bibinfo {author}
  {\bibfnamefont {Y.}~\bibnamefont {Maeno}}, \bibinfo {author} {\bibfnamefont
  {I.~I.}\ \bibnamefont {Mazin}}, \ and\ \bibinfo {author} {\bibfnamefont
  {M.}~\bibnamefont {Braden}},\ }\bibfield  {title} {\enquote {\bibinfo {title}
  {{Spin Fluctuations in Sr$_2$RuO$_4$ from Polarized Neutron Scattering:
  Implications for Superconductivity}},}\ }\href {\doibase
  10.1103/PhysRevLett.122.047004} {\bibfield  {journal} {\bibinfo  {journal}
  {Phys. Rev. Lett.}\ }\textbf {\bibinfo {volume} {122}},\ \bibinfo {pages}
  {047004} (\bibinfo {year} {2019})}\BibitemShut {NoStop}%
\bibitem [{\citenamefont {Ishida}\ \emph {et~al.}(2001)\citenamefont {Ishida},
  \citenamefont {Mukuda}, \citenamefont {Minami}, \citenamefont {Kitaoka},
  \citenamefont {Mao}, \citenamefont {Fukazawa},\ and\ \citenamefont
  {Maeno}}]{ishida_2001}%
  \BibitemOpen
  \bibfield  {author} {\bibinfo {author} {\bibfnamefont {K.}~\bibnamefont
  {Ishida}}, \bibinfo {author} {\bibfnamefont {H.}~\bibnamefont {Mukuda}},
  \bibinfo {author} {\bibfnamefont {Y.}~\bibnamefont {Minami}}, \bibinfo
  {author} {\bibfnamefont {Y.}~\bibnamefont {Kitaoka}}, \bibinfo {author}
  {\bibfnamefont {Z.~Q.}\ \bibnamefont {Mao}}, \bibinfo {author} {\bibfnamefont
  {H.}~\bibnamefont {Fukazawa}}, \ and\ \bibinfo {author} {\bibfnamefont
  {Y.}~\bibnamefont {Maeno}},\ }\bibfield  {title} {\enquote {\bibinfo {title}
  {{Normal-state spin dynamics in the spin-triplet superconductor
  Sr$_2$RuO$_4$}},}\ }\href {\doibase 10.1103/PhysRevB.64.100501} {\bibfield
  {journal} {\bibinfo  {journal} {Phys. Rev. B}\ }\textbf {\bibinfo {volume}
  {64}},\ \bibinfo {pages} {100501} (\bibinfo {year} {2001})}\BibitemShut
  {NoStop}%
\bibitem [{\citenamefont {Imai}\ \emph {et~al.}(1998)\citenamefont {Imai},
  \citenamefont {Hunt}, \citenamefont {Thurber},\ and\ \citenamefont
  {Chou}}]{imai_1998}%
  \BibitemOpen
  \bibfield  {author} {\bibinfo {author} {\bibfnamefont {T.}~\bibnamefont
  {Imai}}, \bibinfo {author} {\bibfnamefont {A.~W.}\ \bibnamefont {Hunt}},
  \bibinfo {author} {\bibfnamefont {K.~R.}\ \bibnamefont {Thurber}}, \ and\
  \bibinfo {author} {\bibfnamefont {F.~C.}\ \bibnamefont {Chou}},\ }\bibfield
  {title} {\enquote {\bibinfo {title} {{$^{17}$O NMR Evidence for Orbital
  Dependent Ferromagnetic Correlations in Sr$_2$RuO$_4$}},}\ }\href {\doibase
  10.1103/PhysRevLett.81.3006} {\bibfield  {journal} {\bibinfo  {journal}
  {Phys. Rev. Lett.}\ }\textbf {\bibinfo {volume} {81}},\ \bibinfo {pages}
  {3006--3009} (\bibinfo {year} {1998})}\BibitemShut {NoStop}%
\bibitem [{\citenamefont {Lou}\ \emph {et~al.}(2003)\citenamefont {Lou},
  \citenamefont {Chang},\ and\ \citenamefont {Wu}}]{lou_2003}%
  \BibitemOpen
  \bibfield  {author} {\bibinfo {author} {\bibfnamefont {Ping}\ \bibnamefont
  {Lou}}, \bibinfo {author} {\bibfnamefont {Ming-Che}\ \bibnamefont {Chang}}, \
  and\ \bibinfo {author} {\bibfnamefont {Wen-Chin}\ \bibnamefont {Wu}},\
  }\bibfield  {title} {\enquote {\bibinfo {title} {{Evidence for the coupling
  between \ensuremath{\gamma}-band carriers and the incommensurate spin
  fluctuations in Sr$_2$RuO$_4$}},}\ }\href {\doibase
  10.1103/PhysRevB.68.012506} {\bibfield  {journal} {\bibinfo  {journal} {Phys.
  Rev. B}\ }\textbf {\bibinfo {volume} {68}},\ \bibinfo {pages} {012506}
  (\bibinfo {year} {2003})}\BibitemShut {NoStop}%
\bibitem [{\citenamefont {Kim}\ \emph {et~al.}(2017)\citenamefont {Kim},
  \citenamefont {Khmelevskyi}, \citenamefont {Mazin}, \citenamefont
  {Agterberg},\ and\ \citenamefont {Franchini}}]{kim_mazin_2017}%
  \BibitemOpen
  \bibfield  {author} {\bibinfo {author} {\bibfnamefont {Bongjae}\ \bibnamefont
  {Kim}}, \bibinfo {author} {\bibfnamefont {Sergii}\ \bibnamefont
  {Khmelevskyi}}, \bibinfo {author} {\bibfnamefont {Igor~I.}\ \bibnamefont
  {Mazin}}, \bibinfo {author} {\bibfnamefont {Daniel~F.}\ \bibnamefont
  {Agterberg}}, \ and\ \bibinfo {author} {\bibfnamefont {Cesare}\ \bibnamefont
  {Franchini}},\ }\bibfield  {title} {\enquote {\bibinfo {title} {{Anisotropy
  of magnetic interactions and symmetry of the order parameter in
  unconventional superconductor Sr$_2$RuO$_4$}},}\ }\href {\doibase
  10.1038/s41535-017-0041-8} {\bibfield  {journal} {\bibinfo  {journal} {npj
  Quantum Materials}\ }\textbf {\bibinfo {volume} {2}},\ \bibinfo {pages} {37}
  (\bibinfo {year} {2017})}\BibitemShut {NoStop}%
\bibitem [{\citenamefont {Vollhardt}(1984)}]{vollhardt_1984}%
  \BibitemOpen
  \bibfield  {author} {\bibinfo {author} {\bibfnamefont {Dieter}\ \bibnamefont
  {Vollhardt}},\ }\bibfield  {title} {\enquote {\bibinfo {title} {{Normal
  $^{3}\mathrm{He}$: an almost localized Fermi liquid}},}\ }\href {\doibase
  10.1103/RevModPhys.56.99} {\bibfield  {journal} {\bibinfo  {journal} {Rev.
  Mod. Phys.}\ }\textbf {\bibinfo {volume} {56}},\ \bibinfo {pages} {99}
  (\bibinfo {year} {1984})}\BibitemShut {NoStop}%
\bibitem [{\citenamefont {Aiura}\ \emph {et~al.}(2004)\citenamefont {Aiura},
  \citenamefont {Yoshida}, \citenamefont {Hase}, \citenamefont {Ikeda},
  \citenamefont {Higashiguchi}, \citenamefont {Cui}, \citenamefont {Shimada},
  \citenamefont {Namatame}, \citenamefont {Taniguchi},\ and\ \citenamefont
  {Bando}}]{Aiura2004}%
  \BibitemOpen
  \bibfield  {author} {\bibinfo {author} {\bibfnamefont {Y}~\bibnamefont
  {Aiura}}, \bibinfo {author} {\bibfnamefont {Y}~\bibnamefont {Yoshida}},
  \bibinfo {author} {\bibfnamefont {I}~\bibnamefont {Hase}}, \bibinfo {author}
  {\bibfnamefont {S~I}\ \bibnamefont {Ikeda}}, \bibinfo {author} {\bibfnamefont
  {M}~\bibnamefont {Higashiguchi}}, \bibinfo {author} {\bibfnamefont {X~Y}\
  \bibnamefont {Cui}}, \bibinfo {author} {\bibfnamefont {K}~\bibnamefont
  {Shimada}}, \bibinfo {author} {\bibfnamefont {H}~\bibnamefont {Namatame}},
  \bibinfo {author} {\bibfnamefont {M}~\bibnamefont {Taniguchi}}, \ and\
  \bibinfo {author} {\bibfnamefont {H}~\bibnamefont {Bando}},\ }\bibfield
  {title} {\enquote {\bibinfo {title} {{Kink in the Dispersion of Layered
  Strontium Ruthenates}},}\ }\href
  {http://link.aps.org/abstract/PRL/v93/e117005} {\bibfield  {journal}
  {\bibinfo  {journal} {Phys. Rev. Lett.}\ }\textbf {\bibinfo {volume} {93}},\
  \bibinfo {pages} {117004} (\bibinfo {year} {2004})}\BibitemShut {NoStop}%
\bibitem [{\citenamefont {Kim}\ \emph {et~al.}(2011)\citenamefont {Kim},
  \citenamefont {Kim}, \citenamefont {Kyung}, \citenamefont {Park},
  \citenamefont {Leem}, \citenamefont {Song}, \citenamefont {Kim},
  \citenamefont {Choi}, \citenamefont {Jung}, \citenamefont {Koh},
  \citenamefont {Choi}, \citenamefont {Yoshida}, \citenamefont {Moore},\ and\
  \citenamefont {Shen}}]{Kim2011}%
  \BibitemOpen
  \bibfield  {author} {\bibinfo {author} {\bibfnamefont {C.}~\bibnamefont
  {Kim}}, \bibinfo {author} {\bibfnamefont {Chul}\ \bibnamefont {Kim}},
  \bibinfo {author} {\bibfnamefont {W.~S.}\ \bibnamefont {Kyung}}, \bibinfo
  {author} {\bibfnamefont {S.~R.}\ \bibnamefont {Park}}, \bibinfo {author}
  {\bibfnamefont {C.~S.}\ \bibnamefont {Leem}}, \bibinfo {author}
  {\bibfnamefont {D.~J.}\ \bibnamefont {Song}}, \bibinfo {author}
  {\bibfnamefont {Y.~K.}\ \bibnamefont {Kim}}, \bibinfo {author} {\bibfnamefont
  {S.~K.}\ \bibnamefont {Choi}}, \bibinfo {author} {\bibfnamefont {W.~S.}\
  \bibnamefont {Jung}}, \bibinfo {author} {\bibfnamefont {Y.~Y.}\ \bibnamefont
  {Koh}}, \bibinfo {author} {\bibfnamefont {H.~Y.}\ \bibnamefont {Choi}},
  \bibinfo {author} {\bibfnamefont {Yoshiyuki}\ \bibnamefont {Yoshida}},
  \bibinfo {author} {\bibfnamefont {R.~G.}\ \bibnamefont {Moore}}, \ and\
  \bibinfo {author} {\bibfnamefont {Z.~X.}\ \bibnamefont {Shen}},\ }\bibfield
  {title} {\enquote {\bibinfo {title} {{Self-energy analysis of
  multiple-bosonic mode coupling in Sr$_2$RuO$_4$}},}\ }\href {\doibase
  10.1016/j.jpcs.2010.10.068} {\bibfield  {journal} {\bibinfo  {journal} {J.
  Phys. Chem. Solids}\ }\textbf {\bibinfo {volume} {72}},\ \bibinfo {pages}
  {556} (\bibinfo {year} {2011})}\BibitemShut {NoStop}%
\bibitem [{\citenamefont {Iwasawa}\ \emph {et~al.}(2013)\citenamefont
  {Iwasawa}, \citenamefont {Yoshida}, \citenamefont {Hase}, \citenamefont
  {Shimada}, \citenamefont {Namatame}, \citenamefont {Taniguchi},\ and\
  \citenamefont {Aiura}}]{Iwasawa2013}%
  \BibitemOpen
  \bibfield  {author} {\bibinfo {author} {\bibfnamefont {Hideaki}\ \bibnamefont
  {Iwasawa}}, \bibinfo {author} {\bibfnamefont {Yoshiyuki}\ \bibnamefont
  {Yoshida}}, \bibinfo {author} {\bibfnamefont {Izumi}\ \bibnamefont {Hase}},
  \bibinfo {author} {\bibfnamefont {Kenya}\ \bibnamefont {Shimada}}, \bibinfo
  {author} {\bibfnamefont {Hirofumi}\ \bibnamefont {Namatame}}, \bibinfo
  {author} {\bibfnamefont {Masaki}\ \bibnamefont {Taniguchi}}, \ and\ \bibinfo
  {author} {\bibfnamefont {Yoshihiro}\ \bibnamefont {Aiura}},\ }\bibfield
  {title} {\enquote {\bibinfo {title} {{`True' bosonic coupling strength in
  strongly correlated superconductors}},}\ }\href
  {https://doi.org/10.1038/srep01930} {\bibfield  {journal} {\bibinfo
  {journal} {Sci. Rep.}\ }\textbf {\bibinfo {volume} {3}},\ \bibinfo {pages}
  {1930} (\bibinfo {year} {2013})}\BibitemShut {NoStop}%
\bibitem [{\citenamefont {Wang}\ \emph {et~al.}(2017)\citenamefont {Wang},
  \citenamefont {Walkup}, \citenamefont {Derry}, \citenamefont {Scaffidi},
  \citenamefont {Rak}, \citenamefont {Vig}, \citenamefont {Kogar},
  \citenamefont {Zeljkovic}, \citenamefont {Husain}, \citenamefont {Santos},
  \citenamefont {Wang}, \citenamefont {Damascelli}, \citenamefont {Maeno},
  \citenamefont {Abbamonte}, \citenamefont {Fradkin},\ and\ \citenamefont
  {Madhavan}}]{Wang2017}%
  \BibitemOpen
  \bibfield  {author} {\bibinfo {author} {\bibfnamefont {Zhenyu}\ \bibnamefont
  {Wang}}, \bibinfo {author} {\bibfnamefont {Daniel}\ \bibnamefont {Walkup}},
  \bibinfo {author} {\bibfnamefont {Philip}\ \bibnamefont {Derry}}, \bibinfo
  {author} {\bibfnamefont {Thomas}\ \bibnamefont {Scaffidi}}, \bibinfo {author}
  {\bibfnamefont {Melinda}\ \bibnamefont {Rak}}, \bibinfo {author}
  {\bibfnamefont {Sean}\ \bibnamefont {Vig}}, \bibinfo {author} {\bibfnamefont
  {Anshul}\ \bibnamefont {Kogar}}, \bibinfo {author} {\bibfnamefont {Ilija}\
  \bibnamefont {Zeljkovic}}, \bibinfo {author} {\bibfnamefont {Ali}\
  \bibnamefont {Husain}}, \bibinfo {author} {\bibfnamefont {Luiz~H.}\
  \bibnamefont {Santos}}, \bibinfo {author} {\bibfnamefont {Yuxuan}\
  \bibnamefont {Wang}}, \bibinfo {author} {\bibfnamefont {Andrea}\ \bibnamefont
  {Damascelli}}, \bibinfo {author} {\bibfnamefont {Yoshiteru}\ \bibnamefont
  {Maeno}}, \bibinfo {author} {\bibfnamefont {Peter}\ \bibnamefont
  {Abbamonte}}, \bibinfo {author} {\bibfnamefont {Eduardo}\ \bibnamefont
  {Fradkin}}, \ and\ \bibinfo {author} {\bibfnamefont {Vidya}\ \bibnamefont
  {Madhavan}},\ }\bibfield  {title} {\enquote {\bibinfo {title} {{Quasiparticle
  interference and strong electron--mode coupling in the quasi-one-dimensional
  bands of Sr$_2$RuO$_4$}},}\ }\href {\doibase 10.1038/nphys4107} {\bibfield
  {journal} {\bibinfo  {journal} {Nat. Phys.}\ }\textbf {\bibinfo {volume}
  {13}},\ \bibinfo {pages} {799} (\bibinfo {year} {2017})}\BibitemShut
  {NoStop}%
\bibitem [{\citenamefont {Akebi}\ \emph {et~al.}(2019)\citenamefont {Akebi},
  \citenamefont {Kondo}, \citenamefont {Nakayama}, \citenamefont {Kuroda},
  \citenamefont {Kunisada}, \citenamefont {Taniguchi}, \citenamefont {Maeno},\
  and\ \citenamefont {Shin}}]{Shuntaro2019}%
  \BibitemOpen
  \bibfield  {author} {\bibinfo {author} {\bibfnamefont {Shuntaro}\
  \bibnamefont {Akebi}}, \bibinfo {author} {\bibfnamefont {Takeshi}\
  \bibnamefont {Kondo}}, \bibinfo {author} {\bibfnamefont {Mitsuhiro}\
  \bibnamefont {Nakayama}}, \bibinfo {author} {\bibfnamefont {Kenta}\
  \bibnamefont {Kuroda}}, \bibinfo {author} {\bibfnamefont {So}~\bibnamefont
  {Kunisada}}, \bibinfo {author} {\bibfnamefont {Haruka}\ \bibnamefont
  {Taniguchi}}, \bibinfo {author} {\bibfnamefont {Yoshiteru}\ \bibnamefont
  {Maeno}}, \ and\ \bibinfo {author} {\bibfnamefont {Shik}\ \bibnamefont
  {Shin}},\ }\bibfield  {title} {\enquote {\bibinfo {title} {{Low-energy
  electron-mode couplings in the surface bands of Sr$_2$RuO$_4$ revealed by
  laser-based angle-resolved photoemission spectroscopy}},}\ }\href {\doibase
  10.1103/PhysRevB.99.081108} {\bibfield  {journal} {\bibinfo  {journal} {Phys.
  Rev. B}\ }\textbf {\bibinfo {volume} {99}},\ \bibinfo {pages} {081108}
  (\bibinfo {year} {2019})}\BibitemShut {NoStop}%
\bibitem [{\citenamefont {Liu}\ \emph {et~al.}(2008)\citenamefont {Liu},
  \citenamefont {Antonov}, \citenamefont {Jepsen},\ and\ \citenamefont
  {Andersen.}}]{liu_prl_2008}%
  \BibitemOpen
  \bibfield  {author} {\bibinfo {author} {\bibfnamefont {Guo-Qiang}\
  \bibnamefont {Liu}}, \bibinfo {author} {\bibfnamefont {V.~N.}\ \bibnamefont
  {Antonov}}, \bibinfo {author} {\bibfnamefont {O.}~\bibnamefont {Jepsen}}, \
  and\ \bibinfo {author} {\bibfnamefont {O.~K.}\ \bibnamefont {Andersen.}},\
  }\bibfield  {title} {\enquote {\bibinfo {title} {{Coulomb-Enhanced Spin-Orbit
  Splitting: The Missing Piece in the Sr$_2$RhO$_4$ Puzzle}},}\ }\href
  {\doibase 10.1103/PhysRevLett.101.026408} {\bibfield  {journal} {\bibinfo
  {journal} {Phys. Rev. Lett.}\ }\textbf {\bibinfo {volume} {101}},\ \bibinfo
  {pages} {026408} (\bibinfo {year} {2008})}\BibitemShut {NoStop}%
\bibitem [{\citenamefont {He}\ \emph {et~al.}(2016)\citenamefont {He},
  \citenamefont {Vishik}, \citenamefont {Yi}, \citenamefont {Yang},
  \citenamefont {Liu}, \citenamefont {Lee}, \citenamefont {Chen}, \citenamefont
  {Rebec}, \citenamefont {Leuenberger}, \citenamefont {Zong}, \citenamefont
  {Jefferson}, \citenamefont {Moore}, \citenamefont {Kirchmann}, \citenamefont
  {Merriam},\ and\ \citenamefont {Shen}}]{He2016}%
  \BibitemOpen
  \bibfield  {author} {\bibinfo {author} {\bibfnamefont {Yu}~\bibnamefont
  {He}}, \bibinfo {author} {\bibfnamefont {Inna~M.}\ \bibnamefont {Vishik}},
  \bibinfo {author} {\bibfnamefont {Ming}\ \bibnamefont {Yi}}, \bibinfo
  {author} {\bibfnamefont {Shuolong}\ \bibnamefont {Yang}}, \bibinfo {author}
  {\bibfnamefont {Zhongkai}\ \bibnamefont {Liu}}, \bibinfo {author}
  {\bibfnamefont {James~J.}\ \bibnamefont {Lee}}, \bibinfo {author}
  {\bibfnamefont {Sudi}\ \bibnamefont {Chen}}, \bibinfo {author} {\bibfnamefont
  {Slavko~N.}\ \bibnamefont {Rebec}}, \bibinfo {author} {\bibfnamefont
  {Dominik}\ \bibnamefont {Leuenberger}}, \bibinfo {author} {\bibfnamefont
  {Alfred}\ \bibnamefont {Zong}}, \bibinfo {author} {\bibfnamefont
  {C.~Michael}\ \bibnamefont {Jefferson}}, \bibinfo {author} {\bibfnamefont
  {Robert~G.}\ \bibnamefont {Moore}}, \bibinfo {author} {\bibfnamefont
  {Patrick~S.}\ \bibnamefont {Kirchmann}}, \bibinfo {author} {\bibfnamefont
  {Andrew~J.}\ \bibnamefont {Merriam}}, \ and\ \bibinfo {author} {\bibfnamefont
  {Zhi-Xun}\ \bibnamefont {Shen}},\ }\bibfield  {title} {\enquote {\bibinfo
  {title} {Invited article: High resolution angle resolved photoemission with
  tabletop 11 ev laser},}\ }\href {\doibase 10.1063/1.4939759} {\bibfield
  {journal} {\bibinfo  {journal} {Rev. Sci. Instrum.}\ }\textbf {\bibinfo
  {volume} {87}},\ \bibinfo {pages} {011301} (\bibinfo {year}
  {2016})}\BibitemShut {NoStop}%
\bibitem [{\citenamefont {Hoesch}\ \emph {et~al.}(2017)\citenamefont {Hoesch},
  \citenamefont {Kim}, \citenamefont {Dudin}, \citenamefont {Wang},
  \citenamefont {Scott}, \citenamefont {Harris}, \citenamefont {Patel},
  \citenamefont {Matthews}, \citenamefont {Hawkins}, \citenamefont {Alcock},
  \citenamefont {Richter}, \citenamefont {Mudd}, \citenamefont {Basham},
  \citenamefont {Pratt}, \citenamefont {Leicester}, \citenamefont {Longhi},
  \citenamefont {Tamai},\ and\ \citenamefont {Baumberger}}]{Hoesch2017}%
  \BibitemOpen
  \bibfield  {author} {\bibinfo {author} {\bibfnamefont {M.}~\bibnamefont
  {Hoesch}}, \bibinfo {author} {\bibfnamefont {T.K.}\ \bibnamefont {Kim}},
  \bibinfo {author} {\bibfnamefont {P.}~\bibnamefont {Dudin}}, \bibinfo
  {author} {\bibfnamefont {H.}~\bibnamefont {Wang}}, \bibinfo {author}
  {\bibfnamefont {S.}~\bibnamefont {Scott}}, \bibinfo {author} {\bibfnamefont
  {P.}~\bibnamefont {Harris}}, \bibinfo {author} {\bibfnamefont
  {S.}~\bibnamefont {Patel}}, \bibinfo {author} {\bibfnamefont
  {M.}~\bibnamefont {Matthews}}, \bibinfo {author} {\bibfnamefont
  {D.}~\bibnamefont {Hawkins}}, \bibinfo {author} {\bibfnamefont {S.G.}\
  \bibnamefont {Alcock}}, \bibinfo {author} {\bibfnamefont {T.}~\bibnamefont
  {Richter}}, \bibinfo {author} {\bibfnamefont {J.J.}\ \bibnamefont {Mudd}},
  \bibinfo {author} {\bibfnamefont {M.}~\bibnamefont {Basham}}, \bibinfo
  {author} {\bibfnamefont {L.}~\bibnamefont {Pratt}}, \bibinfo {author}
  {\bibfnamefont {P.}~\bibnamefont {Leicester}}, \bibinfo {author}
  {\bibfnamefont {E.C.}\ \bibnamefont {Longhi}}, \bibinfo {author}
  {\bibfnamefont {A.}~\bibnamefont {Tamai}}, \ and\ \bibinfo {author}
  {\bibfnamefont {F.}~\bibnamefont {Baumberger}},\ }\bibfield  {title}
  {\enquote {\bibinfo {title} {{A facility for the analysis of the electronic
  structures of solids and their surfaces by synchrotron radiation
  photoelectron spectroscopy}},}\ }\href {\doibase 10.1063/1.4973562}
  {\bibfield  {journal} {\bibinfo  {journal} {Rev. Sci. Instrum.}\ }\textbf
  {\bibinfo {volume} {88}},\ \bibinfo {pages} {013106} (\bibinfo {year}
  {2017})}\BibitemShut {NoStop}%
\bibitem [{\citenamefont {Shen}\ \emph {et~al.}(2001)\citenamefont {Shen},
  \citenamefont {Damascelli}, \citenamefont {Lu}, \citenamefont {Armitage},
  \citenamefont {Ronning}, \citenamefont {Feng}, \citenamefont {Kim},
  \citenamefont {Shen}, \citenamefont {Singh}, \citenamefont {Mazin},
  \citenamefont {Nakatsuji}, \citenamefont {Mao}, \citenamefont {Maeno},
  \citenamefont {Kimura},\ and\ \citenamefont {Tokura}}]{Shen2001}%
  \BibitemOpen
  \bibfield  {author} {\bibinfo {author} {\bibfnamefont {K.~M.}\ \bibnamefont
  {Shen}}, \bibinfo {author} {\bibfnamefont {A.}~\bibnamefont {Damascelli}},
  \bibinfo {author} {\bibfnamefont {D.~H.}\ \bibnamefont {Lu}}, \bibinfo
  {author} {\bibfnamefont {N.~P.}\ \bibnamefont {Armitage}}, \bibinfo {author}
  {\bibfnamefont {F.}~\bibnamefont {Ronning}}, \bibinfo {author} {\bibfnamefont
  {D.~L.}\ \bibnamefont {Feng}}, \bibinfo {author} {\bibfnamefont
  {C.}~\bibnamefont {Kim}}, \bibinfo {author} {\bibfnamefont {Z.~X.}\
  \bibnamefont {Shen}}, \bibinfo {author} {\bibfnamefont {D.~J.}\ \bibnamefont
  {Singh}}, \bibinfo {author} {\bibfnamefont {I.~I.}\ \bibnamefont {Mazin}},
  \bibinfo {author} {\bibfnamefont {S.}~\bibnamefont {Nakatsuji}}, \bibinfo
  {author} {\bibfnamefont {Z.~Q.}\ \bibnamefont {Mao}}, \bibinfo {author}
  {\bibfnamefont {Y.}~\bibnamefont {Maeno}}, \bibinfo {author} {\bibfnamefont
  {T.}~\bibnamefont {Kimura}}, \ and\ \bibinfo {author} {\bibfnamefont
  {Y.}~\bibnamefont {Tokura}},\ }\bibfield  {title} {\enquote {\bibinfo {title}
  {{Surface electronic structure of Sr$_2$RuO$_4$}},}\ }\href
  {http://link.aps.org/abstract/PRB/v64/e180502} {\bibfield  {journal}
  {\bibinfo  {journal} {Phys. Rev. B}\ }\textbf {\bibinfo {volume} {64}},\
  \bibinfo {pages} {180502} (\bibinfo {year} {2001})}\BibitemShut {NoStop}%
\bibitem [{\citenamefont {St{\"{o}}ger}\ \emph {et~al.}(2014)\citenamefont
  {St{\"{o}}ger}, \citenamefont {Hieckel}, \citenamefont {Mittendorfer},
  \citenamefont {Wang}, \citenamefont {Fobes}, \citenamefont {Peng},
  \citenamefont {Mao}, \citenamefont {Schmid}, \citenamefont {Redinger},\ and\
  \citenamefont {Diebold}}]{Stoger2014}%
  \BibitemOpen
  \bibfield  {author} {\bibinfo {author} {\bibfnamefont {Bernhard}\
  \bibnamefont {St{\"{o}}ger}}, \bibinfo {author} {\bibfnamefont {Marcel}\
  \bibnamefont {Hieckel}}, \bibinfo {author} {\bibfnamefont {Florian}\
  \bibnamefont {Mittendorfer}}, \bibinfo {author} {\bibfnamefont {Zhiming}\
  \bibnamefont {Wang}}, \bibinfo {author} {\bibfnamefont {David}\ \bibnamefont
  {Fobes}}, \bibinfo {author} {\bibfnamefont {Jin}\ \bibnamefont {Peng}},
  \bibinfo {author} {\bibfnamefont {Zhiqiang}\ \bibnamefont {Mao}}, \bibinfo
  {author} {\bibfnamefont {Michael}\ \bibnamefont {Schmid}}, \bibinfo {author}
  {\bibfnamefont {Josef}\ \bibnamefont {Redinger}}, \ and\ \bibinfo {author}
  {\bibfnamefont {Ulrike}\ \bibnamefont {Diebold}},\ }\bibfield  {title}
  {\enquote {\bibinfo {title} {{High Chemical Activity of a Perovskite Surface:
  Reaction of CO with Sr$_3$Ru$_2$O$_7$}},}\ }\href {\doibase
  10.1103/PhysRevLett.113.116101} {\bibfield  {journal} {\bibinfo  {journal}
  {Phys. Rev. Lett.}\ }\textbf {\bibinfo {volume} {113}},\ \bibinfo {pages}
  {116101} (\bibinfo {year} {2014})}\BibitemShut {NoStop}%
\bibitem [{\citenamefont {Shen}\ \emph {et~al.}(2007)\citenamefont {Shen},
  \citenamefont {Kikugawa}, \citenamefont {Bergemann}, \citenamefont {Balicas},
  \citenamefont {Baumberger}, \citenamefont {Meevasana}, \citenamefont {Ingle},
  \citenamefont {Maeno}, \citenamefont {Shen},\ and\ \citenamefont
  {Mackenzie}}]{Shen2007}%
  \BibitemOpen
  \bibfield  {author} {\bibinfo {author} {\bibfnamefont {K.~M.}\ \bibnamefont
  {Shen}}, \bibinfo {author} {\bibfnamefont {N.}~\bibnamefont {Kikugawa}},
  \bibinfo {author} {\bibfnamefont {C.}~\bibnamefont {Bergemann}}, \bibinfo
  {author} {\bibfnamefont {L.}~\bibnamefont {Balicas}}, \bibinfo {author}
  {\bibfnamefont {F.}~\bibnamefont {Baumberger}}, \bibinfo {author}
  {\bibfnamefont {W.}~\bibnamefont {Meevasana}}, \bibinfo {author}
  {\bibfnamefont {N.~J.~C.}\ \bibnamefont {Ingle}}, \bibinfo {author}
  {\bibfnamefont {Y.}~\bibnamefont {Maeno}}, \bibinfo {author} {\bibfnamefont
  {Z.~X.}\ \bibnamefont {Shen}}, \ and\ \bibinfo {author} {\bibfnamefont
  {A.~P.}\ \bibnamefont {Mackenzie}},\ }\bibfield  {title} {\enquote {\bibinfo
  {title} {{Evolution of the Fermi Surface and Quasiparticle Renormalization
  through a van Hove Singularity in Sr$_{2-y}$La$_y$RuO$_4$}},}\ }\href
  {http://link.aps.org/abstract/PRL/v99/e187001} {\bibfield  {journal}
  {\bibinfo  {journal} {Phys. Rev. Lett.}\ }\textbf {\bibinfo {volume} {99}},\
  \bibinfo {pages} {187001} (\bibinfo {year} {2007})}\BibitemShut {NoStop}%
\bibitem [{Note1()}]{Note1}%
  \BibitemOpen
  \bibinfo {note} {We note that radial ${\protect \bm {k}}$-space cuts are not
  exactly perpendicular to the Fermi surface. This can cause the velocities
  $v_F$ and $v_b$ given here to deviate by up to 10\% from the Fermi velocity.
  However, since we evaluate the experimental and theoretical dispersion along
  the same ${\protect \bm {k}}$-space cut, this effect cancels in Fig.~\ref
  {fig:f1bis}~(d) and does not affect the self-energy determination in
  Sec.~\ref {sec:self}}\BibitemShut {NoStop}%
\bibitem [{\citenamefont {Baumberger}\ \emph {et~al.}(2006)\citenamefont
  {Baumberger}, \citenamefont {Ingle}, \citenamefont {Meevasana}, \citenamefont
  {Shen}, \citenamefont {Lu}, \citenamefont {Perry}, \citenamefont {Mackenzie},
  \citenamefont {Hussain}, \citenamefont {Singh},\ and\ \citenamefont
  {Shen}}]{Baumberger2006}%
  \BibitemOpen
  \bibfield  {author} {\bibinfo {author} {\bibfnamefont {F.}~\bibnamefont
  {Baumberger}}, \bibinfo {author} {\bibfnamefont {N.~J.~C.}\ \bibnamefont
  {Ingle}}, \bibinfo {author} {\bibfnamefont {W.}~\bibnamefont {Meevasana}},
  \bibinfo {author} {\bibfnamefont {K.~M.}\ \bibnamefont {Shen}}, \bibinfo
  {author} {\bibfnamefont {D.~H.}\ \bibnamefont {Lu}}, \bibinfo {author}
  {\bibfnamefont {R.~S.}\ \bibnamefont {Perry}}, \bibinfo {author}
  {\bibfnamefont {A.~P.}\ \bibnamefont {Mackenzie}}, \bibinfo {author}
  {\bibfnamefont {Z.}~\bibnamefont {Hussain}}, \bibinfo {author} {\bibfnamefont
  {D.~J.}\ \bibnamefont {Singh}}, \ and\ \bibinfo {author} {\bibfnamefont
  {Z.-X.}\ \bibnamefont {Shen}},\ }\bibfield  {title} {\enquote {\bibinfo
  {title} {{Fermi Surface and Quasiparticle Excitations of Sr$_2$RhO$_4$}},}\
  }\href {\doibase 10.1103/PhysRevLett.96.246402} {\bibfield  {journal}
  {\bibinfo  {journal} {Phys. Rev. Lett.}\ }\textbf {\bibinfo {volume} {96}},\
  \bibinfo {pages} {246402} (\bibinfo {year} {2006})}\BibitemShut {NoStop}%
\bibitem [{Note2()}]{Note2}%
  \BibitemOpen
  \bibinfo {note} {We attribute the higher Fermi velocities reported in some
  earlier studies~\cite {Iwasawa2005,Iwasawa2010,Iwasawa2012,Burganov2016} to
  the lower energy resolution, which causes an extended range near the Fermi
  level where the dispersion extracted from fits to individual MDCs is
  artificially enhanced, rendering a precise determination of $v_{F}$
  difficult. The much lower value of $v_{F}^{\beta }$ along $\Gamma $M reported
  in Ref.~\cite {Wang2017} corresponds to the surface $\beta $ band, as shown
  in appendix~\ref {sec:appA}.}\BibitemShut {Stop}%
\bibitem [{\citenamefont {Marzari}\ and\ \citenamefont
  {Vanderbilt}(1997)}]{MLWF1}%
  \BibitemOpen
  \bibfield  {author} {\bibinfo {author} {\bibfnamefont {Nicola}\ \bibnamefont
  {Marzari}}\ and\ \bibinfo {author} {\bibfnamefont {David}\ \bibnamefont
  {Vanderbilt}},\ }\bibfield  {title} {\enquote {\bibinfo {title} {Maximally
  localized generalized wannier functions for composite energy bands},}\ }\href
  {\doibase 10.1103/PhysRevB.56.12847} {\bibfield  {journal} {\bibinfo
  {journal} {Phys. Rev. B}\ }\textbf {\bibinfo {volume} {56}},\ \bibinfo
  {pages} {12847--12865} (\bibinfo {year} {1997})}\BibitemShut {NoStop}%
\bibitem [{\citenamefont {Souza}\ \emph {et~al.}(2001)\citenamefont {Souza},
  \citenamefont {Marzari},\ and\ \citenamefont {Vanderbilt}}]{MLWF2}%
  \BibitemOpen
  \bibfield  {author} {\bibinfo {author} {\bibfnamefont {Ivo}\ \bibnamefont
  {Souza}}, \bibinfo {author} {\bibfnamefont {Nicola}\ \bibnamefont {Marzari}},
  \ and\ \bibinfo {author} {\bibfnamefont {David}\ \bibnamefont {Vanderbilt}},\
  }\bibfield  {title} {\enquote {\bibinfo {title} {Maximally localized wannier
  functions for entangled energy bands},}\ }\href {\doibase
  10.1103/PhysRevB.65.035109} {\bibfield  {journal} {\bibinfo  {journal} {Phys.
  Rev. B}\ }\textbf {\bibinfo {volume} {65}},\ \bibinfo {pages} {035109}
  (\bibinfo {year} {2001})}\BibitemShut {NoStop}%
\bibitem [{\citenamefont {Lechermann}\ \emph {et~al.}(2006)\citenamefont
  {Lechermann}, \citenamefont {Georges}, \citenamefont {Poteryaev},
  \citenamefont {Biermann}, \citenamefont {Posternak}, \citenamefont
  {Yamasaki},\ and\ \citenamefont {Andersen}}]{Lechermann_2006}%
  \BibitemOpen
  \bibfield  {author} {\bibinfo {author} {\bibfnamefont {F.}~\bibnamefont
  {Lechermann}}, \bibinfo {author} {\bibfnamefont {A.}~\bibnamefont {Georges}},
  \bibinfo {author} {\bibfnamefont {A.}~\bibnamefont {Poteryaev}}, \bibinfo
  {author} {\bibfnamefont {S.}~\bibnamefont {Biermann}}, \bibinfo {author}
  {\bibfnamefont {M.}~\bibnamefont {Posternak}}, \bibinfo {author}
  {\bibfnamefont {A.}~\bibnamefont {Yamasaki}}, \ and\ \bibinfo {author}
  {\bibfnamefont {O.~K.}\ \bibnamefont {Andersen}},\ }\bibfield  {title}
  {\enquote {\bibinfo {title} {Dynamical mean-field theory using wannier
  functions: A flexible route to electronic structure calculations of strongly
  correlated materials},}\ }\href {\doibase 10.1103/PhysRevB.74.125120}
  {\bibfield  {journal} {\bibinfo  {journal} {Phys. Rev. B}\ }\textbf {\bibinfo
  {volume} {74}},\ \bibinfo {pages} {125120} (\bibinfo {year}
  {2006})}\BibitemShut {NoStop}%
\bibitem [{Note3()}]{Note3}%
  \BibitemOpen
  \bibinfo {note} {By restricting the Hamiltonian to the $t_{2g}$-subspace we
  neglect the $e_{g}$-$t_{2g}$ coupling terms of $\protect \mathaccentV
  {hat}05E{H}^{\protect \mathrm {SOC}}_{\lambda }$. This approximation is valid
  as long as the $e_{g}$-$t_{2g}$ crystal-field splitting is large in
  comparison to $\lambda $, which is the case for \ce {Sr2RuO4}{}.}\BibitemShut
  {Stop}%
\bibitem [{\citenamefont {Ng}\ and\ \citenamefont
  {Sigrist}(2002)}]{ng_sigrist_2002}%
  \BibitemOpen
  \bibfield  {author} {\bibinfo {author} {\bibfnamefont {K.~K.}\ \bibnamefont
  {Ng}}\ and\ \bibinfo {author} {\bibfnamefont {M.}~\bibnamefont {Sigrist}},\
  }\bibfield  {title} {\enquote {\bibinfo {title} {{The role of spin-orbit
  coupling for the superconducting state in Sr$_2$RuO$_4$}},}\ }\href
  {https://doi.org/10.1209/epl/i2000-00173-x} {\bibfield  {journal} {\bibinfo
  {journal} {Europhys. Lett.}\ }\textbf {\bibinfo {volume} {49}},\ \bibinfo
  {pages} {473} (\bibinfo {year} {2002})}\BibitemShut {NoStop}%
\bibitem [{\citenamefont {Eremin}\ \emph {et~al.}(2002)\citenamefont {Eremin},
  \citenamefont {Manske},\ and\ \citenamefont {Bennemann}}]{eremin_2002}%
  \BibitemOpen
  \bibfield  {author} {\bibinfo {author} {\bibfnamefont {I.}~\bibnamefont
  {Eremin}}, \bibinfo {author} {\bibfnamefont {D.}~\bibnamefont {Manske}}, \
  and\ \bibinfo {author} {\bibfnamefont {K.~H.}\ \bibnamefont {Bennemann}},\
  }\bibfield  {title} {\enquote {\bibinfo {title} {{Electronic theory for the
  normal-state spin dynamics in Sr$_2$RuO$_4$: Anisotropy due to spin-orbit
  coupling}},}\ }\href {\doibase 10.1103/PhysRevB.65.220502} {\bibfield
  {journal} {\bibinfo  {journal} {Phys. Rev. B}\ }\textbf {\bibinfo {volume}
  {65}},\ \bibinfo {pages} {220502} (\bibinfo {year} {2002})}\BibitemShut
  {NoStop}%
\bibitem [{\citenamefont {Haverkort}\ \emph {et~al.}(2008)\citenamefont
  {Haverkort}, \citenamefont {Elfimov}, \citenamefont {Tjeng}, \citenamefont
  {Sawatzky},\ and\ \citenamefont {Damascelli}}]{Haverkort2008}%
  \BibitemOpen
  \bibfield  {author} {\bibinfo {author} {\bibfnamefont {M.}~\bibnamefont
  {Haverkort}}, \bibinfo {author} {\bibfnamefont {I.}~\bibnamefont {Elfimov}},
  \bibinfo {author} {\bibfnamefont {L.}~\bibnamefont {Tjeng}}, \bibinfo
  {author} {\bibfnamefont {G.}~\bibnamefont {Sawatzky}}, \ and\ \bibinfo
  {author} {\bibfnamefont {A.}~\bibnamefont {Damascelli}},\ }\bibfield  {title}
  {\enquote {\bibinfo {title} {{Strong Spin-Orbit Coupling Effects on the Fermi
  Surface of Sr$_2$RuO$_4$ and Sr$_2$RhO$_4$}},}\ }\href {\doibase
  10.1103/PhysRevLett.101.026406} {\bibfield  {journal} {\bibinfo  {journal}
  {Phys. Rev. Lett.}\ }\textbf {\bibinfo {volume} {101}},\ \bibinfo {pages}
  {026406} (\bibinfo {year} {2008})}\BibitemShut {NoStop}%
\bibitem [{\citenamefont {{Puetter}}\ and\ \citenamefont
  {{Kee}}(2012)}]{puetter_kee_2012}%
  \BibitemOpen
  \bibfield  {author} {\bibinfo {author} {\bibfnamefont {C.~M.}\ \bibnamefont
  {{Puetter}}}\ and\ \bibinfo {author} {\bibfnamefont {H.-Y.}\ \bibnamefont
  {{Kee}}},\ }\bibfield  {title} {\enquote {\bibinfo {title} {{Identifying
  spin-triplet pairing in spin-orbit coupled multi-band superconductors}},}\
  }\href {\doibase 10.1209/0295-5075/98/27010} {\bibfield  {journal} {\bibinfo
  {journal} {Europhysics Letters}\ }\textbf {\bibinfo {volume} {98}},\ \bibinfo
  {pages} {27010} (\bibinfo {year} {2012})}\BibitemShut {NoStop}%
\bibitem [{\citenamefont {Veenstra}\ \emph {et~al.}(2014)\citenamefont
  {Veenstra}, \citenamefont {Zhu}, \citenamefont {Raichle}, \citenamefont
  {Ludbrook}, \citenamefont {Nicolaou}, \citenamefont {Slomski}, \citenamefont
  {Landolt}, \citenamefont {Kittaka}, \citenamefont {Maeno}, \citenamefont
  {Dil}, \citenamefont {Elfimov}, \citenamefont {Haverkort},\ and\
  \citenamefont {Damascelli}}]{Veenstra2013}%
  \BibitemOpen
  \bibfield  {author} {\bibinfo {author} {\bibfnamefont {C.~N.}\ \bibnamefont
  {Veenstra}}, \bibinfo {author} {\bibfnamefont {Z.~H.}\ \bibnamefont {Zhu}},
  \bibinfo {author} {\bibfnamefont {M.}~\bibnamefont {Raichle}}, \bibinfo
  {author} {\bibfnamefont {B.~M.}\ \bibnamefont {Ludbrook}}, \bibinfo {author}
  {\bibfnamefont {A.}~\bibnamefont {Nicolaou}}, \bibinfo {author}
  {\bibfnamefont {B.}~\bibnamefont {Slomski}}, \bibinfo {author} {\bibfnamefont
  {G.}~\bibnamefont {Landolt}}, \bibinfo {author} {\bibfnamefont
  {S.}~\bibnamefont {Kittaka}}, \bibinfo {author} {\bibfnamefont
  {Y.}~\bibnamefont {Maeno}}, \bibinfo {author} {\bibfnamefont {J.~H.}\
  \bibnamefont {Dil}}, \bibinfo {author} {\bibfnamefont {I.~S.}\ \bibnamefont
  {Elfimov}}, \bibinfo {author} {\bibfnamefont {M.~W.}\ \bibnamefont
  {Haverkort}}, \ and\ \bibinfo {author} {\bibfnamefont {A.}~\bibnamefont
  {Damascelli}},\ }\bibfield  {title} {\enquote {\bibinfo {title}
  {{Spin-orbital entanglement and the breakdown of Singlets and triplets in
  Sr$_2$RuO$_4$ revealed by spin- and angle-resolved photoemission
  spectroscopy}},}\ }\href {\doibase 10.1103/PhysRevLett.112.127002} {\bibfield
   {journal} {\bibinfo  {journal} {Phys. Rev. Lett.}\ }\textbf {\bibinfo
  {volume} {112}},\ \bibinfo {pages} {127002} (\bibinfo {year}
  {2014})}\BibitemShut {NoStop}%
\bibitem [{\citenamefont {Poteryaev}\ \emph {et~al.}(2007)\citenamefont
  {Poteryaev}, \citenamefont {Tomczak}, \citenamefont {Biermann}, \citenamefont
  {Georges}, \citenamefont {Lichtenstein}, \citenamefont {Rubtsov},
  \citenamefont {Saha-Dasgupta},\ and\ \citenamefont
  {Andersen}}]{Poteryaev2007}%
  \BibitemOpen
  \bibfield  {author} {\bibinfo {author} {\bibfnamefont {Alexander~I.}\
  \bibnamefont {Poteryaev}}, \bibinfo {author} {\bibfnamefont {Jan~M.}\
  \bibnamefont {Tomczak}}, \bibinfo {author} {\bibfnamefont {Silke}\
  \bibnamefont {Biermann}}, \bibinfo {author} {\bibfnamefont {Antoine}\
  \bibnamefont {Georges}}, \bibinfo {author} {\bibfnamefont {Alexander~I.}\
  \bibnamefont {Lichtenstein}}, \bibinfo {author} {\bibfnamefont {Alexey~N.}\
  \bibnamefont {Rubtsov}}, \bibinfo {author} {\bibfnamefont {Tanusri}\
  \bibnamefont {Saha-Dasgupta}}, \ and\ \bibinfo {author} {\bibfnamefont
  {Ole~K.}\ \bibnamefont {Andersen}},\ }\bibfield  {title} {\enquote {\bibinfo
  {title} {{Enhanced crystal-field splitting and orbital-selective coherence
  induced by strong correlations in V$_2$O$_3$}},}\ }\href {\doibase
  10.1103/PhysRevB.76.085127} {\bibfield  {journal} {\bibinfo  {journal} {Phys.
  Rev. B}\ }\textbf {\bibinfo {volume} {76}},\ \bibinfo {pages} {085127}
  (\bibinfo {year} {2007})}\BibitemShut {NoStop}%
\bibitem [{\citenamefont {Borisenko}\ \emph {et~al.}(2015)\citenamefont
  {Borisenko}, \citenamefont {Evtushinsky}, \citenamefont {Liu}, \citenamefont
  {Morozov}, \citenamefont {Kappenberger}, \citenamefont {Wurmehl},
  \citenamefont {B{\"{u}}chner}, \citenamefont {Yaresko}, \citenamefont {Kim},
  \citenamefont {Hoesch}, \citenamefont {Wolf},\ and\ \citenamefont
  {Zhigadlo}}]{Borisenko2015}%
  \BibitemOpen
  \bibfield  {author} {\bibinfo {author} {\bibfnamefont {S.~V.}\ \bibnamefont
  {Borisenko}}, \bibinfo {author} {\bibfnamefont {D.~V.}\ \bibnamefont
  {Evtushinsky}}, \bibinfo {author} {\bibfnamefont {Z.-H.}\ \bibnamefont
  {Liu}}, \bibinfo {author} {\bibfnamefont {I.}~\bibnamefont {Morozov}},
  \bibinfo {author} {\bibfnamefont {R.}~\bibnamefont {Kappenberger}}, \bibinfo
  {author} {\bibfnamefont {S.}~\bibnamefont {Wurmehl}}, \bibinfo {author}
  {\bibfnamefont {B.}~\bibnamefont {B{\"{u}}chner}}, \bibinfo {author}
  {\bibfnamefont {A.~N.}\ \bibnamefont {Yaresko}}, \bibinfo {author}
  {\bibfnamefont {T.~K.}\ \bibnamefont {Kim}}, \bibinfo {author} {\bibfnamefont
  {M.}~\bibnamefont {Hoesch}}, \bibinfo {author} {\bibfnamefont
  {T.}~\bibnamefont {Wolf}}, \ and\ \bibinfo {author} {\bibfnamefont {N.~D.}\
  \bibnamefont {Zhigadlo}},\ }\bibfield  {title} {\enquote {\bibinfo {title}
  {{Direct observation of spin--orbit coupling in iron-based
  superconductors}},}\ }\href {\doibase 10.1038/nphys3594} {\bibfield
  {journal} {\bibinfo  {journal} {Nat. Phys.}\ }\textbf {\bibinfo {volume}
  {12}},\ \bibinfo {pages} {311--317} (\bibinfo {year} {2015})}\BibitemShut
  {NoStop}%
\bibitem [{Note4()}]{Note4}%
  \BibitemOpen
  \bibinfo {note} {Ref.~\cite {Rozbicki2011b} noted that an enhance effective
  SOC also improves the description of de Haas van Alphen data.}\BibitemShut
  {Stop}%
\bibitem [{Note5()}]{Note5}%
  \BibitemOpen
  \bibinfo {note} {Using the free electron final state approximation, we obtain
  $k_z = (2m_e\hbar ^{-2}(h\nu +U-\Phi ) - k_{\parallel }^2)^{1/2}\approx
  0.4(0.12) \pi /c$ assuming an inner potential relative to $E_F$ of $U-\Phi
  =8.5(1.0)$~eV.}\BibitemShut {Stop}%
\bibitem [{\citenamefont {Hengsberger}\ \emph {et~al.}(1999)\citenamefont
  {Hengsberger}, \citenamefont {Purdie}, \citenamefont {Segovia}, \citenamefont
  {Garnier},\ and\ \citenamefont {Baer}}]{Hengsberger1999}%
  \BibitemOpen
  \bibfield  {author} {\bibinfo {author} {\bibfnamefont {M.}~\bibnamefont
  {Hengsberger}}, \bibinfo {author} {\bibfnamefont {D.}~\bibnamefont {Purdie}},
  \bibinfo {author} {\bibfnamefont {P.}~\bibnamefont {Segovia}}, \bibinfo
  {author} {\bibfnamefont {M.}~\bibnamefont {Garnier}}, \ and\ \bibinfo
  {author} {\bibfnamefont {Y.}~\bibnamefont {Baer}},\ }\bibfield  {title}
  {\enquote {\bibinfo {title} {Photoemission study of a strongly coupled
  electron-phonon system},}\ }\href {\doibase 10.1103/PhysRevLett.83.592}
  {\bibfield  {journal} {\bibinfo  {journal} {Phys. Rev. Lett.}\ }\textbf
  {\bibinfo {volume} {83}},\ \bibinfo {pages} {592--595} (\bibinfo {year}
  {1999})}\BibitemShut {NoStop}%
\bibitem [{\citenamefont {Lanzara}\ \emph {et~al.}(2001)\citenamefont
  {Lanzara}, \citenamefont {Bogdanov}, \citenamefont {Zhou}, \citenamefont
  {Kellar}, \citenamefont {Feng}, \citenamefont {Lu}, \citenamefont {Yoshida},
  \citenamefont {Eisaki}, \citenamefont {Fujimori}, \citenamefont {Kishio},
  \citenamefont {Shimoyama}, \citenamefont {Noda}, \citenamefont {Uchida},
  \citenamefont {Hussain},\ and\ \citenamefont {Shen}}]{Lanzara2001}%
  \BibitemOpen
  \bibfield  {author} {\bibinfo {author} {\bibfnamefont {A.}~\bibnamefont
  {Lanzara}}, \bibinfo {author} {\bibfnamefont {P.~V.}\ \bibnamefont
  {Bogdanov}}, \bibinfo {author} {\bibfnamefont {X.~J.}\ \bibnamefont {Zhou}},
  \bibinfo {author} {\bibfnamefont {S.~A.}\ \bibnamefont {Kellar}}, \bibinfo
  {author} {\bibfnamefont {D.~L.}\ \bibnamefont {Feng}}, \bibinfo {author}
  {\bibfnamefont {E.~D.}\ \bibnamefont {Lu}}, \bibinfo {author} {\bibfnamefont
  {T.}~\bibnamefont {Yoshida}}, \bibinfo {author} {\bibfnamefont
  {H.}~\bibnamefont {Eisaki}}, \bibinfo {author} {\bibfnamefont
  {A.}~\bibnamefont {Fujimori}}, \bibinfo {author} {\bibfnamefont
  {K.}~\bibnamefont {Kishio}}, \bibinfo {author} {\bibfnamefont {J.~I.}\
  \bibnamefont {Shimoyama}}, \bibinfo {author} {\bibfnamefont {T.}~\bibnamefont
  {Noda}}, \bibinfo {author} {\bibfnamefont {S.}~\bibnamefont {Uchida}},
  \bibinfo {author} {\bibfnamefont {Z.}~\bibnamefont {Hussain}}, \ and\
  \bibinfo {author} {\bibfnamefont {Z.~X.}\ \bibnamefont {Shen}},\ }\bibfield
  {title} {\enquote {\bibinfo {title} {{Evidence for ubiquitous strong
  electron-phonon coupling in high-temperature superconductors}},}\ }\href@noop
  {} {\bibfield  {journal} {\bibinfo  {journal} {Nature}\ }\textbf {\bibinfo
  {volume} {412}},\ \bibinfo {pages} {510} (\bibinfo {year}
  {2001})}\BibitemShut {NoStop}%
\bibitem [{\citenamefont {Tamai}\ \emph {et~al.}(2013)\citenamefont {Tamai},
  \citenamefont {Meevasana}, \citenamefont {King}, \citenamefont {Nicholson},
  \citenamefont {de~la Torre}, \citenamefont {Rozbicki},\ and\ \citenamefont
  {Baumberger}}]{Tamai2013}%
  \BibitemOpen
  \bibfield  {author} {\bibinfo {author} {\bibfnamefont {A.}~\bibnamefont
  {Tamai}}, \bibinfo {author} {\bibfnamefont {W.}~\bibnamefont {Meevasana}},
  \bibinfo {author} {\bibfnamefont {P.~D.~C.}\ \bibnamefont {King}}, \bibinfo
  {author} {\bibfnamefont {C.~W.}\ \bibnamefont {Nicholson}}, \bibinfo {author}
  {\bibfnamefont {A.}~\bibnamefont {de~la Torre}}, \bibinfo {author}
  {\bibfnamefont {E.}~\bibnamefont {Rozbicki}}, \ and\ \bibinfo {author}
  {\bibfnamefont {F.}~\bibnamefont {Baumberger}},\ }\bibfield  {title}
  {\enquote {\bibinfo {title} {{Spin-orbit splitting of the Shockley surface
  state on Cu(111)}},}\ }\href {\doibase 10.1103/PhysRevB.87.075113} {\bibfield
   {journal} {\bibinfo  {journal} {Phys. Rev. B}\ }\textbf {\bibinfo {volume}
  {87}},\ \bibinfo {pages} {075113} (\bibinfo {year} {2013})}\BibitemShut
  {NoStop}%
\bibitem [{\citenamefont {Byczuk}\ \emph {et~al.}(2007)\citenamefont {Byczuk},
  \citenamefont {Kollar}, \citenamefont {Held}, \citenamefont {Yang},
  \citenamefont {Nekrasov}, \citenamefont {Pruschke},\ and\ \citenamefont
  {Vollhardt}}]{Byczuk2007}%
  \BibitemOpen
  \bibfield  {author} {\bibinfo {author} {\bibfnamefont {K.}~\bibnamefont
  {Byczuk}}, \bibinfo {author} {\bibfnamefont {M.}~\bibnamefont {Kollar}},
  \bibinfo {author} {\bibfnamefont {K.}~\bibnamefont {Held}}, \bibinfo {author}
  {\bibfnamefont {Y.~F.}\ \bibnamefont {Yang}}, \bibinfo {author}
  {\bibfnamefont {I.~A.}\ \bibnamefont {Nekrasov}}, \bibinfo {author}
  {\bibfnamefont {Th.}\ \bibnamefont {Pruschke}}, \ and\ \bibinfo {author}
  {\bibfnamefont {D.}~\bibnamefont {Vollhardt}},\ }\bibfield  {title} {\enquote
  {\bibinfo {title} {{Kinks in the dispersion of strongly correlated
  electrons}},}\ }\href {http://dx.doi.org/10.1038/nphys538
  http://www.nature.com/nphys/journal/v3/n3/suppinfo/nphys538{\_}S1.html}
  {\bibfield  {journal} {\bibinfo  {journal} {Nat Phys}\ }\textbf {\bibinfo
  {volume} {3}},\ \bibinfo {pages} {168--171} (\bibinfo {year}
  {2007})}\BibitemShut {NoStop}%
\bibitem [{\citenamefont {Raas}\ \emph {et~al.}(2009)\citenamefont {Raas},
  \citenamefont {Grete},\ and\ \citenamefont {Uhrig}}]{Raas2009}%
  \BibitemOpen
  \bibfield  {author} {\bibinfo {author} {\bibfnamefont {Carsten}\ \bibnamefont
  {Raas}}, \bibinfo {author} {\bibfnamefont {Patrick}\ \bibnamefont {Grete}}, \
  and\ \bibinfo {author} {\bibfnamefont {G\"otz~S.}\ \bibnamefont {Uhrig}},\
  }\bibfield  {title} {\enquote {\bibinfo {title} {Emergent collective modes
  and kinks in electronic dispersions},}\ }\href {\doibase
  10.1103/PhysRevLett.102.076406} {\bibfield  {journal} {\bibinfo  {journal}
  {Phys. Rev. Lett.}\ }\textbf {\bibinfo {volume} {102}},\ \bibinfo {pages}
  {076406} (\bibinfo {year} {2009})}\BibitemShut {NoStop}%
\bibitem [{\citenamefont {Held}\ \emph {et~al.}(2013)\citenamefont {Held},
  \citenamefont {Peters},\ and\ \citenamefont {Toschi}}]{Held2013}%
  \BibitemOpen
  \bibfield  {author} {\bibinfo {author} {\bibfnamefont {K.}~\bibnamefont
  {Held}}, \bibinfo {author} {\bibfnamefont {R.}~\bibnamefont {Peters}}, \ and\
  \bibinfo {author} {\bibfnamefont {A.}~\bibnamefont {Toschi}},\ }\bibfield
  {title} {\enquote {\bibinfo {title} {Poor man's understanding of kinks
  originating from strong electronic correlations},}\ }\href {\doibase
  10.1103/PhysRevLett.110.246402} {\bibfield  {journal} {\bibinfo  {journal}
  {Phys. Rev. Lett.}\ }\textbf {\bibinfo {volume} {110}},\ \bibinfo {pages}
  {246402} (\bibinfo {year} {2013})}\BibitemShut {NoStop}%
\bibitem [{\citenamefont {Deng}\ \emph {et~al.}(2013)\citenamefont {Deng},
  \citenamefont {Mravlje}, \citenamefont {\ifmmode~\check{Z}\else
  \v{Z}\fi{}itko}, \citenamefont {Ferrero}, \citenamefont {Kotliar},\ and\
  \citenamefont {Georges}}]{Deng2013}%
  \BibitemOpen
  \bibfield  {author} {\bibinfo {author} {\bibfnamefont {Xiaoyu}\ \bibnamefont
  {Deng}}, \bibinfo {author} {\bibfnamefont {Jernej}\ \bibnamefont {Mravlje}},
  \bibinfo {author} {\bibfnamefont {Rok}\ \bibnamefont {\ifmmode~\check{Z}\else
  \v{Z}\fi{}itko}}, \bibinfo {author} {\bibfnamefont {Michel}\ \bibnamefont
  {Ferrero}}, \bibinfo {author} {\bibfnamefont {Gabriel}\ \bibnamefont
  {Kotliar}}, \ and\ \bibinfo {author} {\bibfnamefont {Antoine}\ \bibnamefont
  {Georges}},\ }\bibfield  {title} {\enquote {\bibinfo {title} {How bad metals
  turn good: Spectroscopic signatures of resilient quasiparticles},}\ }\href
  {\doibase 10.1103/PhysRevLett.110.086401} {\bibfield  {journal} {\bibinfo
  {journal} {Phys. Rev. Lett.}\ }\textbf {\bibinfo {volume} {110}},\ \bibinfo
  {pages} {086401} (\bibinfo {year} {2013})}\BibitemShut {NoStop}%
\bibitem [{\citenamefont {\ifmmode~\check{Z}\else \v{Z}\fi{}itko}\ \emph
  {et~al.}(2013)\citenamefont {\ifmmode~\check{Z}\else \v{Z}\fi{}itko},
  \citenamefont {Hansen}, \citenamefont {Perepelitsky}, \citenamefont
  {Mravlje}, \citenamefont {Georges},\ and\ \citenamefont
  {Shastry}}]{Zitko2013}%
  \BibitemOpen
  \bibfield  {author} {\bibinfo {author} {\bibfnamefont {R.}~\bibnamefont
  {\ifmmode~\check{Z}\else \v{Z}\fi{}itko}}, \bibinfo {author} {\bibfnamefont
  {D.}~\bibnamefont {Hansen}}, \bibinfo {author} {\bibfnamefont
  {E.}~\bibnamefont {Perepelitsky}}, \bibinfo {author} {\bibfnamefont
  {J.}~\bibnamefont {Mravlje}}, \bibinfo {author} {\bibfnamefont
  {A.}~\bibnamefont {Georges}}, \ and\ \bibinfo {author} {\bibfnamefont
  {B.~S.}\ \bibnamefont {Shastry}},\ }\bibfield  {title} {\enquote {\bibinfo
  {title} {Extremely correlated fermi liquid theory meets dynamical mean-field
  theory: Analytical insights into the doping-driven mott transition},}\ }\href
  {\doibase 10.1103/PhysRevB.88.235132} {\bibfield  {journal} {\bibinfo
  {journal} {Phys. Rev. B}\ }\textbf {\bibinfo {volume} {88}},\ \bibinfo
  {pages} {235132} (\bibinfo {year} {2013})}\BibitemShut {NoStop}%
\bibitem [{\citenamefont {Braden}\ \emph {et~al.}(2007)\citenamefont {Braden},
  \citenamefont {Reichardt}, \citenamefont {Sidis}, \citenamefont {Mao},\ and\
  \citenamefont {Maeno}}]{Braden2007}%
  \BibitemOpen
  \bibfield  {author} {\bibinfo {author} {\bibfnamefont {M.}~\bibnamefont
  {Braden}}, \bibinfo {author} {\bibfnamefont {W.}~\bibnamefont {Reichardt}},
  \bibinfo {author} {\bibfnamefont {Y.}~\bibnamefont {Sidis}}, \bibinfo
  {author} {\bibfnamefont {Z.}~\bibnamefont {Mao}}, \ and\ \bibinfo {author}
  {\bibfnamefont {Y.}~\bibnamefont {Maeno}},\ }\bibfield  {title} {\enquote
  {\bibinfo {title} {{Lattice dynamics and electron-phonon coupling in
  Sr$_2$RuO$_4$: Inelastic neutron scattering and shell-model calculations}},}\
  }\href {\doibase 10.1103/PhysRevB.76.014505} {\bibfield  {journal} {\bibinfo
  {journal} {Phys. Rev. B}\ }\textbf {\bibinfo {volume} {76}},\ \bibinfo
  {pages} {014505} (\bibinfo {year} {2007})}\BibitemShut {NoStop}%
\bibitem [{\citenamefont {Kugel}\ and\ \citenamefont
  {Khomskii}(1982)}]{kugel_khomskii_1982}%
  \BibitemOpen
  \bibfield  {author} {\bibinfo {author} {\bibfnamefont {K.I.}\ \bibnamefont
  {Kugel}}\ and\ \bibinfo {author} {\bibfnamefont {D.I.}\ \bibnamefont
  {Khomskii}},\ }\bibfield  {title} {\enquote {\bibinfo {title} {{The
  Jahn-Teller effect and magnetism: transition metal compounds}},}\ }\href
  {https://iopscience.iop.org/article/10.1070/PU1982v025n04ABEH004537/meta}
  {\bibfield  {journal} {\bibinfo  {journal} {Sov. Phys. Usp.}\ }\textbf
  {\bibinfo {volume} {25}},\ \bibinfo {pages} {231} (\bibinfo {year}
  {1982})}\BibitemShut {NoStop}%
\bibitem [{\citenamefont {Tokura}\ and\ \citenamefont
  {Nagaosa}(2000)}]{tokura_nagaosa_orbital_2000}%
  \BibitemOpen
  \bibfield  {author} {\bibinfo {author} {\bibfnamefont {Y.}~\bibnamefont
  {Tokura}}\ and\ \bibinfo {author} {\bibfnamefont {N.}~\bibnamefont
  {Nagaosa}},\ }\bibfield  {title} {\enquote {\bibinfo {title} {Orbital physics
  in transition-metal oxides},}\ }\href {\doibase 10.1126/science.288.5465.462}
  {\bibfield  {journal} {\bibinfo  {journal} {Science}\ }\textbf {\bibinfo
  {volume} {288}},\ \bibinfo {pages} {462--468} (\bibinfo {year}
  {2000})}\BibitemShut {NoStop}%
\bibitem [{\citenamefont {Miao}\ \emph {et~al.}(2016)\citenamefont {Miao},
  \citenamefont {Yin}, \citenamefont {Wu}, \citenamefont {Li}, \citenamefont
  {Ma}, \citenamefont {Lv}, \citenamefont {Wang}, \citenamefont {Qian},
  \citenamefont {Richard}, \citenamefont {Xing}, \citenamefont {Wang},
  \citenamefont {Jin}, \citenamefont {Haule}, \citenamefont {Kotliar},\ and\
  \citenamefont {Ding}}]{Miao2016}%
  \BibitemOpen
  \bibfield  {author} {\bibinfo {author} {\bibfnamefont {H.}~\bibnamefont
  {Miao}}, \bibinfo {author} {\bibfnamefont {Z.~P.}\ \bibnamefont {Yin}},
  \bibinfo {author} {\bibfnamefont {S.~F.}\ \bibnamefont {Wu}}, \bibinfo
  {author} {\bibfnamefont {J.~M.}\ \bibnamefont {Li}}, \bibinfo {author}
  {\bibfnamefont {J.}~\bibnamefont {Ma}}, \bibinfo {author} {\bibfnamefont
  {B.-Q.}\ \bibnamefont {Lv}}, \bibinfo {author} {\bibfnamefont {X.~P.}\
  \bibnamefont {Wang}}, \bibinfo {author} {\bibfnamefont {T.}~\bibnamefont
  {Qian}}, \bibinfo {author} {\bibfnamefont {P.}~\bibnamefont {Richard}},
  \bibinfo {author} {\bibfnamefont {L.-Y.}\ \bibnamefont {Xing}}, \bibinfo
  {author} {\bibfnamefont {X.-C.}\ \bibnamefont {Wang}}, \bibinfo {author}
  {\bibfnamefont {C.~Q.}\ \bibnamefont {Jin}}, \bibinfo {author} {\bibfnamefont
  {K.}~\bibnamefont {Haule}}, \bibinfo {author} {\bibfnamefont
  {G.}~\bibnamefont {Kotliar}}, \ and\ \bibinfo {author} {\bibfnamefont
  {H.}~\bibnamefont {Ding}},\ }\bibfield  {title} {\enquote {\bibinfo {title}
  {Orbital-differentiated coherence-incoherence crossover identified by
  photoemission spectroscopy in lifeas},}\ }\href {\doibase
  10.1103/PhysRevB.94.201109} {\bibfield  {journal} {\bibinfo  {journal} {Phys.
  Rev. B}\ }\textbf {\bibinfo {volume} {94}},\ \bibinfo {pages} {201109}
  (\bibinfo {year} {2016})}\BibitemShut {NoStop}%
\bibitem [{\citenamefont {Sprau}\ \emph {et~al.}(2017)\citenamefont {Sprau},
  \citenamefont {Kostin}, \citenamefont {Kreisel}, \citenamefont {B{\"o}hmer},
  \citenamefont {Taufour}, \citenamefont {Canfield}, \citenamefont {Mukherjee},
  \citenamefont {Hirschfeld}, \citenamefont {Andersen},\ and\ \citenamefont
  {Davis}}]{Sprau75}%
  \BibitemOpen
  \bibfield  {author} {\bibinfo {author} {\bibfnamefont {P.~O.}\ \bibnamefont
  {Sprau}}, \bibinfo {author} {\bibfnamefont {A.}~\bibnamefont {Kostin}},
  \bibinfo {author} {\bibfnamefont {A.}~\bibnamefont {Kreisel}}, \bibinfo
  {author} {\bibfnamefont {A.~E.}\ \bibnamefont {B{\"o}hmer}}, \bibinfo
  {author} {\bibfnamefont {V.}~\bibnamefont {Taufour}}, \bibinfo {author}
  {\bibfnamefont {P.~C.}\ \bibnamefont {Canfield}}, \bibinfo {author}
  {\bibfnamefont {S.}~\bibnamefont {Mukherjee}}, \bibinfo {author}
  {\bibfnamefont {P.~J.}\ \bibnamefont {Hirschfeld}}, \bibinfo {author}
  {\bibfnamefont {B.~M.}\ \bibnamefont {Andersen}}, \ and\ \bibinfo {author}
  {\bibfnamefont {J.~C.~S{\'e}amus}\ \bibnamefont {Davis}},\ }\bibfield
  {title} {\enquote {\bibinfo {title} {{Discovery of orbital-selective Cooper
  pairing in FeSe}},}\ }\href {\doibase 10.1126/science.aal1575} {\bibfield
  {journal} {\bibinfo  {journal} {Science}\ }\textbf {\bibinfo {volume}
  {357}},\ \bibinfo {pages} {75--80} (\bibinfo {year} {2017})}\BibitemShut
  {NoStop}%
\bibitem [{\citenamefont {Braden}\ \emph {et~al.}(2002)\citenamefont {Braden},
  \citenamefont {Friedt}, \citenamefont {Sidis}, \citenamefont {Bourges},
  \citenamefont {Minakata},\ and\ \citenamefont
  {Maeno}}]{braden_impurities_2002}%
  \BibitemOpen
  \bibfield  {author} {\bibinfo {author} {\bibfnamefont {M.}~\bibnamefont
  {Braden}}, \bibinfo {author} {\bibfnamefont {O.}~\bibnamefont {Friedt}},
  \bibinfo {author} {\bibfnamefont {Y.}~\bibnamefont {Sidis}}, \bibinfo
  {author} {\bibfnamefont {P.}~\bibnamefont {Bourges}}, \bibinfo {author}
  {\bibfnamefont {M.}~\bibnamefont {Minakata}}, \ and\ \bibinfo {author}
  {\bibfnamefont {Y.}~\bibnamefont {Maeno}},\ }\bibfield  {title} {\enquote
  {\bibinfo {title} {{Incommensurate Magnetic Ordering in
  Sr$_2$Ru$_{1-x}$Ti$_x$O$_4$}},}\ }\href {\doibase
  10.1103/PhysRevLett.88.197002} {\bibfield  {journal} {\bibinfo  {journal}
  {Phys. Rev. Lett.}\ }\textbf {\bibinfo {volume} {88}},\ \bibinfo {pages}
  {197002} (\bibinfo {year} {2002})}\BibitemShut {NoStop}%
\bibitem [{\citenamefont {{Ortmann}}\ \emph {et~al.}(2013)\citenamefont
  {{Ortmann}}, \citenamefont {{Liu}}, \citenamefont {{Hu}}, \citenamefont
  {{Zhu}}, \citenamefont {{Peng}}, \citenamefont {{Matsuda}}, \citenamefont
  {{Ke}},\ and\ \citenamefont {{Mao}}}]{ortmann_impurities_2013}%
  \BibitemOpen
  \bibfield  {author} {\bibinfo {author} {\bibfnamefont {J.~E.}\ \bibnamefont
  {{Ortmann}}}, \bibinfo {author} {\bibfnamefont {J.~Y.}\ \bibnamefont
  {{Liu}}}, \bibinfo {author} {\bibfnamefont {J.}~\bibnamefont {{Hu}}},
  \bibinfo {author} {\bibfnamefont {M.}~\bibnamefont {{Zhu}}}, \bibinfo
  {author} {\bibfnamefont {J.}~\bibnamefont {{Peng}}}, \bibinfo {author}
  {\bibfnamefont {M.}~\bibnamefont {{Matsuda}}}, \bibinfo {author}
  {\bibfnamefont {X.}~\bibnamefont {{Ke}}}, \ and\ \bibinfo {author}
  {\bibfnamefont {Z.~Q.}\ \bibnamefont {{Mao}}},\ }\bibfield  {title} {\enquote
  {\bibinfo {title} {{Competition Between Antiferromagnetism and Ferromagnetism
  in Sr$_{2}$RuO$_{4}$ Probed by Mn and Co Doping}},}\ }\href {\doibase
  10.1038/srep02950} {\bibfield  {journal} {\bibinfo  {journal} {Scientific
  Reports}\ }\textbf {\bibinfo {volume} {3}},\ \bibinfo {eid} {2950} (\bibinfo
  {year} {2013})}\BibitemShut {NoStop}%
\bibitem [{\citenamefont {Wang}\ \emph {et~al.}(2004)\citenamefont {Wang},
  \citenamefont {Yang}, \citenamefont {Sekharan}, \citenamefont {Ding},
  \citenamefont {Engelbrecht}, \citenamefont {Dai}, \citenamefont {Wang},
  \citenamefont {Kaminski}, \citenamefont {Valla}, \citenamefont {Kidd},
  \citenamefont {Fedorov},\ and\ \citenamefont {Johnson}}]{Wang2004}%
  \BibitemOpen
  \bibfield  {author} {\bibinfo {author} {\bibfnamefont {S.~C.}\ \bibnamefont
  {Wang}}, \bibinfo {author} {\bibfnamefont {H.~B.}\ \bibnamefont {Yang}},
  \bibinfo {author} {\bibfnamefont {A.~K.~P.}\ \bibnamefont {Sekharan}},
  \bibinfo {author} {\bibfnamefont {H.}~\bibnamefont {Ding}}, \bibinfo {author}
  {\bibfnamefont {J.~R.}\ \bibnamefont {Engelbrecht}}, \bibinfo {author}
  {\bibfnamefont {X.}~\bibnamefont {Dai}}, \bibinfo {author} {\bibfnamefont
  {Z.}~\bibnamefont {Wang}}, \bibinfo {author} {\bibfnamefont {A.}~\bibnamefont
  {Kaminski}}, \bibinfo {author} {\bibfnamefont {T.}~\bibnamefont {Valla}},
  \bibinfo {author} {\bibfnamefont {T.}~\bibnamefont {Kidd}}, \bibinfo {author}
  {\bibfnamefont {A.~V.}\ \bibnamefont {Fedorov}}, \ and\ \bibinfo {author}
  {\bibfnamefont {P.~D.}\ \bibnamefont {Johnson}},\ }\bibfield  {title}
  {\enquote {\bibinfo {title} {{Quasiparticle Line Shape of Sr$_2$RuO$_4$ and
  Its Relation to Anisotropic Transport}},}\ }\href
  {http://link.aps.org/abstract/PRL/v92/e137002} {\bibfield  {journal}
  {\bibinfo  {journal} {Phys. Rev. Lett.}\ }\textbf {\bibinfo {volume} {92}},\
  \bibinfo {pages} {137002} (\bibinfo {year} {2004})}\BibitemShut {NoStop}%
\bibitem [{\citenamefont {Kidd}\ \emph {et~al.}(2005)\citenamefont {Kidd},
  \citenamefont {Valla}, \citenamefont {Fedorov}, \citenamefont {Johnson},
  \citenamefont {Cava},\ and\ \citenamefont {Haas}}]{Kidd2005}%
  \BibitemOpen
  \bibfield  {author} {\bibinfo {author} {\bibfnamefont {T.~E.}\ \bibnamefont
  {Kidd}}, \bibinfo {author} {\bibfnamefont {T.}~\bibnamefont {Valla}},
  \bibinfo {author} {\bibfnamefont {A.~V.}\ \bibnamefont {Fedorov}}, \bibinfo
  {author} {\bibfnamefont {P.~D.}\ \bibnamefont {Johnson}}, \bibinfo {author}
  {\bibfnamefont {R.~J.}\ \bibnamefont {Cava}}, \ and\ \bibinfo {author}
  {\bibfnamefont {M.~K.}\ \bibnamefont {Haas}},\ }\bibfield  {title} {\enquote
  {\bibinfo {title} {{Orbital Dependence of the Fermi Liquid State in
  Sr$_2$RuO$_4$}},}\ }\href {http://link.aps.org/abstract/PRL/v94/e107003}
  {\bibfield  {journal} {\bibinfo  {journal} {Phys. Rev. Lett.}\ }\textbf
  {\bibinfo {volume} {94}},\ \bibinfo {pages} {107003} (\bibinfo {year}
  {2005})}\BibitemShut {NoStop}%
\bibitem [{\citenamefont {Kondo}\ \emph {et~al.}(2016)\citenamefont {Kondo},
  \citenamefont {Ochi}, \citenamefont {Nakayama}, \citenamefont {Taniguchi},
  \citenamefont {Akebi}, \citenamefont {Kuroda}, \citenamefont {Arita},
  \citenamefont {Sakai}, \citenamefont {Namatame}, \citenamefont {Taniguchi},
  \citenamefont {Maeno}, \citenamefont {Arita},\ and\ \citenamefont
  {Shin}}]{Kondo2016}%
  \BibitemOpen
  \bibfield  {author} {\bibinfo {author} {\bibfnamefont {Takeshi}\ \bibnamefont
  {Kondo}}, \bibinfo {author} {\bibfnamefont {M.}~\bibnamefont {Ochi}},
  \bibinfo {author} {\bibfnamefont {M.}~\bibnamefont {Nakayama}}, \bibinfo
  {author} {\bibfnamefont {H.}~\bibnamefont {Taniguchi}}, \bibinfo {author}
  {\bibfnamefont {S.}~\bibnamefont {Akebi}}, \bibinfo {author} {\bibfnamefont
  {K.}~\bibnamefont {Kuroda}}, \bibinfo {author} {\bibfnamefont
  {M.}~\bibnamefont {Arita}}, \bibinfo {author} {\bibfnamefont
  {S.}~\bibnamefont {Sakai}}, \bibinfo {author} {\bibfnamefont
  {H.}~\bibnamefont {Namatame}}, \bibinfo {author} {\bibfnamefont
  {M.}~\bibnamefont {Taniguchi}}, \bibinfo {author} {\bibfnamefont
  {Y.}~\bibnamefont {Maeno}}, \bibinfo {author} {\bibfnamefont
  {R.}~\bibnamefont {Arita}}, \ and\ \bibinfo {author} {\bibfnamefont
  {S.}~\bibnamefont {Shin}},\ }\bibfield  {title} {\enquote {\bibinfo {title}
  {{Orbital-Dependent Band Narrowing Revealed in an Extremely Correlated Hund's
  Metal Emerging on the Topmost Layer of Sr$_2$RuO$_4$}},}\ }\href {\doibase
  10.1103/PhysRevLett.117.247001} {\bibfield  {journal} {\bibinfo  {journal}
  {Phys. Rev. Lett.}\ }\textbf {\bibinfo {volume} {117}},\ \bibinfo {pages}
  {247001} (\bibinfo {year} {2016})}\BibitemShut {NoStop}%
\bibitem [{\citenamefont {Blaha}\ \emph {et~al.}(2001)\citenamefont {Blaha},
  \citenamefont {Schwarz}, \citenamefont {Madsen}, \citenamefont {Kvasnicka},\
  and\ \citenamefont {Luitz}}]{Blaha2001}%
  \BibitemOpen
  \bibfield  {author} {\bibinfo {author} {\bibfnamefont {P}~\bibnamefont
  {Blaha}}, \bibinfo {author} {\bibfnamefont {K}~\bibnamefont {Schwarz}},
  \bibinfo {author} {\bibfnamefont {G.}~\bibnamefont {Madsen}}, \bibinfo
  {author} {\bibfnamefont {D}~\bibnamefont {Kvasnicka}}, \ and\ \bibinfo
  {author} {\bibfnamefont {J}~\bibnamefont {Luitz}},\ }\href@noop {} {\emph
  {\bibinfo {title} {WIEN2k, An Augmented Plane Wave + Local Orbitals Program
  for Calculating Crystal Properties}}}\ (\bibinfo  {publisher} {K. Schwarz,
  Tech. Univ. Wien, Austria},\ \bibinfo {year} {2001})\BibitemShut {NoStop}%
\bibitem [{\citenamefont {Perdew}\ \emph {et~al.}(1996)\citenamefont {Perdew},
  \citenamefont {Burke},\ and\ \citenamefont {Ernzerhof}}]{PBE}%
  \BibitemOpen
  \bibfield  {author} {\bibinfo {author} {\bibfnamefont {John~P.}\ \bibnamefont
  {Perdew}}, \bibinfo {author} {\bibfnamefont {Kieron}\ \bibnamefont {Burke}},
  \ and\ \bibinfo {author} {\bibfnamefont {Matthias}\ \bibnamefont
  {Ernzerhof}},\ }\bibfield  {title} {\enquote {\bibinfo {title} {Generalized
  gradient approximation made simple},}\ }\href {\doibase
  10.1103/PhysRevLett.77.3865} {\bibfield  {journal} {\bibinfo  {journal}
  {Phys. Rev. Lett.}\ }\textbf {\bibinfo {volume} {77}},\ \bibinfo {pages}
  {3865--3868} (\bibinfo {year} {1996})}\BibitemShut {NoStop}%
\bibitem [{\citenamefont {Kune{\v s}}\ \emph {et~al.}(2010)\citenamefont
  {Kune{\v s}}, \citenamefont {Arita}, \citenamefont {Wissgott}, \citenamefont
  {Toschi}, \citenamefont {Ikeda},\ and\ \citenamefont {Held}}]{wien2wannier}%
  \BibitemOpen
  \bibfield  {author} {\bibinfo {author} {\bibfnamefont {Jan}\ \bibnamefont
  {Kune{\v s}}}, \bibinfo {author} {\bibfnamefont {Ryotaro}\ \bibnamefont
  {Arita}}, \bibinfo {author} {\bibfnamefont {Philipp}\ \bibnamefont
  {Wissgott}}, \bibinfo {author} {\bibfnamefont {Alessandro}\ \bibnamefont
  {Toschi}}, \bibinfo {author} {\bibfnamefont {Hiroaki}\ \bibnamefont {Ikeda}},
  \ and\ \bibinfo {author} {\bibfnamefont {Karsten}\ \bibnamefont {Held}},\
  }\bibfield  {title} {\enquote {\bibinfo {title} {{Wien2wannier: From
  linearized augmented plane waves to maximally localized Wannier
  functions}},}\ }\href {\doibase http://dx.doi.org/10.1016/j.cpc.2010.08.005}
  {\bibfield  {journal} {\bibinfo  {journal} {Comput. Phys. Commun.}\ }\textbf
  {\bibinfo {volume} {181}},\ \bibinfo {pages} {1888 -- 1895} (\bibinfo {year}
  {2010})}\BibitemShut {NoStop}%
\bibitem [{\citenamefont {Mostofi}\ \emph {et~al.}(2008)\citenamefont
  {Mostofi}, \citenamefont {Yates}, \citenamefont {Lee}, \citenamefont {Souza},
  \citenamefont {Vanderbilt},\ and\ \citenamefont {Marzari}}]{wannier90}%
  \BibitemOpen
  \bibfield  {author} {\bibinfo {author} {\bibfnamefont {Arash~A.}\
  \bibnamefont {Mostofi}}, \bibinfo {author} {\bibfnamefont {Jonathan~R.}\
  \bibnamefont {Yates}}, \bibinfo {author} {\bibfnamefont {Young-Su}\
  \bibnamefont {Lee}}, \bibinfo {author} {\bibfnamefont {Ivo}\ \bibnamefont
  {Souza}}, \bibinfo {author} {\bibfnamefont {David}\ \bibnamefont
  {Vanderbilt}}, \ and\ \bibinfo {author} {\bibfnamefont {Nicola}\ \bibnamefont
  {Marzari}},\ }\bibfield  {title} {\enquote {\bibinfo {title} {{wannier90: A
  tool for obtaining maximally-localised Wannier functions}},}\ }\href
  {\doibase http://dx.doi.org/10.1016/j.cpc.2007.11.016} {\bibfield  {journal}
  {\bibinfo  {journal} {Comput. Phys. Commun.}\ }\textbf {\bibinfo {volume}
  {178}},\ \bibinfo {pages} {685} (\bibinfo {year} {2008})}\BibitemShut
  {NoStop}%
\bibitem [{\citenamefont {Vogt}\ and\ \citenamefont
  {Buttrey}(1995)}]{Vogt1995}%
  \BibitemOpen
  \bibfield  {author} {\bibinfo {author} {\bibfnamefont {T.}~\bibnamefont
  {Vogt}}\ and\ \bibinfo {author} {\bibfnamefont {D.~J.}\ \bibnamefont
  {Buttrey}},\ }\bibfield  {title} {\enquote {\bibinfo {title}
  {{Low-temperature structural behavior of Sr$_2$RuO$_4$}},}\ }\href {\doibase
  10.1103/PhysRevB.52.R9843} {\bibfield  {journal} {\bibinfo  {journal} {Phys.
  Rev. B}\ }\textbf {\bibinfo {volume} {52}},\ \bibinfo {pages} {R9843}
  (\bibinfo {year} {1995})}\BibitemShut {NoStop}%
\bibitem [{\citenamefont {Parcollet}\ \emph {et~al.}(2015)\citenamefont
  {Parcollet}, \citenamefont {Ferrero}, \citenamefont {Ayral}, \citenamefont
  {Hafermann}, \citenamefont {Krivenko}, \citenamefont {Messio},\ and\
  \citenamefont {Seth}}]{TRIQS}%
  \BibitemOpen
  \bibfield  {author} {\bibinfo {author} {\bibfnamefont {Olivier}\ \bibnamefont
  {Parcollet}}, \bibinfo {author} {\bibfnamefont {Michel}\ \bibnamefont
  {Ferrero}}, \bibinfo {author} {\bibfnamefont {Thomas}\ \bibnamefont {Ayral}},
  \bibinfo {author} {\bibfnamefont {Hartmut}\ \bibnamefont {Hafermann}},
  \bibinfo {author} {\bibfnamefont {Igor}\ \bibnamefont {Krivenko}}, \bibinfo
  {author} {\bibfnamefont {Laura}\ \bibnamefont {Messio}}, \ and\ \bibinfo
  {author} {\bibfnamefont {Priyanka}\ \bibnamefont {Seth}},\ }\bibfield
  {title} {\enquote {\bibinfo {title} {{TRIQS: A toolbox for research on
  interacting quantum systems}},}\ }\href {\doibase
  http://dx.doi.org/10.1016/j.cpc.2015.04.023} {\bibfield  {journal} {\bibinfo
  {journal} {Comput. Phys. Commun.}\ }\textbf {\bibinfo {volume} {196}},\
  \bibinfo {pages} {398 -- 415} (\bibinfo {year} {2015})}\BibitemShut {NoStop}%
\bibitem [{\citenamefont {{Beach}}(2004)}]{MaxEntBeach}%
  \BibitemOpen
  \bibfield  {author} {\bibinfo {author} {\bibfnamefont {K.~S.~D.}\
  \bibnamefont {{Beach}}},\ }\bibfield  {title} {\enquote {\bibinfo {title}
  {{Identifying the maximum entropy method as a special limit of stochastic
  analytic continuation}},}\ }\href@noop {} {\bibfield  {journal} {\bibinfo
  {journal} {ArXiv e-prints}\ } (\bibinfo {year} {2004})},\ \Eprint
  {http://arxiv.org/abs/cond-mat/0403055} {arXiv:cond-mat/0403055
  [cond-mat.str-el]} \BibitemShut {NoStop}%
\bibitem [{TRI()}]{TRIQS/MAXENT}%
  \BibitemOpen
  \href@noop {} {\enquote {\bibinfo {title} {{TRIQS/maxent package}},}\
  }\bibinfo {howpublished} {\url{https://triqs.github.io/maxent}}\BibitemShut
  {NoStop}%
\bibitem [{\citenamefont {Aichhorn}\ \emph {et~al.}(2016)\citenamefont
  {Aichhorn}, \citenamefont {Pourovskii}, \citenamefont {Seth}, \citenamefont
  {Vildosola}, \citenamefont {Zingl}, \citenamefont {Peil}, \citenamefont
  {Deng}, \citenamefont {Mravlje}, \citenamefont {Kraberger}, \citenamefont
  {Martins}, \citenamefont {Ferrero},\ and\ \citenamefont
  {Parcollet}}]{TRIQS/DFTTOOLS}%
  \BibitemOpen
  \bibfield  {author} {\bibinfo {author} {\bibfnamefont {M.}~\bibnamefont
  {Aichhorn}}, \bibinfo {author} {\bibfnamefont {L.}~\bibnamefont
  {Pourovskii}}, \bibinfo {author} {\bibfnamefont {P.}~\bibnamefont {Seth}},
  \bibinfo {author} {\bibfnamefont {V.}~\bibnamefont {Vildosola}}, \bibinfo
  {author} {\bibfnamefont {M.}~\bibnamefont {Zingl}}, \bibinfo {author}
  {\bibfnamefont {O.~E.}\ \bibnamefont {Peil}}, \bibinfo {author}
  {\bibfnamefont {X.}~\bibnamefont {Deng}}, \bibinfo {author} {\bibfnamefont
  {J.}~\bibnamefont {Mravlje}}, \bibinfo {author} {\bibfnamefont {Gernot~J.}\
  \bibnamefont {Kraberger}}, \bibinfo {author} {\bibfnamefont {Cyril}\
  \bibnamefont {Martins}}, \bibinfo {author} {\bibfnamefont {Michel}\
  \bibnamefont {Ferrero}}, \ and\ \bibinfo {author} {\bibfnamefont {Olivier}\
  \bibnamefont {Parcollet}},\ }\bibfield  {title} {\enquote {\bibinfo {title}
  {{TRIQS/DFTTools: A \{TRIQS\} application for ab initio calculations of
  correlated materials}},}\ }\href {\doibase 10.1016/j.cpc.2016.03.014}
  {\bibfield  {journal} {\bibinfo  {journal} {Comput. Phys. Commun.}\ }\textbf
  {\bibinfo {volume} {204}},\ \bibinfo {pages} {200 -- 208} (\bibinfo {year}
  {2016})}\BibitemShut {NoStop}%
\bibitem [{\citenamefont {Seth}\ \emph {et~al.}(2016)\citenamefont {Seth},
  \citenamefont {Krivenko}, \citenamefont {Ferrero},\ and\ \citenamefont
  {Parcollet}}]{TRIQS/CTHYB}%
  \BibitemOpen
  \bibfield  {author} {\bibinfo {author} {\bibfnamefont {Priyanka}\
  \bibnamefont {Seth}}, \bibinfo {author} {\bibfnamefont {Igor}\ \bibnamefont
  {Krivenko}}, \bibinfo {author} {\bibfnamefont {Michel}\ \bibnamefont
  {Ferrero}}, \ and\ \bibinfo {author} {\bibfnamefont {Olivier}\ \bibnamefont
  {Parcollet}},\ }\bibfield  {title} {\enquote {\bibinfo {title} {{TRIQS/CTHYB:
  A continuous-time quantum Monte Carlo hybridisation expansion solver for
  quantum impurity problems}},}\ }\href
  {http://www.sciencedirect.com/science/article/pii/S001046551500404X}
  {\bibfield  {journal} {\bibinfo  {journal} {Comput. Phys. Commun.}\ }\textbf
  {\bibinfo {volume} {200}},\ \bibinfo {pages} {274 -- 284} (\bibinfo {year}
  {2016})}\BibitemShut {NoStop}%
\bibitem [{\citenamefont {Rozbicki}\ \emph {et~al.}(2011)\citenamefont
  {Rozbicki}, \citenamefont {Annett}, \citenamefont {Souquet},\ and\
  \citenamefont {Mackenzie}}]{Rozbicki2011b}%
  \BibitemOpen
  \bibfield  {author} {\bibinfo {author} {\bibfnamefont {Emil~J.}\ \bibnamefont
  {Rozbicki}}, \bibinfo {author} {\bibfnamefont {James~F.}\ \bibnamefont
  {Annett}}, \bibinfo {author} {\bibfnamefont {Jean-Ren{\'e}}\ \bibnamefont
  {Souquet}}, \ and\ \bibinfo {author} {\bibfnamefont {Andrew~P.}\ \bibnamefont
  {Mackenzie}},\ }\bibfield  {title} {\enquote {\bibinfo {title} {{Spin--orbit
  coupling and $k$-dependent Zeeman splitting in strontium ruthenate}},}\
  }\href {http://stacks.iop.org/0953-8984/23/i=9/a=094201} {\bibfield
  {journal} {\bibinfo  {journal} {J. Phys.: Condens. Matter}\ }\textbf
  {\bibinfo {volume} {23}},\ \bibinfo {pages} {094201} (\bibinfo {year}
  {2011})}\BibitemShut {NoStop}%
\end{thebibliography}

%

\end{document}